\documentclass[a4paper, fleqn, usenatbib]{mnras}

\usepackage{ae} 
\usepackage{aecompl}
\usepackage{afterpage}
\usepackage{amsfonts}
\usepackage{amsmath}
\usepackage{amssymb}
\usepackage{array}
\usepackage{bm} 
\usepackage{cancel}
\usepackage[usenames, dvipsnames]{color}
\usepackage{cuted}
\usepackage{dcolumn} 
\usepackage{epsfig}
\usepackage{float}
\usepackage[T1]{fontenc}
\usepackage{geometry}
\usepackage{graphicx} 
\usepackage{hyperref}
\usepackage{lipsum}
\usepackage{mathtools}
\usepackage{multicol}
\usepackage{newtxmath}
\usepackage{newtxtext}
\usepackage{pdflscape}
\usepackage{placeins}
\usepackage{rotating}
\usepackage{soul}
\usepackage{subfigure}
\usepackage{tikz}
	\usetikzlibrary{arrows, shapes}

\newcommand{\stkout}[1]{\ifmmode\text{\sout{\ensuremath{#1}}}\else\sout{#1}\fi}

\title[Pulsar Beam Evolution near a Kerr  Black Hole]{Gravitomagnetism and Pulsar Beam Precession near a Kerr Black Hole}

\author[Kocherlakota et al.]{
Prashant Kocherlakota,$^1$ \thanks{E-mail: k.prashant@tifr.res.in}
Pankaj S Joshi,$^{1,2}$ \thanks{E-mail: psj@tifr.res.in}
Sudip Bhattacharyya,$^1$ \thanks{E-mail: sudip@tifr.res.in} 
\newauthor
Chandrachur Chakraborty,$^3$ \thanks{E-mail: chandra@pku.edu.cn}
Alak Ray$^{1,4}$ \thanks{E-mail: akr@tifr.res.in}
and Sounak Biswas$^5$ 
\\
$^1$Department of Astronomy and Astrophysics, Tata Institute of Fundamental Research, Mumbai 400005, India\\ 
$^2$International Center for Cosmology, Charusat University, Anand 388421, India\\ 
$^3$Kavli Institute for Astronomy and Astrophysics, Peking University, Beijing 100871, China\\
$^4$Homi Bhabha Centre for Science Education, Tata Institute of Fundamental Research, Mumbai 400088, India\\ 
$^5$Department of Theoretical Physics, Tata Institute of Fundamental Research, Mumbai 400005, India}

\date{Accepted XXX. Received YYY; in original form ZZZ}

\pubyear{2018}

\begin{document}
\label{firstpage}
\pagerange{\pageref{firstpage}--\pageref{lastpage}}
\maketitle

\begin{abstract}
A rotating black hole causes the spin-axis of a nearby pulsar to precess due to geodetic and gravitomagnetic frame-dragging effects. The aim of our theoretical work here is to explore how this spin-precession can modify the rate at which pulses are received on earth. Towards this end, we obtain the complete evolution of the beam vectors of pulsars moving on equatorial circular orbits in the Kerr spacetime, relative to asymptotic fixed observers. We proceed to establish that such spin-precession effects can significantly modify observed pulse frequencies and, in specific, we find that the observed pulse frequency rises sharply as the orbit shrinks, potentially providing a new way to locate horizons of Kerr black holes, even if observed for a very short time period. We also discuss implications for detections of sub-millisecond pulsars, pulsar nulling, quasi-periodic oscillations, multiply-peaked pulsar Fourier profiles and how Kerr black holes can potentially be distinguished from naked singularities.
\end{abstract}



\section{Introduction}
A neutron star (or white dwarf) that emits a beam of electromagnetic radiation and spins at remarkably precise, stable frequencies is typically called a pulsar since this radiation can only be observed on earth when the beam of emission crosses our line of sight, causing them to appear to be pulsating. Since their discovery \citep{Hewish+68}, they have served as important astrophysical tools in the study of a wide range of fundamental physics and boast an impressive r{\'e}sum{\'e} \citep{Lorimer08}. The first indirect confirmation of gravitational radiation \citep{Taylor_Weisberg82, Taylor_Weisberg89}, the discovery of the first extra-solar planetary system \citep{Wolszczan_Frail92} and the first detection of gas in a globular cluster \citep{Freire+01} are some of the successes of pulsar physics. The detection of a highly relativistic double neutron star binary system \citep{Burgay+03} played a crucial role in predicting the success of gravitational wave detectors by providing a reliable estimate of the merger rate of such systems.

An active pursuit in astronomy has been to locate pulsars in the vicinity of black holes (BHs). The interest in such systems is fuelled by the hope that by studying deviations in the observed frequency of pulses due to strong gravitational fields, one could study the properties of BHs and eventually even test general relativity (GR) in these strong-curvature regimes \citep{Kramer+04}. For example, we know from monitoring the motion of stars in the central region of our Galaxy that a large amount of mass, roughly about $4.2 \times 10^6 M_\odot$ \citep{Ghez+08, Gillessen+09}, is enclosed within a volume of radius smaller than $0.01$ pc (see \citealt{Eckart_Genzel97, Ghez+98}), and it is widely believed that our Galaxy houses a massive compact object Sgr$A^\star$ at its centre \citep{Schodel+02, Ghez+08, Gillessen+09}, possibly a supermassive black hole; further, we have recently identified a magnetar (neutron stars with large magnetic fields and higher rotation frequencies) SGR J1745--2900 in the proximity of Sgr$A^\star$ \citep{Bower+15}. And, this detection has yielded rich dividends: it offers an unparalleled tool for probing the ionized interstellar medium toward the galactic center \citep{Eatough+13} and a possible avenue with which to test quantum gravity effects \citep{Pen_Broderick14}.

Furthermore, one of the desired objectives of X-ray and gamma-ray space telescopes like \textit{Fermi} \citep{Atwood+09}, and of large radio telescopes such as the upcoming Square Kilometre Array (\href{https://www.skatelescope.org/books/}{SKA}; \citealt{Braun+14, Konar+16}) and the Five-hundred-meter Aperture Spherical radio Telescope (FAST, \citealt{Nan06, Nan+11}) has been to detect more such pulsar-BH systems. This is true also of gravitational wave observatories like the Laser Interferometer Gravitational-Wave Observatory (\href{https://www.lsc-group.phys.uwm.edu/ppcomm/Papers.html}{LIGO}; \citealt{Abbott+16}) and the Virgo interferometer \citep{Accadia+12}, and the next generation Laser Interferometric Space Antenna (\href{https://www.lisamission.org/news/Papers}{LISA}; see for example \citealt{Amaro-Seoane+07}). We discuss where one may find such systems in later sections.

A pulsar, which can be treated effectively as a test spinning particle near a sufficiently massive BH (for black hole masses of about $10^2 M_\odot$ or greater; see for example \citealt{Singh+14}), is affected in characteristically disctinct ways depending on whether or not the BH posseses angular momentum. For example, particle orbits precess due to geodetic precession around a non-spinning Schwarzschild BH \citep{deSitter16}, and the rate of the advance of the periapsis depends only on the mass of the central black hole. The precession of particle orbits near a spinning Kerr BH was studied by \cite{Bardeen+72}, and it was found there that an additional Lense-Thirring precession \citep{Lense_Thirring18} piece arises due to the coupling of the orbital angular momentum of the particle with the angular momentum of the Kerr BH. These orbit-precession effects would be experienced by both spinning and non-spinning test particles, and can be neatly thought of as causing the rotation of the associated Laplace-Runge-Lenz vector of the test object (see for example \citealt{Brill_Goel99}). The Lense-Thirring or gravitomagnetic frame-dragging precession of orbits has been explored in a variety of astrophysically important contexts including for accretion disk matter \citep{Bardeen_Petterson75, Chakraborty_Bhattacharyya17, Banerjee+19} and for pulsars present in the vicinity of spinning BHs \citep{Kramer98, Weisberg_Taylor02}. 

In addition to these orbit-precession effects, the spin-axis of a test spinning object could precess, relative to a fixed observer at infinity. This is evident from the evolution of its intrinsic spin angular momentum, which is governed by the Fermi-Walker transport law (\citealt{Fermi22, Walker35}; see also \citealt{Ciufolini_Wheeler95}). In our work here, we will focus on this particular curious effect of \textit{spin}-precession on observations of pulsars in binaries with massive BHs of masses $\gtrsim 10^2M_\odot$, on \textit{circular} orbits. Spin-precession experienced by test spinning objects present near a Schwarzschild BH was studied by \cite{Sakina_Chiba79} and others, and it was found that when these objects remain at fixed spatial locations, they do not experience these effects. However, when pulsars move through the curved space (on geodesics) near a Schwarzschild BH, they experience non-zero spin-precession. Therefore, spin-precession in a static spacetime is a kinetic effect, and this geodetic spin-precession depends only on the mass of the central non-spinning BH \citep{deSitter16} and the properties of motion of the pulsar. However, when in the vicinity of a (rotating) Kerr black hole, even pulsars that remain spatially-fixed experience non-zero spin-precession \citep{Iyer_Vishveshwara93}. This is due to the intrinsic gravitomagnetic frame-dragging effects, characteristic of a spinning central mass, that are encoded into the metric of a stationary spacetime. Essentially, the intrinsic spin angular momentum of the pulsar couples to the gravitomagnetic field associated with the non-zero spin of the central BH \citep{Damour+08, Barausse+09, Bini+17}; for a discussion on gravitomagnetism, one can see \cite{Mashhoon+01}, \cite{Clark_Tucker00} or \cite{Natario07}. 

\cite{Schiff60a, Schiff60b} proposed searching for these characteristic general relativistic couplings by measuring the precession properties of a gyroscope in earth's gravitational field. This search was successfully conducted by Gravity Probe B (GPB, \citealt{Everitt+15}) and data from four gyroscopes of the GPB mission measured (one sigma results) both a geodetic and a gravitomagnetic frame-dragging frequency of 6601.8 $\pm$ 18.3 marc-s yr$^{-1}$ and 37.2 $\pm$ 7.2 marc-s yr$^{-1}$ respectively (1 marc-s = $4.848 \times 10^{-9}$ radians). To compare, the GR predicted values are 6606.1 marc-s yr$^{-1}$ and 39.2 marc-s yr$^{-1}$. In recent decades, precession experienced by both isolated pulsars due to their internal structure \citep{Stairs+00}, and those in binaries has been explored \citep{Stairs+04, Clifton_Weisberg08}. However, the effect of gravitomagnetic spin-precession in pulsar-BH systems on pulsar beam profiles has not been sufficiently well characterized, and this forms the focus of the current article. This problem is complicated by difficulties in (a) evaluating the precession frequency experienced by pulsars on arbitrary time-like orbits, and in (b) relating statements made in the pulsar frame to the frame of an asymptotic spatially-fixed observer. The spin-precession frequency, which is the rate at which the intrinsic spin-angular momentum or the spin-axis of a pulsar precesses, depends on both the characteristics of the central object, like its mass and angular momentum, and the properties of motion of the pulsar, like its distance from the central collapsed object and its four-velocity. Employing a completely covariant formalism, \cite{Iyer_Vishveshwara93} calculated the spin-precession frequency experienced by gyroscopes moving on Killing orbits in arbitrary stationary spacetimes, which will be the orbits of choice in the present study, as we discuss below. Major leaps in overcoming both of the hurdles mentioned above have come recently in two seminal papers by \cite{Bini+16, Bini+17}, where gyroscopic precession along unbound equatorial plane orbits and along general timelike geodesics in the Kerr spacetime respectively was analysed.

We model a pulsar here via the standard lighthouse model, i.e. the pulsar spins around its own axis with a conserved intrinsic spin-angular momentum, at a constant spin angular frequency. We assume that this angular frequency doesn't change considerably even when the pulsar enters strong gravitational fields. The direction in which the beam of radiation is emitted rotates around its spin-axis at a fixed angle, at this frequency. The large mass ratio between intermediate-mass BHs ($10^2\!\!-\!\!10^4 M_\odot$; IMBHs) or super-massive BHs ($10^5\!\!-\!\!10^9 M_\odot$; SMBHs) and pulsars ($\sim\!\! 1.5 M_\odot$; see for example \citealt{Lorimer08, Steiner+10, Lattimer11}) allows for a test spinning particle approximation for the pulsars. We will discuss this in some more detail in Section \ref{sec:PulsarPrecession_Static}. Further, here we neglect spin-curvature coupling, which causes deviations in the world-lines of spinning particles from the world-lines of non-spinning particles \citep{Wald72, Barker_OConnell79}. That is, we model pulsar motion by the motion of a test particle in the Kerr spacetime and simply account for the evolution of its intrinsic spin angular momentum along its world-line via the Fermi-Walker transport law. This enables us to bypass solving the full Mathisson-Papapetrou-Dixon equations \citep{Mathisson37, Papapetrou51, Dixon74} that govern the dynamics of spinning objects in general relativity, and this effect will be incorporated elsewhere. 

We will not restrict our study to pulsars that only move along geodesics for two reasons: firstly, the magnitude of the precession frequency is larger for accelerating pulsars and it would be astrophysically significant if one could observe such pulsars even for a short period of time, as we argue below. Next, we want to understand the effects of acceleration on their appearance since in realistic astrophysical scenarios there can be various sources of accelerations for pulsars. For example, gravitational waves emitted by a nearby compact binary could cause a pulsar to accelerate. Other causes could be due to interactions with other astrophysical objects via scattering processes in dense regions like globular clusters or active galactic nuclei, or due to the supernova kicks that birth them.

Another important source of acceleration is the aforementioned coupling of the spin angular momentum to the background tidal tensor. This acceleration is proportional to the magnitude of the spin of the test object and to the rate of the fall-off of the strength of the gravitational field across it, which is captured by the tidal tensor $R^{\mu}_{\ \nu\rho\sigma}$. \cite{Singh+14} demonstrated recently that the motion of a millisecond (ms)-pulsar (which have spins of $\approx\!\! .3787\text{m}^2$, using a spherical model for the pulsar) present close to a BH ($r \lesssim 50 M_{\text{BH}}$), with mass in the range $M_{\text{BH}} \approx \! 10^3-10^6 M_\odot$, exhibits deviations from geodesic motion due to the acceleration arising from spin-curvature coupling. Strikingly, it is found there that the motion of the pulsar becomes non-planar due to spin-curvature coupling, and the pulsar leaves the $x \! - \! y$ plane. They show how the complexity of the orbit increases with increase in spin and with decrease in mass of the BH. Notably, the `$z$-signal' is potentially observable through variation in pulse arrival time over arrival times that would result if the pulsar stayed in a planar orbit. They argue that it is reasonable to expect the path length of the ray from the pulsar to the Earth to vary by $\pm 5\ $km from the path length of a ray from a pulsar in an otherwise similar planar orbit, which translates to a timing change estimate of the order of $\pm 10 \mu s$. 
In a subsequent study \citep{Li+19}, the orbit-precession of these ms-pulsars (non-zero spin-curvature coupling) was analysed and the usual de Sitter and Lense-Thirring pieces were obtained. Additional contributions arising from spin-curvature coupling and the coupling of the pulsar's spin to its orbital angular momentum (see for example eq. 18 therein) were also reported, a consequence of which is that orbit-precession is enhanced. Further, spin-precession experienced by such a pulsar was also carefully analysed numerically, 
and it is shown that the spin-precession frequency due to spin-orbit and spin-spin coupling decays with an increase in orbit-radius of the pulsar as $\propto M_{\text{BH}}/r$. In particular, they point out that these couplings cannot be ignored for systems that qualify as intermediate-mass-ratio binaries i.e., $M_{\text{BH}}/m_{\text{p}} \sim 10^2 - 10^3$, with $m_{\text{p}}$ being the mass of the pulsar. The upshot of their findings are that spin-precession causes the times of arrival and widths of pulses to change. For example, for a ms-pulsar present about $20M_{\text{BH}}$ away from a Schwarzschild BH, with mass in the range $10^3 \! - \! 10^6 M_\odot$, shifts in the times of arrival of pulses accumulate to about $2.5\ \mu$s every $\sim \!\! 35$s.

On a related note, considering the effect of spin-curvature on the dynamics of a spinning object to be negligible (as is done in the present work) is an excellent approximation for pulsars that are sufficiently slowly-spinning \citep{Ehlers_Geroch04} or when they are present in regions where the background tidal tensor is very small. It is useful to remember that the tidal tensor decays with increase in distance from and mass of the BH. Therefore, our calculations apply very well for normal pulsars around massive BHs, and even for ms-pulsars when they are either about $r \gtrsim 50 M$ away from the BH or if they are near a BH with mass larger than about $10^6 M_\odot$ \citep{Singh+14, Li+19}.

Now, here we consider pulsars to be moving along the world-lines of equatorial Killing observers in the Kerr spacetime. The trends in the accelerations experienced by such observers, with changes in orbit-radius, BH parameters etc. is carefully discussed in Appendix \ref{sec:Reversals}. Further, the precession frequencies experienced by spinning objects on these orbits do not vary with time, which greatly simplifies our calculations. Therefore, the observers we will be interested in are: (a) \textit{static Killing observers}: observers whose spatial position remains unchanged over time, and (b) \textit{stationary Killing observers}: observers that move on equatorial circular orbits. Although the former class of static observers form a proper subset of the latter class of stationary observers, it will be useful for our purposes to demarcate the two. Further, the class of stationary orbits contains the astrophysically important set of equatorial circular geodesics. Now, we find here, from a purely analytical calculation, that as the spin-axis of a pulsar precesses, it pulls the beam vector along with it, and this leads to a modification in the pulse frequency as seen from the earth. Thus, due to stationary gravitational fields, this frequency is no longer simply equal to the intrinsic spin angular frequency of the pulsar about its axis. In our concluding section, we point out how this is a promising first step in this new story of pulsars and black holes involving spin-precession, and discuss more physically interesting extensions. 

Apart from the few we have already mentioned, there have been other significant initiatives to advance our understanding about strong field gravity effects in recent years like the event horizon telescope (\href{https://eventhorizontelescope.org/}{EHT}; \citealt{Doeleman+08, Doeleman+12}), which aims to probe the physics of very strong gravity regions near the event horizon. Notably, EHT reported constructing the first event-horizon-scale image of the supermassive BH candidate thought to be present at the center of the M87 galaxy \citep{Akiyama+19}. From such a perspective, we investigate here effects of gravitomagnetism on observed pulsar periods in the strong field region of the Kerr spacetime and use primarily the strong field results of \cite{Iyer_Vishveshwara93}, \cite{Chakraborty+17a} and \cite{Chakraborty+17b} to obtain the frequency of precession for pulsars located deep inside strong gravitational fields. 

The outline of the paper is as follows. In Section \ref{sec:SpinPrecession}, we briefly outline the physics of  spin-precession in stationary spacetimes. In Section \ref{sec:SpinPrecession_Killing}, we review the properties of the spin-precession frequency for Killing observers in the Kerr spacetime. In Section \ref{sec:AdaptedKerr}, we will discuss the adapted-Kerr spacetime, which is obtained by choosing coordinates co-moving with a particular observer moving on a circular orbit, and demonstrate that one can treat stationary observers in the Kerr spacetime as static observers in the adapted-Kerr spacetime. We will argue that this simplifies the problem of obtaining the beam evolution for stationary observers relative to asymptotic fixed observers. In Section \ref{sec:PulsarPrecession_Static}, we work out the complete time evolution of the beam vector for static pulsars and obtain their observed pulse frequencies, without approximation. In Section \ref{sec:PulsarPrecession_Stationary}, we extend this analysis to pulsars moving on equatorial circular orbits around the central object at constant angular speeds. In Section \ref{sec:Results} and Section \ref{sec:Astrophysical_Implications}, we explore the astrophysical consequences of our work and conclude with a quick summary and a few comments on possible future extensions. 

Conventions: Greek and Latin indices (with the exception of $i, j, k$) run from 0--3. $i, j, k$ take values 1--3. Hatted indices represent components projected onto a tetrad. In Section \ref{sec:PulsarPrecession_Static}, we will switch to the Euclidean three-vector notation since all entities will be calculated relative to a spatial triad, i.e. $a^i \rightarrow \vec{a}$. Also, when $b^i$ denotes a unit three-vector, we shall write $b^i \rightarrow \hat{b}$. 

\section{Spin-Precession: Physics \& Formalism} \label{sec:SpinPrecession}
The evolution of the intrinsic spin-angular momentum $\mathcal{S}$ of a test spinning object, with negligible spin-curvature interaction, that moves with a time-like four-velocity $u$, is given by the Fermi-Walker (FW) transport law (\citealt{Walker35}; also see for example \citealt{Misner+73}),
\begin{equation} \label{eq:FW_S}
\mathbb{F}_u \mathcal{S} = 0,
\end{equation}
where in the above we have introduced the Fermi derivative of a vector field $X$ along $u$ as \citep{Fermi22},
\begin{equation} \label{eq:FermiDerivative}
\mathbb{F}_u X = \nabla_u X - (X \! \cdot \! \alpha) u + (X \! \cdot \! u) \alpha,
\end{equation}
and in the above, $\alpha = \nabla_u u$ is the four-acceleration of the spinning object and ($\cdot$) represents the inner product. It is clear that along a geodesic i.e. for $\alpha = 0$, Fermi-Walker transport reduces to parallel transport.

The above system of equations (\ref{eq:FW_S}) is not closed and one is required to provide additional algebraic constraints so that given an initial state of the spin four-vector, one can determine its intantaneous state at any later time. Several supplementary conditions have been used for this purpose, and for a detailed discussion on the necessity of such conditions and their physical content one can see \cite{Semerak99}, \cite{Kyrian_Semerak07} or \cite{Costa+16}. Here we shall exclusively use the Pirani condition \citep{Pirani56}, 
\begin{equation} \label{eq:Pirani}
\mathcal{S} \cdot u = 0,
\end{equation}
that is, the spin four-vector will have support only in directions transverse to the four-velocity $u$ of the spinning object. 

The advantage of employing the Pirani condition (\ref{eq:Pirani}) is that we can now restrict to analysing the evolution of the space-like spin three-vector $S$. This is best accomplished by constructing an orthonormal frame spanned by three space-like vectors, which are also orthogonal to the four-velocity $u$, all along the world-line of the spinning object and projecting the evolution equation (\ref{eq:FW_S}) onto this triad. The Frenet-Serret (FS) tetrad comprises of precisely such a set of four orthonormal vectors, three spacelike with the timelike leg being the four-velocity of the spinning object. The FS tetrad is also one of the most natural frames associated with a given curve because it is invariant under reparametrization and captures inherent differential geometric properties of the curve, namely its generalized curvatures, which play a fundamental role in the analysis of the evolution of their spin vectors, as we shall discuss below. 

Furthermore, since in the present paper we will consider only the evolution of spin vectors carried by Killing observers of the Kerr spacetime, it is important to note that this choice of tetrad is particularly convenient: it was pointed out by \cite{Iyer_Vishveshwara93} that for Killing observers in arbitrary stationary spacetimes the associated FS tetrad and generalised curvatures are all time-independent, simplifying our analysis greatly. Furthermore, in Section \ref{sec:SpinPrecession_Killing}, after a review of the properties of Killing observers of the Kerr spacetime, we will discuss also why our choice to study pulse profiles of pulsar moving on Killing orbits is of fundamental physical importance.

The FS tetrad attached to an observer that moves along an arbitrary time-like world-line comprises of a set of four orthonormal vector fields $\{e_{\hat{\alpha}}, (\hat{\alpha} \! = \! 0 \! - \! 3)\}$ and is constructed as follows. The timelike leg is simply defined to be the four-velocity along the world-line, $e_{\hat{0}} = u$. Next, $e_{\hat{1}}$ will be defined to be the normal to the curve and to find it, we introduce the directional derivative along the four-velocity, denoted by an overdot, $\dot{} = \nabla_u = d/d\tau$. That is,
\begin{equation}
\dot{e}_{\hat{0}} = \kappa e_{\hat{1}},
\end{equation}
where $\kappa = (\dot{e}_{\hat{0}} \! \cdot \! e_{\hat{1}})$ measures the curvature of the world-line relative to the osculating plane spanned by $e_{\hat{0}}$ and $e_{\hat{1}}$.  Note that since $e_{\hat{0}}$ is normalised, $\dot{e}_{\hat{0}}$, and hence $e_{\hat{1}}$, is orthogonal to it. This is clear if we recognise that $\alpha = \kappa e_{\hat{1}}$ is simply the four-acceleration of the observer, and that $\nabla_u (u.u) = 0 \Rightarrow \alpha.u = 0$. Now, we turn to $\dot{e}_{\hat{1}}$. This vector will be a linear combination of $e_{\hat{0}}, e_{\hat{1}}$ and a unit vector $e_{\hat{2}}$ orthogonal to the osculating plane. Yet again, since $e_{\hat{1}}$ has unit-norm, we can write,
\begin{equation}
\dot{e}_{\hat{1}} = \kappa e_{\hat{0}} + \sigma_1 e_{\hat{2}},
\end{equation}
where $\sigma_1 = 	(\dot{e}_{\hat{1}} \! \cdot \! e_{\hat{2}})$ is called the first torsion. 
Proceeding similarly to the above, the following picture emerges. The tetrad legs $\{e_{\hat{\alpha}}, (\hat{\alpha} \! = \! 0 \! - \! 3)\}$ are respectively the tangent, normal, binormal and trinormal vectors along the world-line of the observer and satisfy,
\begin{equation} \label{eq:FSTetrad_Orthonormality}
(e_{\hat{\alpha}} \! \cdot \! e_{\hat{\beta}}) = \eta_{\hat{\alpha}\hat{\beta}}.
\end{equation}
In terms of the generalized curvatures ($\kappa, \sigma_1, \sigma_2$), the evolution equations of the tetrad legs along the world-line can be written out succinctly as,
\begin{equation} \label{eq:FS_EvolutionEquations}
\begin{bmatrix}
\dot{e}_{\hat{0}} \\
\dot{e}_{\hat{1}} \\
\dot{e}_{\hat{2}} \\
\dot{e}_{\hat{3}}
\end{bmatrix} =
\begin{bmatrix}
0 & \kappa & 0 & 0 \\
\kappa & 0 & \sigma_1 & 0 \\
0 & -\sigma_1 & 0 & \sigma_2 \\
0 & 0 & -\sigma_2 & 0
\end{bmatrix}
\begin{bmatrix}
e_{\hat{0}} \\
e_{\hat{1}} \\
e_{\hat{2}} \\
e_{\hat{3}}
\end{bmatrix} ,
\end{equation}
where in the above we introduced $\sigma_2 = (\dot{e}_{\hat{2}} \! \cdot \! e_{\hat{3}})$ the second torsion. $\sigma_1, \sigma_2$ measure the deviation of the world-line from being a planar curve restricted to the osculating plane. The curvature $\kappa$ has been identified as being the particle acceleration and the two torsions are directly related to spin-precession, as we shall see below.

Since we want to analyse the evolution of the spin vector in this tetrad, and we know that it is FW-transported along its world-line (\ref{eq:FW_S}), we now study the FW-transport of the FS tetrad. We obtain immediately that (see \citealt{Straumann09}),
\begin{equation} \label{eq:FW_FSFrame}
\mathbb{F}_u e_{\hat{\alpha}} = \omega_{\hat{\alpha}}^{\ {\hat{\beta}}}e_{\hat{\beta}},
\end{equation}
where $\omega_{\hat{\alpha}}^{\ {\hat{\beta}}}$ is given as,
\begin{equation} \label{eq:Vorticity4Tensor}
\omega_{\hat{\alpha}}^{\ \hat{\beta}} = 
\begin{bmatrix}
0 & 0 & 0 & 0 \\
0 & 0 & \sigma_1 & 0 \\
0 & -\sigma_1 & 0 & \sigma_2 \\
0 & 0 & -\sigma_2 & 0
\end{bmatrix} ,
\end{equation}
The FW-transport of the FS tetrad (\ref{eq:FW_FSFrame}) along with the equation of motion of the spin vector (\ref{eq:FW_S}) imply that the spin three-vector $S = S^{\hat{i}}e_{\hat{i}}$ satisfies%
\footnote{$\mathbb{F}_u S^{\hat{i}} = \dot{S}^{\hat{i}}$. See Section 2.10.3 of \cite{Straumann09} for a quick review on the properties of the Fermi derivative.},
\begin{equation}
\dot{S}^{\hat{i}}e_{\hat{i}} + S^{\hat{j}}\mathbb{F}_u e_{\hat{j}} = 0,
\end{equation}
and we have,
\begin{equation} \label{eq:Spin_EoM1}
\dot{S}^{\hat{i}} = \omega^{\hat{i}}_{\ \hat{j}}S^{\hat{j}}.
\end{equation}
That is, the spin of the gyroscope precesses relative to the FS spatial triad $\{e_{\hat{i}}, \hat{i} \! = \! 1, 2, 3\}$ with an angular velocity $\Omega_{\text{p}}$,
\begin{equation} \label{eq:Vorticity_Precession}
\omega_{\hat{i}\hat{j}} = \epsilon_{\hat{i}\hat{j}\hat{k}}\Omega^{\hat{k}}_{\text{p}},
\end{equation}
and we can write,
\begin{equation} \label{eq:Omega_p}
\Omega_{\text{p}} = -(\sigma_2 e_{\hat{1}} + \sigma_1 e_{\hat{3}}),
\end{equation}
when the FS spatial triad is right-handed \citep{Iyer_Vishveshwara93}. Then, employing standard three-dimensional vector notation, in the \textit{Euclidean} FS triad, we can write (\ref{eq:Spin_EoM1}) as,
\begin{equation} \label{eq:Spin_EoM}
\dot{\vec{S}} = \vec{S}\times\vec{\Omega}_{\text{p}}.
\end{equation}
In the above discussion, we have not imposed any restrictions on $u$ barring that it be timelike. In the rest of the paper, as mentioned above, we shall restrict $u$ to be tangent to arbitrary Killing orbits of the Kerr spacetime and we now move to a discussion of the precession properties of these observers. 

\section{Spin-Precession for Killing Observers in the Kerr Spacetime} \label{sec:SpinPrecession_Killing}
The Kerr metric in the standard Boyer-Lindquist (BL) coordinates $x^\mu = (t, r, \theta, \phi)$ is given as,
\begin{align} \label{eq:KerrMetricBL} 
ds^2 =& -\left(1- \frac{2 M r}{\rho^2}\right)dt^2 -  \frac{4Mar\sin^2\theta}{\rho^2}dtd\phi + \frac{\Pi\sin^2\theta}{\rho^2} d\phi^2 \nonumber \\
& \ \ \ \ + \frac{\rho^2}{\Delta}dr^2 + \rho^2 d\theta^2 \\
=& g_{00}dt^2 + 2g_{03}dt~d\phi + g_{33}d\phi^2 + g_{11}dr^2 + g_{22}d\theta^2. \nonumber
\end{align}
Here we have used geometrized units ($G \! = \! c \! = \! 1$) and $a = J/M$ is the specific angular momentum. $J, M$ are the angular momentum and mass of the Kerr collapsed object respectively, with $\Delta=r^2-2Mr+a^2$, $\rho^2=r^2+a^2 \cos^2\theta$ and $\Pi = (r^2 + a^2)^2 -a^2 \Delta\sin^2\theta$.

The Kerr spacetime possesses a time-like Killing vector, $\xi = \partial_0$, corresponding to time-translational invariance, and a space-like Killing vector, $\psi = \partial_3$, corresponding to azimuthal rotational invariance. A constant-coefficient linear combination of these two is the most general Killing vector, $\xi^\prime  = \xi + \Omega~\psi$. The time-translational Killing vector becomes null on the ergosurface of the Kerr spacetime, which is characterised by $g_{00} = 0$, i.e. on the ergosurface, $\xi^2 = (\xi \cdot \xi) \! = \! g_{00} = 0$. Similarly, it can be shown that $\xi^{\prime 2} = 0$ is the location of the event horizon. The locations of the ergosurface $r=r_+$ and the event horizon $r=r_{\text{H}}$ are given respectively as,
\begin{equation} \label{eq:Ergo_Horizon}
r_+ = M + \sqrt{M^2 - a^2\cos^2\theta}, r_{\text{H}} = M + \sqrt{M^2 - a^2}.
\end{equation}
In a stationary spacetime, observers that remain fixed spatially with respect to infinity are called static Killing observers \citep{Misner+73} and their four-velocities are given as,
\begin{equation} \label{eq:StaticObserver}
u = \frac{\xi}{\sqrt{-\xi \cdot \xi}}.
\end{equation}
Since $g_{00} < 0$ outside the ergosurface, %
$u$ is time-like only outside it. Observers that move on circular orbits around the axis of the black hole at fixed angular speeds $\Omega$ in planes parallel to the equatorial plane are called stationary Killing observers and their four-velocities are given by,
\begin{equation} \label{eq:StationaryObserver}
u^\prime = \frac{\xi^\prime}{\sqrt{-\xi^\prime \cdot \xi^\prime}}.
\end{equation}
Since we will only be interested in timelike stationary observers, we require $(\xi^\prime \cdot \xi^\prime) < 0$, which has the consequence that $\Omega_- < \Omega < \Omega_+$,
\begin{equation} \label{eq:Omegapm}
\Omega_\pm = \frac{-g_{03} \pm \sqrt{g_{03}^2 - g_{00}g_{33}}}{g_{33}} = \frac{2 M a r\sin\theta \pm \rho^2\sqrt{\Delta}}{\Pi\sin\theta}.
\end{equation}
For $\Delta \geq 0$, we require the radial coordinate to satisfy $r \geq r_{\text{H}}$, i.e. stationary Killing observers exist only outside the horizon. Clearly, static observers form the subclass of stationary observers satisfying $\Omega = 0$. Also, we note that stationary observers with $\Omega = \Omega_\pm$ are stationary null observers.

\cite{Iyer_Vishveshwara93} explored the relation between the FW spin-precession frequency experienced by a static Killing observer and the vorticity of the static Killing congruence, which characterizes the local rotation of nearby world lines in the congruence. Remarkably, they found that these two quantities were equal (see \citealt{Straumann09} for a nice demonstration of this statement). Moreover, since this congruence is also rigid, the FS frame associated with static Killing observers of a spacetime acquires the interpretation of being axes-at-rest relative to asymptotic static Killing observers (`fixed stars'). Basically, the projection of a connecting null vector of the static Killing congruence in the FS triad of a static Killing observer is simply a constant vector. This implies that photons shot out at different times along the same direction relative to the FS frame attached to a static Killing observer in a stationary spacetime all reach the same asymptotic observer. Therefore, the legs of such a FS triad are usually called `optical axes.' These axes can be physically constructed by placing three telescopes pointing towards three (orthogonal) non-planar asymptotic spatially-fixed stars. Therefore, measuring the change of the spin vector relative to this triad gives us the change in the spin relative to fixed asymptotic observers (see for example Section  II.C of \citealt{Costa+16}). One can see Section 2 of \cite{Bini+99} for a more general discussion on the relation between the FW spin-precession frequency along arbitrary time-like observers and the vorticity of their congruence. 

The take-away here is that, in non-static stationary spacetimes, like the Kerr spacetime, the Frenet-Serret frames associated with the congruence of static Killing observers (\ref{eq:StaticObserver}) play a fundamental role in translating physical statements made relative to a particular observer's comoving tetrad frame into the more desirous statements that are made relative to asymptotically spatially-fixed observers. This is of critical importance if, for example, one would like to study the frequency at which the magnetic-axis of a pulsar, present in the vicinity of a Kerr black hole, crosses the line of sight towards earth.

Finally, it is important to note the following. In static spacetimes the precession frequency of static Killing observers identically vanishes $\Omega_{\text{p}} = 0$. This is evident from (\ref{eq:FS_Invariants_AdaptedKerr}) for static Killing observers ($\Omega = 0$) in the Schwarzcshild spacetime ($a=0$), or more explicitly for equatorial static Killing observers ($\theta = \pi/2$) see (\ref{eq:Precession_Frequency_Static}). Also, the congruence of such observers is irrotational. Therefore, if one initially arranged a set of telescopes to point along three non-planar stars and set atop them three gyroscopes that also pointed along these stars, the gyroscopes and telescopes would remain aligned at all times. This does not hold true for non-static stationary spacetimes and the gyroscopes would precess relative to the telescopes.

\begin{figure}
\centering
\includegraphics[scale=.82]{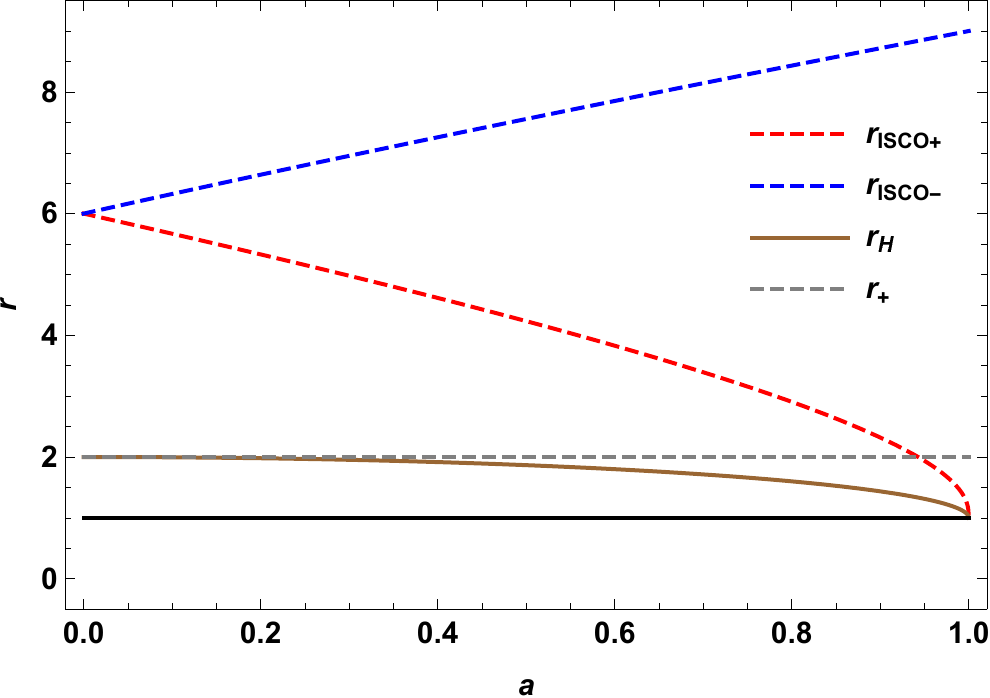}
\caption{We plot here in dashed-red, dashed-blue, brown and dashed-gray respectively the radii of the co-rotating (+) and counter-rotating (-) innermost stable circular orbits (ISCOs) $r_{\text{ISCO}\pm}$ (outside which Kepler observers are time-like and hence allowed), the radius of the horizon $r_{\text{H}}$ and the radius of the ergosurface $r_+$ in the equatorial plane as functions of the spin parameter of the black hole $a$, both in units of $M$. It should be noted that $r_{\text{ISCO}+}$ lies inside the ergoregion for $a > 0.943 M$ (see for example \citealt{Chakraborty+17b}).}
\label{fig:ISCO_Horizon_Ergo}
\end{figure}

We have discussed why the class of static Killing observers is of fundamental importance and a study of the pulse profiles of pulsars moving along such orbits will be presented in Section \ref{sec:PulsarPrecession_Static}. However, from an astrophysical standpoint, it would be even more interesting if one could extend this study to a description of the more general class of stationary Killing observers, and this is discussed in Section \ref{sec:PulsarPrecession_Stationary}. This class of observers contains within it, for example, the set of observers that move on time-like equatorial circular geodesics in the Kerr spacetime, which satisfy,
\begin{equation}
\nabla_{u^\prime} u^\prime = 0.
\end{equation}
This above condition just imposes a constraint on the allowed orbital angular frequencies, and such observers move at Kepler frequencies, $\Omega = \Omega_{\text{K}\pm}$. The $+$ and $-$ signs are associated with co-rotating and counter-rotating equatorial Kepler observers and we have, 
\begin{equation} \label{eq:Kepler_Frequencies}
\Omega_{\text{K}\pm} = \frac{M^{1/2}}{a M^{1/2} \pm r^{3/2}}.
\end{equation}
It is important to note that stable timelike co-rotating and counter-rotating Kepler observers exist only outside the respective innermost stable circular orbits (ISCOs), and the expressions for the ISCO radii can be found in \cite{Bardeen+72}. See Fig. \ref{fig:ISCO_Horizon_Ergo} for how the ISCO radii for co-rotating ($r_{\text{ISCO}+}$) and counter-rotating ($r_{\text{ISCO}-}$) equatorial Kepler observers vary with change in the spin parameter $a$ of the Kerr black hole. As can be seen from the figure the ISCO for the equatorial co-rotating Kepler observers lies inside the ergoregion ($r_{\text{ISCO}+} < 2 M$) for $a > 0.943 M$ (see for example \citealt{Chakraborty+17b}). In this paper, whenever we consider observers moving on equatorial circular geodesics, we will exclusively consider only those that are on stable orbits i.e., those that satisfy $r \geq r_{\text{ISCO}\pm}$ for co-rotating and counter-rotating orbits respectively.

Yet another set of astrophysically important observers are the zero angular momentum observers (ZAMOs). \cite{Bardeen+72} showed that the frame attached to a ZAMO is a powerful tool in the analysis of physical processes near astrophysical objects and these are observers whose world-lines are normal to the $t =$ const. hypersurfaces. They fall within the class of stationary Killing observers and move at angular speeds of,
\begin{equation} \label{eq:ZAMO}
\Omega_{\text{Z}} = \frac{\Omega_+ + \Omega_-}{2} = -\frac{g_{03}}{g_{33}} = \frac{2 M a r}{(r^2 + a^2)^2 - a^2\Delta\sin^2\theta}.
\end{equation}

\begin{figure}
\centering
\includegraphics[scale=.59]{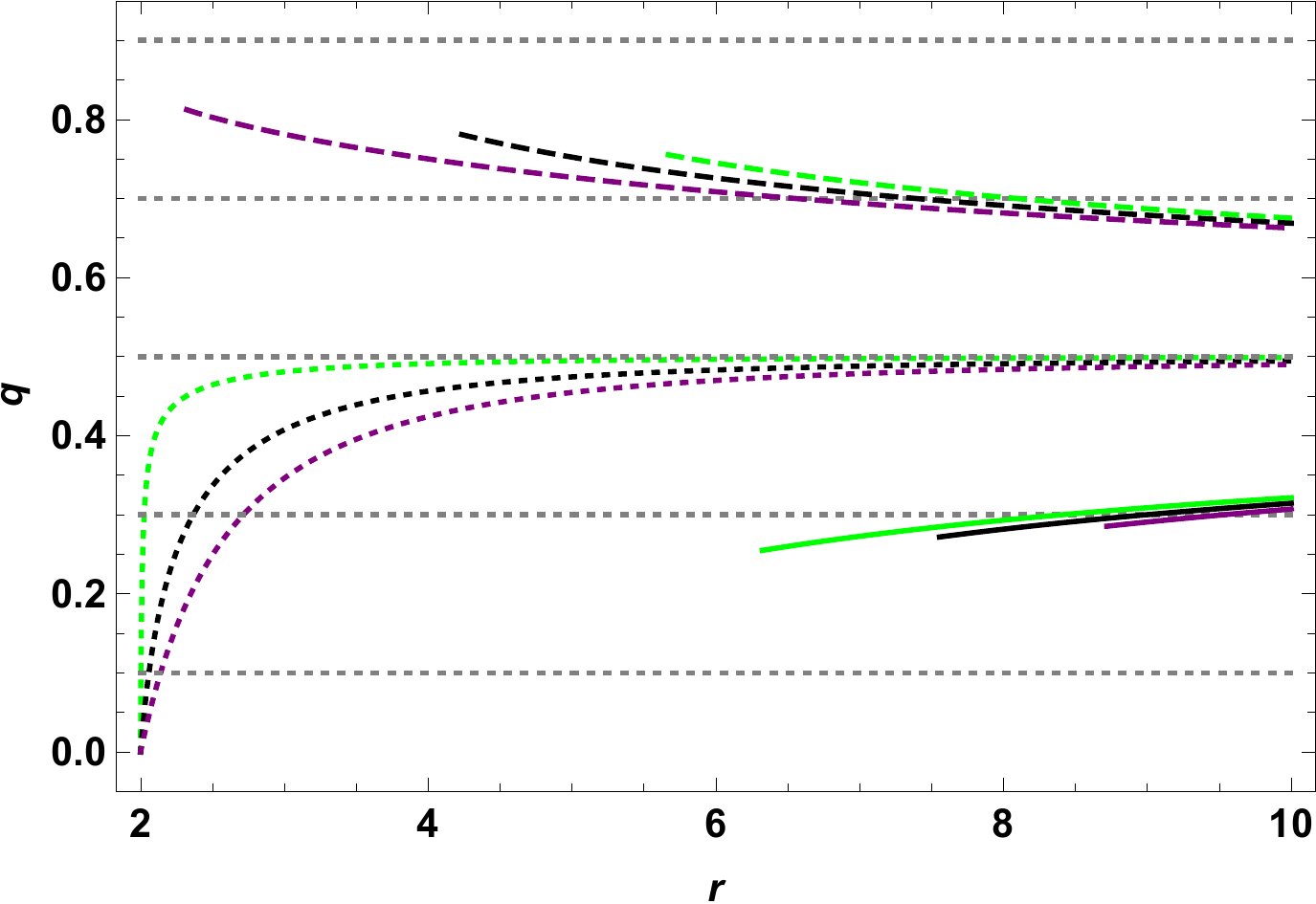}
\caption{Here we plot the values of $q_{\text{static}}$ in dotted lines between $r = 2-10 M$, in regions of strong gravitational fields, and $q_{\text{K}\pm}$ in dashed and solid lines respectively between $r = r_{\text{ISCO}\pm} - 10 M$. The different colors represent black hole spin parameters $a=.1 M, .5 M$ and $.9 M$ in green, black and purple respectively. The dashed-gray lines correspond to the grid-lines $q=.1, .3, .5, .7, .9$. See Fig.\ref{fig:ISCO_Horizon_Ergo} for how the ISCO radii for co-rotating (+) and counter-rotating (-) equatorial Kepler observers $r_{\text{ISCO}\pm}$ vary with the spin parameter $a$.}
\label{fig:qStatic_qKep+-}
\end{figure}

In following sections, we shall parametrize the orbital angular velocity $\Omega$ of the pulsar by $q$ as \citep{Chakraborty+17b},
\begin{equation} \label{eq:q}
q(\Omega) = \frac{\Omega - \Omega_-}{\Omega_+ - \Omega_-},
\end{equation}
with $0 < q < 1$. Let us denote the $q$-values of static observers, ZAMOs and of co-rotating and counter-rotating Kepler observers as $q_{\text{static}}, q_{\text{Z}}$ and $q_{\text{K}\pm}$ respectively. Then,
\begin{equation} \label{eq:Static_ZAMO_Kepler}
q_{\text{static}} = -\frac{\Omega_-}{\Omega_+ - \Omega_-},\ \ q_{\text{Z}} = .5,\ \ q_{\text{K}\pm} = \frac{\Omega_{\text{K}\pm} - \Omega_-}{\Omega_+ - \Omega_-}.
\end{equation}
See Fig. \ref{fig:qStatic_qKep+-} for how $q_{\text{static}}$ and $q_{\text{K}\pm}$ change with radius for different black hole spin parameters $a=.1 M, .5M$ and $.9 M$ denoted in green, black and purple respectively. 

\subsection{The Adapted-Kerr Metric} \label{sec:AdaptedKerr}
Following \cite{Iyer_Vishveshwara93}, we demonstrate below that stationary Killing observers of the Kerr metric are equivalent to static Killing observers in the `adapted-Kerr metric,' which is simply the Kerr metric expressed in coordinates comoving with Kerr stationary Killing observers with a particular orbital angular frequency $\Omega$. We argue therefore that by studying the properties of the precession frequency experienced by adapted-Kerr static Killing observers, one can obtain a complete description of precession effects experienced by the full class of Killing observers of the Kerr spacetime. Another advantage of shifting to the adapted-Kerr spacetime is that one can obtain the evolution of the line-of-sight to earth in the comoving frame of a stationary Killing observer easily, as discussed in Appendix \ref{sec:Connecting_Vector_aFS}.

The relevant change of coordinates from the usual Boyer-Lindquist coordinates $x^\mu$ to those comoving with a Kerr stationary Killing observer with an orbital angular speed of $\Omega$, which we shall call adapted-BL coordinates $x^{\bar{\mu}} = (\bar{t}, \bar{r}, \bar{\theta}, \bar{\phi})$ is given by, 
\begin{equation}  \label{eq:CT}
dt = d\bar{t}, \ d\phi = d\bar{\phi} + \Omega~d\bar{t}.
\end{equation}
$r, \theta$ are left unchanged and the Kerr metric (\ref{eq:KerrMetricBL}) now becomes,
\begin{align} \label{eq:adaptedKerr}
ds^2 =& \left[g_{00} + 2\Omega g_{03} + \Omega^2 g_{33}\right] d\bar{t}^2 + 2\left[g_{03} + \Omega g_{33}\right]d\bar{t}d\bar{\phi} \nonumber \\
& \quad\quad\quad\quad + g_{33}d\bar{\phi}^2 + g_{11}dr^2 + g_{22}d\theta^2, \nonumber \\
= & g_{\bar{0}\bar{0}}d\bar{t}^2 + 2g_{\bar{0}\bar{3}}d\bar{t}~d\bar{\phi} + g_{33}d\bar{\phi}^2 + g_{11}dr^2 + g_{22}d\theta^2.
\end{align}
We shall refer to this metric as the adapted-Kerr metric and the Jacobian associated with this coordinate transformation is,
\begin{equation} \label{eq:Jacobian_BL_aBL}
J^{\bar{\mu}}_{\ \nu} = \frac{dx^{\bar{\mu}}}{dx^\nu} = 
\begin{bmatrix}
1 & 0 & 0 & 0\\
0 & 1 & 0 & 0\\
0 & 0 & 1 & 0\\
-\Omega & 0 & 0 & 1
\end{bmatrix} .
\end{equation}
Under the coordinate transformation, $\xi^{\prime\mu}$ transforms to $\xi^{\prime\bar{\mu}} = J^{\bar{\mu}}_{\ \nu}\xi^{\prime\nu}$ so that,
\begin{equation} \label{eq:aKerr_Static}
\xi^{\prime\bar{\mu}} = (1, 0, 0, 0).
\end{equation}
That is, $\xi^\prime = \frac{\partial}{\partial \bar{t}} = \partial_{\bar{0}}$. Therefore, Kerr stationary observers become adapted-Kerr static observers. Also, Kerr static observers become adapted-Kerr stationary observers, although with the sign of the angular speed reversed. This is of importance because earth, which will naturally be modelled to be an asymptotic static Killing observer in the Kerr spacetime, becomes a stationary Killing observer of the adapted-Kerr metric, i.e., its four-velocity becomes,
\begin{equation}
\xi^{\bar{\mu}} =  \left(1, 0, 0, -\Omega\right).
\end{equation}
A corollary of the above discussion is that in the original Boyer-Lindquist coordinates, $\xi^\prime$ became null-like ($\xi^\prime \! \cdot \! \xi^\prime \! = \! 0$) on the event horizon. Now since we have also $\xi^\prime \! \cdot \! \xi^\prime \! = \! g_{\bar{0}\bar{0}}$, the coordinate transformation (\ref{eq:CT}) maps the event horizon of the Kerr metric onto the ergosurface of the adapted-Kerr metric, as it should.

\subsection{Accelerations and Precession Frequencies along Equatorial Kerr Killing Orbits} \label{sec:Precession_Freq_Statio}
As discussed above, the Frenet-Serret generalised curvature invariants and tetrad associated with Kerr Killing observers can be obtained by treating them as adapted-Kerr static Killing observers (\ref{eq:aKerr_Static}), and from \cite{Iyer_Vishveshwara93} we have,
\begin{align} \label{eq:FS_Invariants_AdaptedKerr}
\bar{\kappa}^2 =& \frac{1}{4 g_{\bar{0}\bar{0}}^2}\left[\frac{g_{\bar{0}\bar{0},1}^2}{g_{11}} + \frac{g_{\bar{0}\bar{0},2}^2}{g_{22}}\right], \\
\bar{\sigma}_1^2 =& - \! \frac{g_{\bar{0}\bar{3}}^2g_{11}g_{22}}{4 \Delta_{03}}\frac{\left[\frac{g_{\bar{0}\bar{0},1}}{g_{11}} \! \left(\frac{g_{\bar{0}\bar{3},1}}{g_{\bar{0}\bar{3}}} \! - \!  \frac{g_{\bar{0}\bar{0},1}}{g_{\bar{0}\bar{0}}}\right) \! + \! \frac{g_{\bar{0}\bar{0},2}}{g_{22}} \! \left(\frac{g_{\bar{0}\bar{3},2}}{g_{\bar{0}\bar{3}}} \! -  \!\frac{g_{\bar{0}\bar{0},2}}{g_{\bar{0}\bar{0}}}\right)\right]^2}{\left[g_{\bar{0}\bar{0},1}^2 g_{22} + g_{\bar{0}\bar{0},2}^2 g_{11}\right]}, \nonumber \\
\bar{\sigma}_2^2 =& -\frac{1}{4 \Delta_{03}}\frac{\left[g_{\bar{0}\bar{0},1}g_{\bar{0}\bar{3},2} - g_{\bar{0}\bar{0},2}g_{\bar{0}\bar{3},1}\right]^2}{\left[g_{\bar{0}\bar{0},1}^2 g_{22} + g_{\bar{0}\bar{0},2}^2 g_{11}\right]}, \nonumber
\end{align}
where we have introduced the determinant 
$\Delta_{03} = g_{\bar{0}\bar{0}}g_{33} - g_{\bar{0}\bar{3}}^2 = -\Delta\sin^2\theta$. Note that $\Delta_{03}$ is the same in both the BL and adapted-BL coordinate systems. Also, the vectors that form the Frenet-Serret tetrad associated with adapted-Kerr static Killing observers, written out in terms of the adapted-BL coordinate basis $(\partial_{\bar{0}}, \partial_1, \partial_2, \partial_{\bar{3}})$, are given as,
\begin{align} \label{eq:FS_AdaptedKerr_Static}
e_{\hat{\bar{0}}}^{\bar{\nu}} =& \left(\frac{1}{\sqrt{-g_{\bar{0}\bar{0}}}}, 0, 0, 0\right), \\
e_{\hat{\bar{1}}}^{\bar{\nu}} =& \frac{1}{2\bar{\kappa} g_{\bar{0}\bar{0}}}\left(0, \frac{g_{\bar{0}\bar{0},1}}{\sqrt{g_{11}}}\frac{1}{\sqrt{g_{11}}}, \frac{g_{\bar{0}\bar{0},2}}{\sqrt{g_{22}}}\frac{1}{\sqrt{g_{22}}}, 0\right), \nonumber \\
e_{\hat{\bar{2}}}^{\bar{\nu}} =& \frac{1}{\sqrt{g_{\bar{0}\bar{0}}\Delta_{03}}}\left(-g_{\bar{0}\bar{3}}, 0, 0, g_{\bar{0}\bar{0}}\right), \nonumber \\
e_{\hat{\bar{3}}}^{\bar{\nu}} =& \frac{1}{2\bar{\kappa} g_{\bar{0}\bar{0}}}\left(0, \frac{g_{\bar{0}\bar{0},2}}{\sqrt{g_{22}}}\frac{1}{\sqrt{g_{11}}}, -\frac{g_{\bar{0}\bar{0},1}}{\sqrt{g_{11}}}\frac{1}{\sqrt{g_{22}}}, 0\right). \nonumber
\end{align}
The accent choices for the indices are as follows: $\bar{\nu}$ and $\hat{\bar{b}}$ are for objects represented in the adapted-BL coordinate basis and the FS tetrad associated with adapted-Kerr static Killing observers respectively. Therefore $e_{\hat{\bar{0}}}^{\bar{\nu}}$ represents the components of the time-like leg of the FS frame of an adapted-Kerr static Killing observer defined relative to the adapted-BL coordinate basis,
\begin{equation}
e_{\hat{\bar{0}}} = e_{\hat{\bar{0}}}^{\bar{\nu}}e_{\bar{\nu}}.
\end{equation}
For conversions between the BL coordinate basis, the adapted-BL coordinate basis, the Kerr static Killing FS tetrad and the adapted-Kerr static Killing FS tetrad, see Appendix \ref{sec:Conversions_Between_Frames}. Also, an alternative derivation of the above quantities (\ref{eq:FS_Invariants_AdaptedKerr}, \ref{eq:FS_AdaptedKerr_Static}) via a neat differential geometric approach may be found in \cite{Straumann09} and, in particular, for the Kerr spacetime in \cite{Chakraborty_Majumdar14}.

Also, of the one-parameter class of coordinate transformations given in (\ref{eq:CT}), the (identity) transformation corresponding to $\Omega \! = \! 0$ leaves the Kerr metric components unchanged and one can obtain the corresponding FS entities associated with Kerr static Killing observers simply by replacing the barred adapted-Kerr metric components $g_{\bar{\mu}\bar{\nu}}$ with the (unbarred) Kerr metric components $g_{\mu\nu}$ in the above (\ref{eq:FS_Invariants_AdaptedKerr}, \ref{eq:FS_AdaptedKerr_Static}). 

Now, since of most astrophysical relevance due to gravitational dynamics are pulsars moving on equatorial circular orbits at constant angular speeds around a Kerr black hole (\citealt{Bardeen+72}, \citealt{Laurentis+18}; also see \citealt{Gonzalez_Lopez-Suspes11} and references therein), we shall henceforth restrict our discussion to such pulsars. Note that we will not introduce a subscript to denote that we will be displaying equatorial plane quantities. 

If we adopt the usual convention $\bar{\kappa} \geq 0$, then clearly $e_{\hat{\bar{1}}}$ and $e_{\hat{\bar{3}}}$ change signs at $\Omega_{\text{K}\pm}$, outside the horizon ($r_{\text{H}} \leq r$). And with the introduction of $\epsilon_{\bar{3}}$ (see Section 3 of \citealt{Bini+99b}),
\begin{equation}
\epsilon_{\bar{3}} = \frac{-g_{\bar{0}\bar{3}}\partial_{\bar{0}} + g_{\bar{0}\bar{0}}\partial_{\bar{3}}}{\sqrt{g_{\bar{0}\bar{0}}\Delta_{03}}},
\end{equation}
we can represent the right-handed FS triad for these observers in the equatorial plane succinctly as,
\begin{align} \label{eq:FS_Tetrad_Final}
\left\{e_{\hat{\bar{1}}}, e_{\hat{\bar{2}}}, e_{\hat{\bar{3}}}\right\} =
\begin{cases}
\left\{-\frac{\partial_1}{\sqrt{g_{11}}}, \epsilon_{\bar{3}}, \frac{\partial_2}{g_{22}}\right\}, & \text{for}\ \Omega_ - < \Omega \leq \Omega_{\text{K}-}, \\
\left\{\frac{\partial_1}{\sqrt{g_{11}}}, \epsilon_{\bar{3}}, -\frac{\partial_2}{g_{22}}\right\}, & \text{for}\ \Omega_{\text{K}-} < \Omega \leq \Omega_{\text{K}+}, \\
\left\{-\frac{\partial_1}{\sqrt{g_{11}}}, \epsilon_{\bar{3}}, \frac{\partial_2}{g_{22}}\right\}, & \text{for}\ \Omega_ {\text{K}+} < \Omega < \Omega_+.
\end{cases}
\end{align}
For a discussion on defining the FS tetrads appropriately in regions where $\Omega_{\text{K}\pm}$ are not allowed orbital angular frequencies, i.e. inside the respective ISCOs, see Appendix \ref{sec:Reversals}. 

Accelerations experienced by these observers is given as,
\begin{equation} \label{eq:Acceleration_aFS_Eq}
\alpha^\prime = \bar{\kappa} e_{\hat{\bar{1}}} = \left|\frac{\sqrt{\Delta}(a^2 M - r^3)(\Omega-\Omega_{\text{K}+})(\Omega-\Omega_{\text{K}-})}{r^3\left[1 \! - \! (r^2 + a^2)\Omega^2 \! - \! \frac{2M(a\Omega - 1)^2}{r}\right]}\right| e_{\hat{\bar{1}}}.
\end{equation}
Outside the horizon, the acceleration changes sign at $\Omega = \Omega_{\text{K}\pm}$. Physically, this means that the sense of the centrifugal force reverses at the locations of the Kepler orbits, for which $\alpha^\prime = 0$ \citep{Nayak_Vishveshwara96}. A detailed discussion on this is presented in Appendix \ref{sec:Reversals} and we partition the regions in the Kerr spacetime where accelerations (and centrifugal forces) experienced by observers moving on equatorial circular orbits are directed along $\pm\partial_1$ respectively in Fig. \ref{fig:Acceleration_Handedness}.

Also, since in the equatorial plane $\bar{\sigma}_2 = 0$ and the spin-precession frequency for these observers is given as,
\begin{align} \label{eq:Precession_Frequency_aFS_Eq}
\Omega_{\text{p}}^\prime =& - \bar{\sigma}_1 e_{\hat{\bar{3}}}, \\
\bar{\sigma}_1 =& \frac{\Omega r^3 + 3 M \Omega r^2(a\Omega - 1) + a M (a\Omega - 1)^2}{r^3\left[1 \! - \! (r^2 + a^2)\Omega^2 \! - \! \frac{2M(a\Omega - 1)^2}{r}\right]}, \nonumber
\end{align}
it is clear that $\Omega_{\text{p}}^\prime$ changes sign at the zeroes of
\begin{equation} \label{eq:Zeroes_Precession_Frequency}
\Omega r^3 + 3 M \Omega r^2(a\Omega - 1) + a M (a\Omega - 1)^2 = 0.
\end{equation}
It must be noted that the orbits where the reversal of the precession frequency occurs do not, in general, coincide with the Kepler orbits where the centrifugal force reverses. For a detailed discussion on the sense of the precession frequency see Appendix \ref{sec:Reversals}. In Fig. \ref{fig:Precession_Frequency_Sense}, we display the regions in the Kerr spacetime where observers moving on equatorial circular geodesics experience positive and negative precession frequencies relative to the $e_{\hat{\bar{3}}}$ leg of their respective right-handed FS tetrads.

In particular, the precession frequencies experienced by equatorial Kerr static observers $\Omega_{\text{p}} = \Omega_{\text{p}}^\prime(\Omega = 0)$ never changes sign and is given as,
\begin{equation} \label{eq:Precession_Frequency_Static}
\Omega_{\text{p}}= \frac{a M}{r^2(r - 2 M)}\frac{\partial_2}{\sqrt{g_{22}}}.
\end{equation}
The precession frequencies experienced by the equatorial ZAMOs are given as,
\begin{equation}
\Omega_{\text{p}}^\prime\left(\Omega = \Omega_{\text{Z}}\right) = \frac{a M(a^2 + 3 r^2)}{r^2(r^3 + a^2(r + 2M))}\frac{\partial_2}{\sqrt{g_{22}}}.
\end{equation}
Note that the zero angular momentum observers experience precession in the same sense as static observers. Also, it is interesting that for both co-rotating and counter-rotating equatorial Kepler observers of the Kerr spacetime, the magnitude of their precession frequencies are independent of the spin-parameter,
\begin{equation} \label{eq:Kepler_Precession}
|\Omega_{\text{p}}^\prime|\left(\Omega = \Omega_{\text{K}\pm}\right) = \frac{M^{1/2}}{r^{3/2}}.
\end{equation}
Since this form is reminiscent of pure geodetic precession \citep{Sakina_Chiba79}, we think it relevant to mention here that, as can be seen from (\ref{eq:FermiDerivative}), when a pulsar moves on an equatorial circular geodesic ($\alpha = 0$), its spin vector is simply parallel transported. 

\begin{figure*}
\centering
\subfigure[~$a = .1 M$, Magnitude of Accelerations $\bar{\kappa}$]
{\includegraphics[scale=.83]{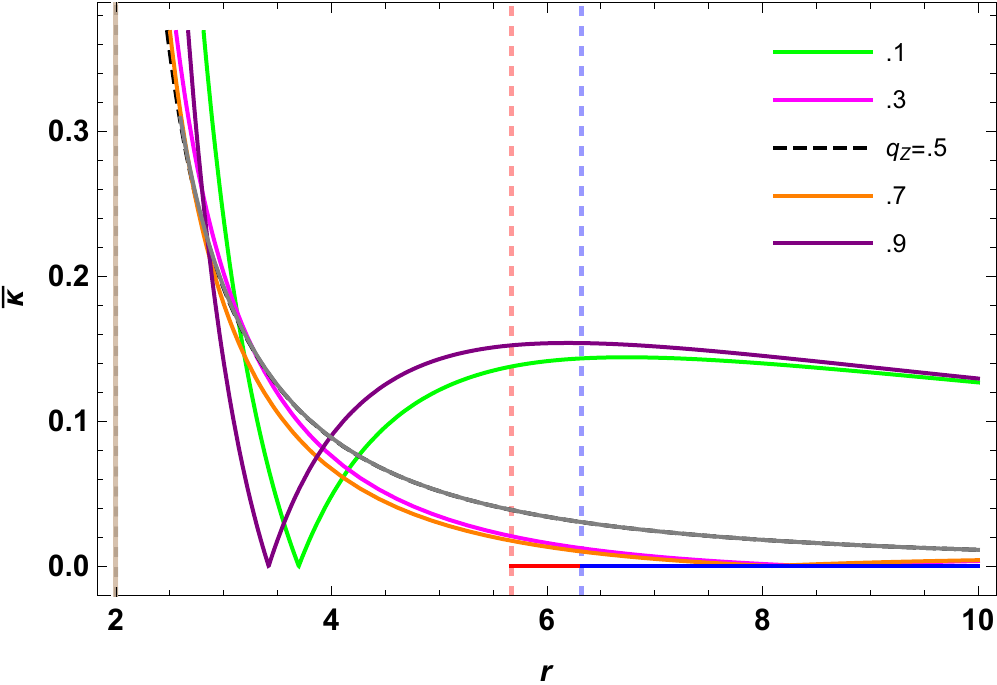}}
\hspace{0.1cm}
\subfigure[~$a = .1 M$, Magnitude of Precession Frequencies $|\bar{\sigma}_1|$]
{\includegraphics[scale=.83]{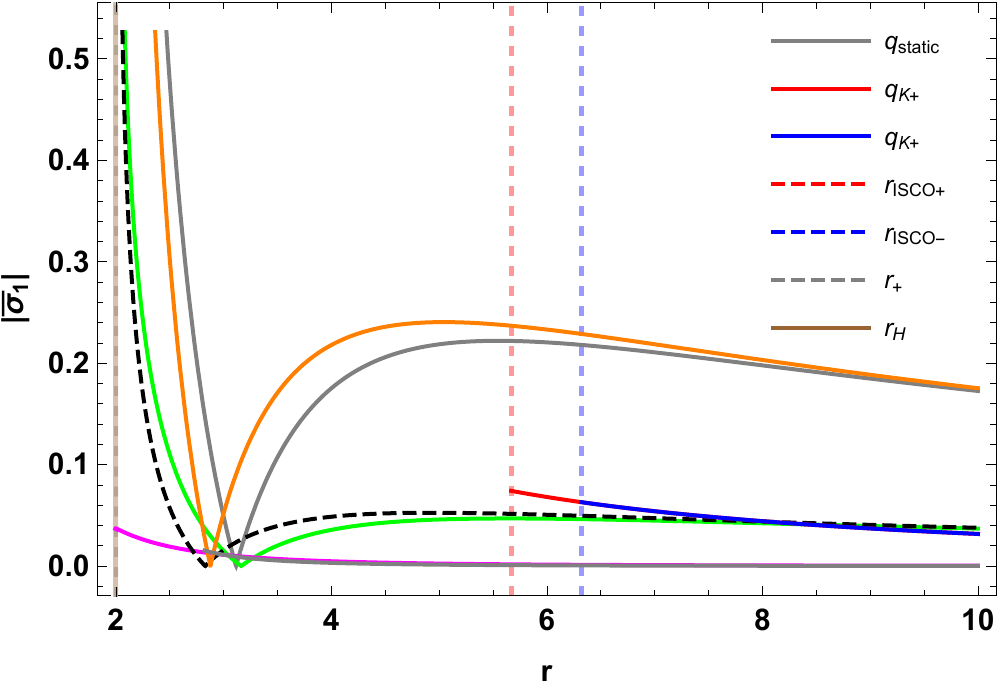}}
\hspace{0.1cm}
\subfigure[~$a=.5 M$, Magnitude of Accelerations $\bar{\kappa}$]
{\includegraphics[scale=.6]{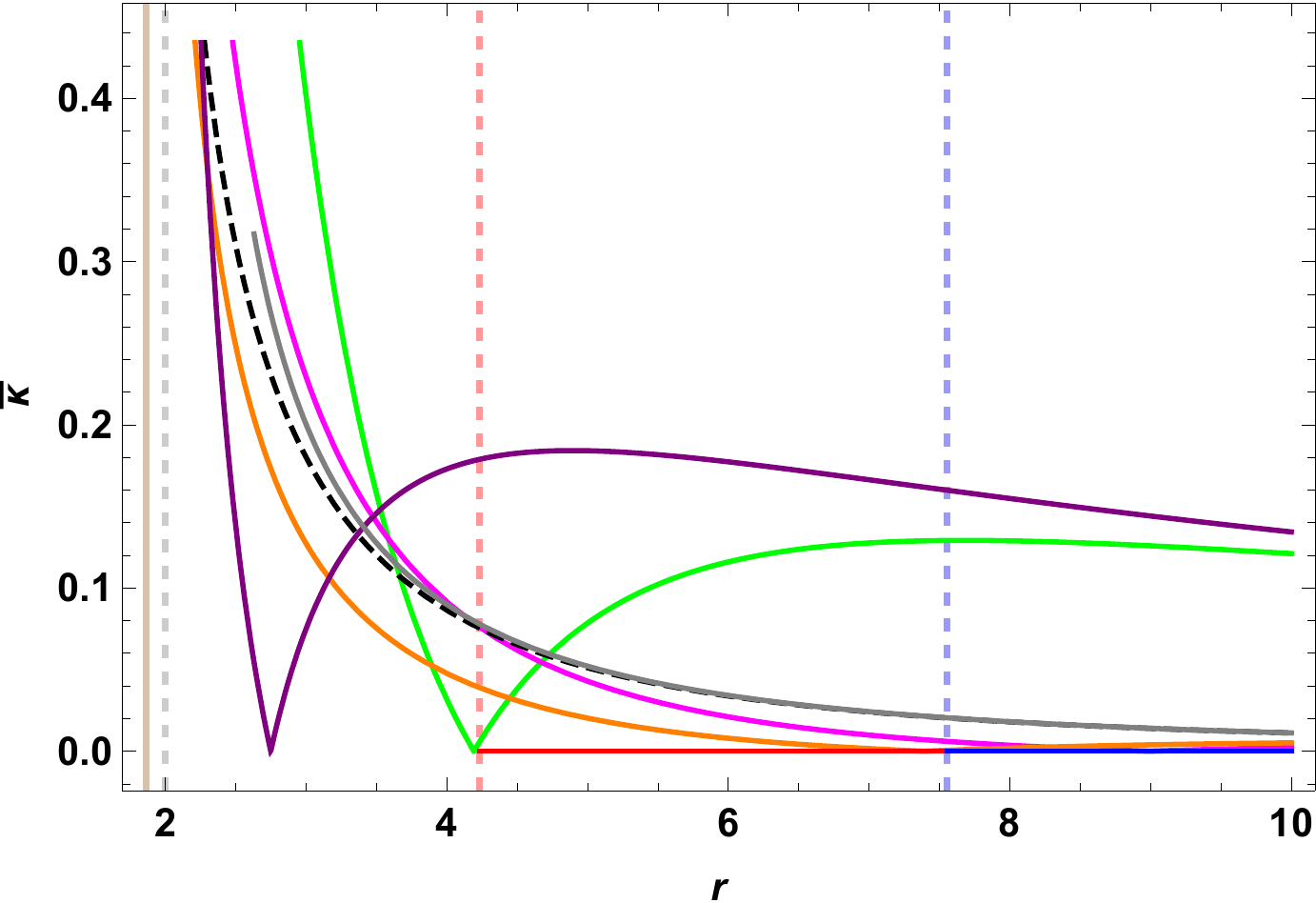}}
\hspace{0.1cm}
\subfigure[~$a=.5 M$, Magnitude of Precession Frequencies $|\bar{\sigma}_1|$]
{\includegraphics[scale=.6]{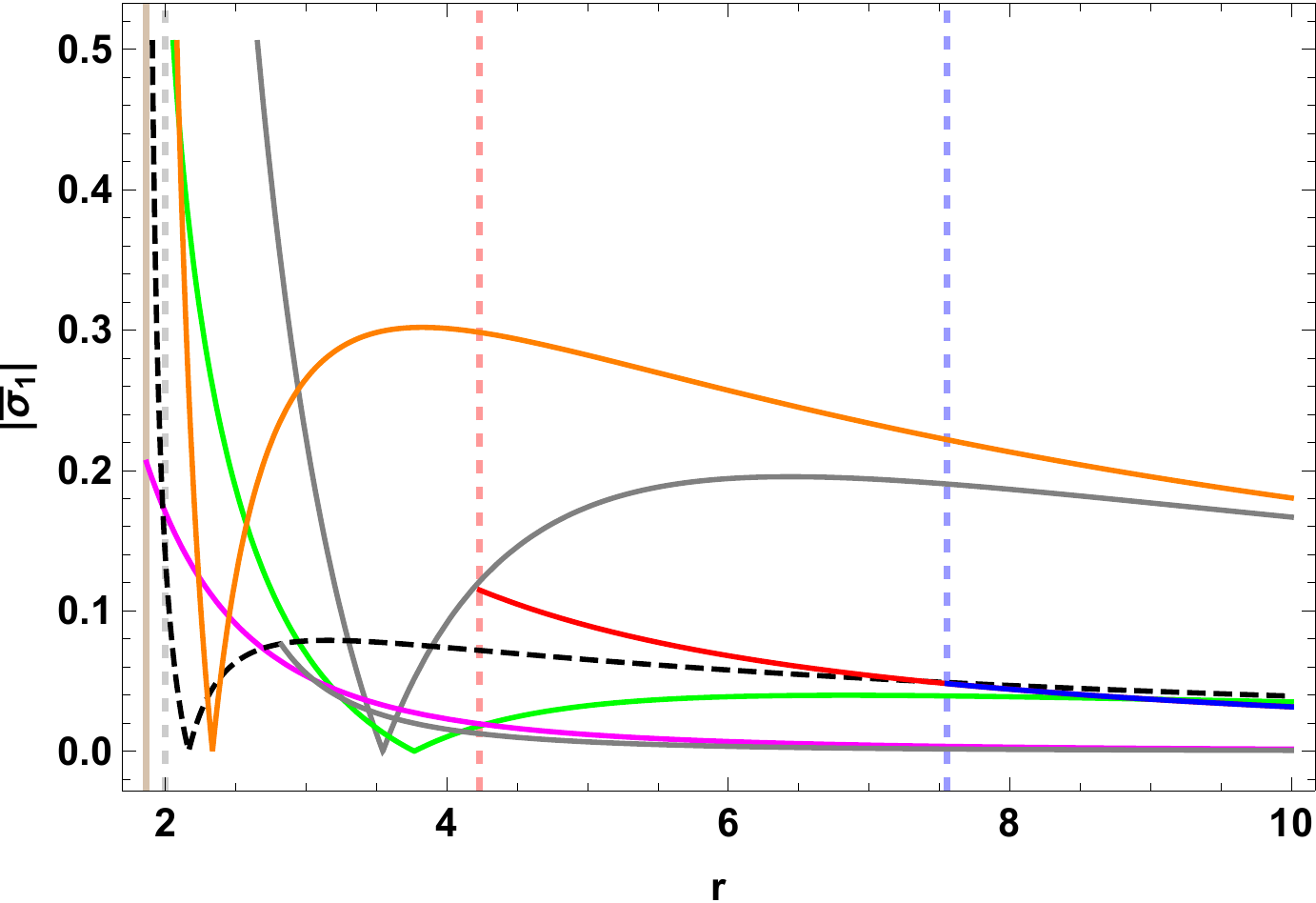}}
\hspace{0.1cm}
\subfigure[~$a=.9 M$, Magnitude of Accelerations $\bar{\kappa}$]
{\includegraphics[scale=.6]{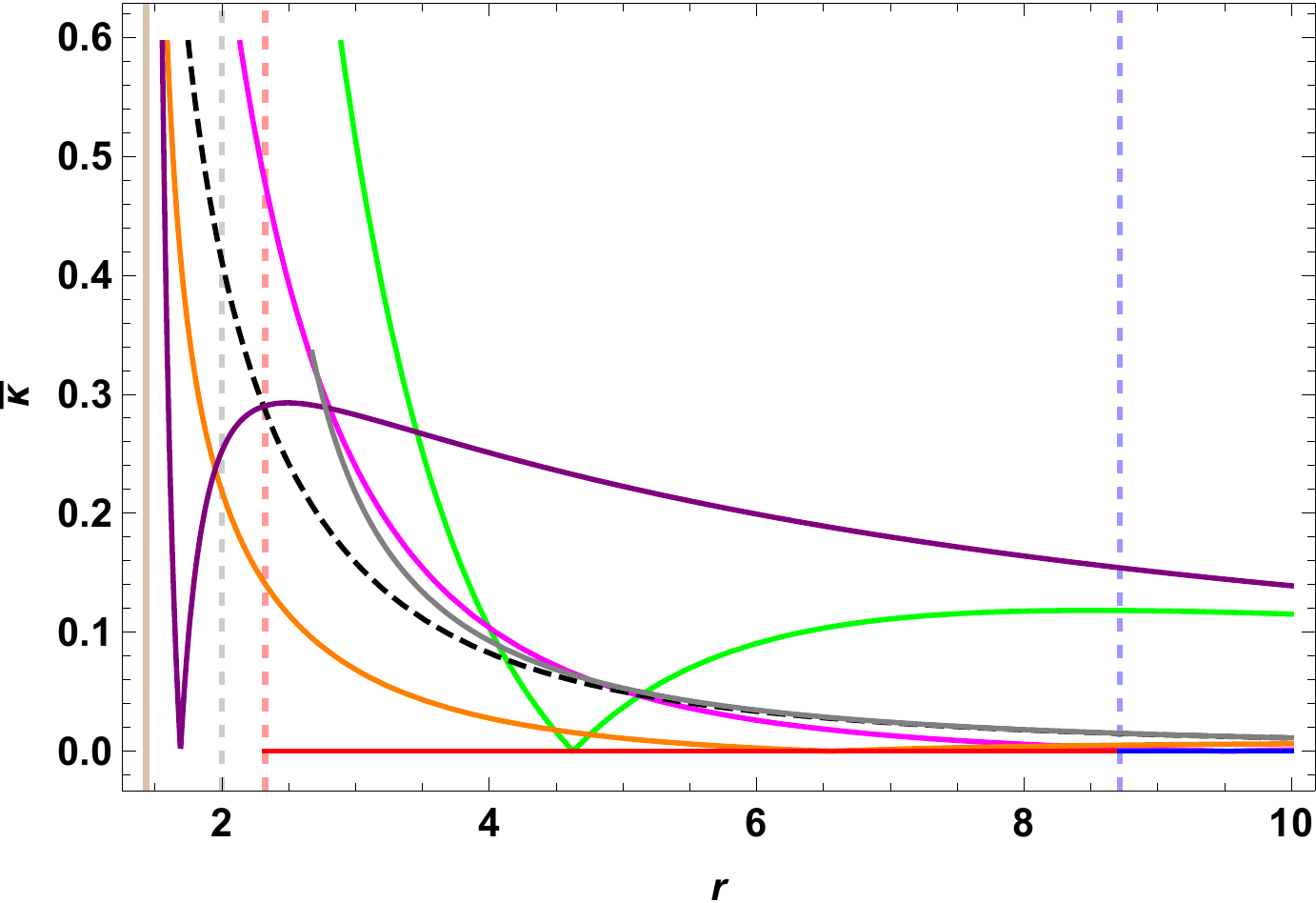}}
\hspace{0.1cm}
\subfigure[~$a=.9 M$, Magnitude of Precession Frequencies $|\bar{\sigma}_1|$]
{\includegraphics[scale=.6]{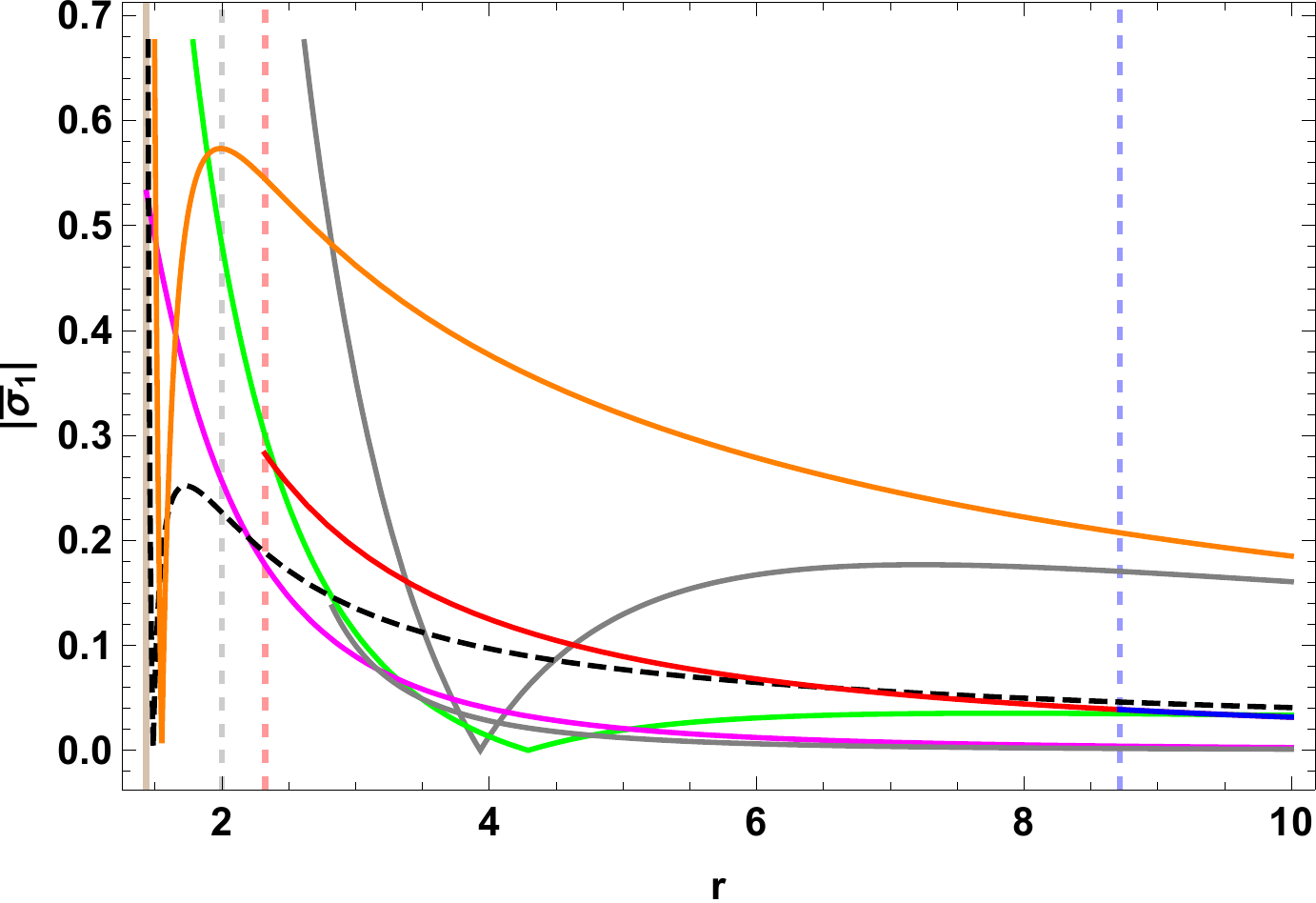}}
\caption{We show here the absolute values of the accelerations and precession frequencies $\bar{\kappa}$ and $|\bar{\sigma}_1|$ experienced by observers moving on equatorial circular orbits at different Boyer-Lindquist orbital radii $r$, around Kerr black holes with specific angular momenta $a = .1 M, .5 M, .9 M$. Such observers have restrictions on the values of allowed angular velocities at any given $r$, i.e. $\Omega_- < \Omega < \Omega_+$. We parametrize $\Omega$ using $0 < q < 1$ as $\Omega = q\Omega_+ + (1-q)\Omega_-$, and plot here $q = .1, .3, .5, .7, .9$ in  green, magenta, dashed-black, orange and purple respectively. In particular, $q = .5$ corresponds to the zero angular momentum observer (ZAMO). Gray represents Kerr static Killing observers with $\Omega = 0$ and red and blue represent non-accelerating co-rotating (+) and counter-rotating (-) Kepler observers, $\Omega = \Omega_{\text{K}\pm}$ respectively. The vertical dashed-red, dashed-blue, dashed-gray and brown lines indicate the location of the innermost stable circular orbits for co-rotating and counter-rotating observers $r_{\text{ISCO}\pm}$, the ergoradius in the equatorial plane $r_+(\theta = \pi/2)$ and the location of the horizon $r_{\text{H}}$ respectively. In Geometrized units ($G=c=1$), $r, \bar{\kappa}$ and $\bar{\sigma}_1$ scale with the mass of the black hole as $M, M^{-1}$ and $M^{-1}$ respectively. Therefore, the larger the mass of the black hole, larger are the sizes of the orbits and smaller are the accelerations and precession frequencies associated with these world-lines. In the above plots, $r$ runs from the location of the horizon $r_{\text{H}}$ to $10 M$. The conversion to physical units for $r$ is $1 M_\odot = 1.5$ km, $\bar{\kappa}$ is $1/M_\odot =  6.0 \times 10^{10}$ km/s$^2$ and $\bar{\sigma}_1$ is $1/M_\odot =  1.3 \times 10^5$ rad/s. Also, in panel (a) the gray line, corresponding to the Kerr static observer, lies very close to the dashed-black line, corresponding to the ZAMO. This can be cross-checked from Fig. \ref{fig:qStatic_qKep+-} from which it is evident that $q_{\text{static}} \approx .5$ in this case.}
\label{fig:Curvatures_AdaptedKerr_Static}
\end{figure*}

The eventual analysis presented in Section \ref{sec:Results} of this paper will largely depend on the strength of the precession frequency, and so in Fig. \ref{fig:Curvatures_AdaptedKerr_Static}, we discuss the trends in the absolute values of the accelerations $\bar{\kappa}$ and precession frequencies $|\bar{\sigma}_1|$ experienced by pulsars that move along equatorial circular orbits around Kerr black holes, which we define from (\ref{eq:Acceleration_aFS_Eq}, \ref{eq:Precession_Frequency_aFS_Eq}) as,
\begin{align}
\bar{\kappa} =& \left|\frac{\sqrt{\Delta}(a^2 M - r^3)(\Omega-\Omega_{\text{K}+})(\Omega-\Omega_{\text{K}-})}{r^3\left[1 \! - \! (r^2 + a^2)\Omega^2 \! - \! \frac{2M(a\Omega - 1)^2}{r}\right]}\right|,\\
|\bar{\sigma}_1| =& \left|\frac{\Omega r^3 + 3 M \Omega r^2(a\Omega - 1) + a M (a\Omega - 1)^2}{r^3\left[1 \! - \! (r^2 + a^2)\Omega^2 \! - \! \frac{2M(a\Omega - 1)^2}{r}\right]}\right|. \nonumber
\end{align}
We display these quantities for BHs of mass $M = 100M_\odot$ with different spin-parameters $a=.1M, .5M, .9M$, at different radii $r$ and orbital angular velocities $\Omega$. On an orbit of any particular radius $r$, a pulsar can move at angular speeds $\Omega_-(r) < \Omega(r) < \Omega_+(r)$, and we parametrize $\Omega$ by its $q$-value (\ref{eq:q}). We consider pulsars present in the strong gravitational field regime of the Kerr spacetime, at distances of $r=r_{\text{H}}-10 M$ and consider $q =  .1, .3, .5, .7, .9, q_{\text{static}}, q_{\text{K}\pm}$. See (\ref{eq:Static_ZAMO_Kepler}) for the definitions of $q_{\text{static}}, q_{\text{K}\pm}$ and Fig. \ref{fig:qStatic_qKep+-} for how they vary with $a, r$. It is useful to remember that static observers with $q = q_{\text{static}}$ and \textit{stable} Kepler observers with $q = q_{\text{K}\pm}$ are allowed only outside $r > 2 M$ and $r \geq r_{\text{ISCO}\pm}$ respectively. In Geometrized units ($G=c=1$), $r, \bar{\kappa}$ and $\bar{\sigma}_1$ scale with the mass of the black hole as $M, M^{-1}$ and $M^{-1}$ respectively. Therefore, the larger the mass of the black hole, larger are the sizes of the orbits and smaller are the accelerations and precession frequencies associated with these world-lines. The sharp changes at $\bar{\kappa} = 0$ and $|\bar{\sigma}_1|=0$ are an artefact of the modulus and actually correspond to a smooth change in the sign of the acceleration or precession frequency respectively. $\bar{\kappa} = 0$ occurs at the Kepler orbits and the zeroes of the precession frequency occur at the solutions of the equation (\ref{eq:Zeroes_Precession_Frequency}). Note that we have displayed $\bar{\kappa}$ and $\bar{\sigma}_1$ in these plots only for co-rotating and counter-rotating Kepler observers only outside their respective ISCOs (i.e. for stable Kepler observers). For a detailed discussion on the reversals of the accelerations and the precession frequencies, see Appendix \ref{sec:Reversals}. And for a more detailed analysis of the trends in the modulii of the precession frequencies experienced by pulsars moving on Killing orbits around both Kerr black holes and naked singularities, we direct the reader to see \cite{Chakraborty+17b}. 

It is also important to note that the $r$ coordinate used in all these expressions is the Boyer-Lindquist radial coordinate. The actual physical radius of the orbit would be given by the Kerr-Schild radial coordinate $\tilde{r}$. The relation between the Kerr-Schild and the BL radial coordinates, in the equatorial plane, is given as (see for example \citealt{Visser08}),
\begin{equation} \label{eq:KS_and_BL}
\tilde{r} = \sqrt{r^2 + a^2}.
\end{equation}
So, for example, the Kerr ergosurface in the equatorial plane is at $r_+ = 2M$ in the Boyer-Lindquist chart but at $\tilde{r}_+ = \sqrt{4 M^2 + a^2}$ in Kerr-Schild coordinates. To not complicate matters, we shall exclusively use the Boyer-Lindquist radial coordinate $r$ until Section \ref{sec:Results}.

\subsection{Earth Line of Sight in the Frenet-Serret Triads of Kerr and adapted-Kerr Static Observers} \label{eq:Earth_LoS}
As previously argued in Section \ref{sec:SpinPrecession_Killing}, the FS triad associated with Kerr static Killing observers can be interpreted as axes-at-rest relative to asymptotic fixed observers. Hence, the projection of the tangent to the null-geodesic connecting the pulsar's spatial location to an arbitrary asymptotic fixed observer's position, like earth (when it is causally connected with the pulsar), onto this FS triad always remains a constant, and we can write,
\begin{equation}
n^{\hat{i}} = \left(\sin\theta_{\text{E}}\cos{\phi_{\text{E}}}, \sin\theta_{\text{E}}\sin{\phi_{\text{E}}}, \cos\theta_{\text{E}}\right),
\end{equation}
For simplicity, we set $\phi_{\text{E}} = 0$,
\begin{equation} \label{eq:Null_Tangent_Static}
n_{\text{E}}^{\hat{i}} = \left(\sin\theta_{\text{E}}, 0, \cos\theta_{\text{E}}\right).
\end{equation}
Further, in the FS triad of an equatorial adapted-Kerr static Killing observer, we can write (see Appendix \ref{sec:Connecting_Vector_aFS}),%
\begin{equation} \label{eq:Null_Tangent_Stationary}
n_{\text{E}}^{\hat{\bar{j}}}(\tau) = \left(\cos{(\Omega \tau)}\sin{\bar{\theta}_{\text{E}}}, \mp\sin{(\Omega \tau)}\sin{\bar{\theta}_{\text{E}}}, \cos{\bar{\theta}_{\text{E}}}\right),
\end{equation}
where the upper (-) and lower (+) signs apply to observers whose Frenet-Serret trinormal leg $e_{\hat{\bar{3}}}$ points along and opposite to the $z$-axis (or direction the black hole spin), denoted by $\hat{z}$ respectively. That is the signs $\mp$ correspond to the cases (see eq. \ref{eq:FS_AdaptedKerr_Static} and also the discussion in Appendix \ref{sec:Connecting_Vector_aFS}),
\begin{equation} \label{eq:e3_relative_z}
e_{\hat{\bar{3}}} = \pm \hat{z} = \mp \frac{\partial_2}{\sqrt{g_{22}}}.
\end{equation}

\section{Effect of Gravitomagnetism on Pulsar Beam Evolution} \label{sec:PulsarPrecession_Static}
We shall treat pulsars to be test spinning objects, i.e. as small but extended `pole-dipole' test particles (see for example \citealt{Dixon74}). For the test spinning object approximation to hold, the primary requirement is that the background gravitational field due to the black hole must not vary much over the spatial size of the pulsar. That is, for a pulsar of mass and radius $m_p, R_p$ present in the vicinity of a Kerr black hole whose mass and angular momentum are $M, J$, and for a radial separation between their centre of masses given by $r$, when $m_p \! \ll M \! < r$ and $R_p \! \ll \! r$, the interaction of the pulsar's spin \textit{quadrupole} moment with inhomogeneities of the gravitational field can be neglected \citep{Singh+14}, it can be treated as such a test spinning object. The dynamics of the pulsar which is governed by the conservation equation,
\begin{equation}
\nabla_{\mu} T^{\mu\nu} = 0,
\end{equation}
is then dominated by the two lowest $T^{\mu\nu}$-moments, namely its monopole $p^\mu$ and its dipole $S^{\mu\nu}$, which are its four-momentum and intrinsic spin-angular momentum tensor respectively \citep{Dixon74}. Once one chooses the Pirani condition (\ref{eq:Pirani}), one can associate with the dipole tensor an intrinsic spin-angular momentum four-vector $\mathcal{S}^\mu$ and subsequently an intrinsic spin angular momentum three-vector $S^\mu$, which we introduced in Section \ref{sec:SpinPrecession}. As mentioned before, we ignore the spin-curvature coupling in this work and calculations incorporating this effect will be reported elsewhere. Therefore, the equation of motion for the spin angular momentum reduces to the by Fermi-Walker transport law given in eq. (\ref{eq:FW_S}); see for example \cite{Semerak99}.


All calculations henceforth will be performed in orthonormal right-handed Euclidean Frenet-Serret spatial triads, which for a Kerr static Killing observer ($\Omega = 0$) is given as (\ref{eq:FS_Tetrad_Final}),
\begin{equation} \label{eq:FS_Static_Kerr}
\left\{\frac{\partial_1}{\sqrt{g_{11}}}, \frac{g_{00}\partial_3 - g_{03}\partial_0}{\sqrt{g_{00}\Delta_{03}}}, \frac{-\partial_2}{\sqrt{g_{22}}}\right\}.
\end{equation}
Therefore, we can switch to the Euclidean three-vector notation and the FS frame written out above (\ref{eq:FS_Static_Kerr}) will be denoted by $\{\hat{e}_1, \hat{e}_2, \hat{e}_3\}$. That is, for a vector $a = a^i e_{\hat{i}}$ defined in the FS frame, we will use $\vec{a} = (a^1, a^2, a^3)$, and for a unit vector, we will use $\hat{b}$. In Section \ref{sec:PulsarPrecession_Stationary}, we will generalize our calculations to describe the full class of equatorial Kerr stationary Killing observers.

The world-line of a pulsar that remains at a fixed spatial location, or that moves at very low orbital angular velocities $\Omega \approx 0$, over the period of observation can be approximated to be a static Killing orbit with four-velocity given in (\ref{eq:StaticObserver}). For such a pulsar, let us denote the instantaneous unit vector along which the spin-axis (or the intrinsic spin angular momentum) lies by $\hat{S}(\tau)$. In this notation, we can now rewrite the time dependence of the spin-axis $\hat{S}$ that arises from the precession due to its coupling with the gravitomagnetic field of the Kerr black hole (\ref{eq:Spin_EoM}) as,
\begin{equation}
\dot{\hat{S}} = -\sigma_1\hat{S} \times \hat{e}_3, \nonumber
\end{equation}
where the $\dot{}$ represents a derivative with respect to the proper time $\tau$ in the pulsar's comoving FS frame and $\sigma_1 = \bar{\sigma}_1(\Omega = 0)$ (see eq. \ref{eq:Precession_Frequency_aFS_Eq}). Therefore, $\hat{S}$ moves on a cone of half-opening angle $\beta$ around the precession-axis $\hat{e}_3$ at the precession frequency $-\sigma_1$. 

\begin{figure}
\centering
\includegraphics[scale=0.4]{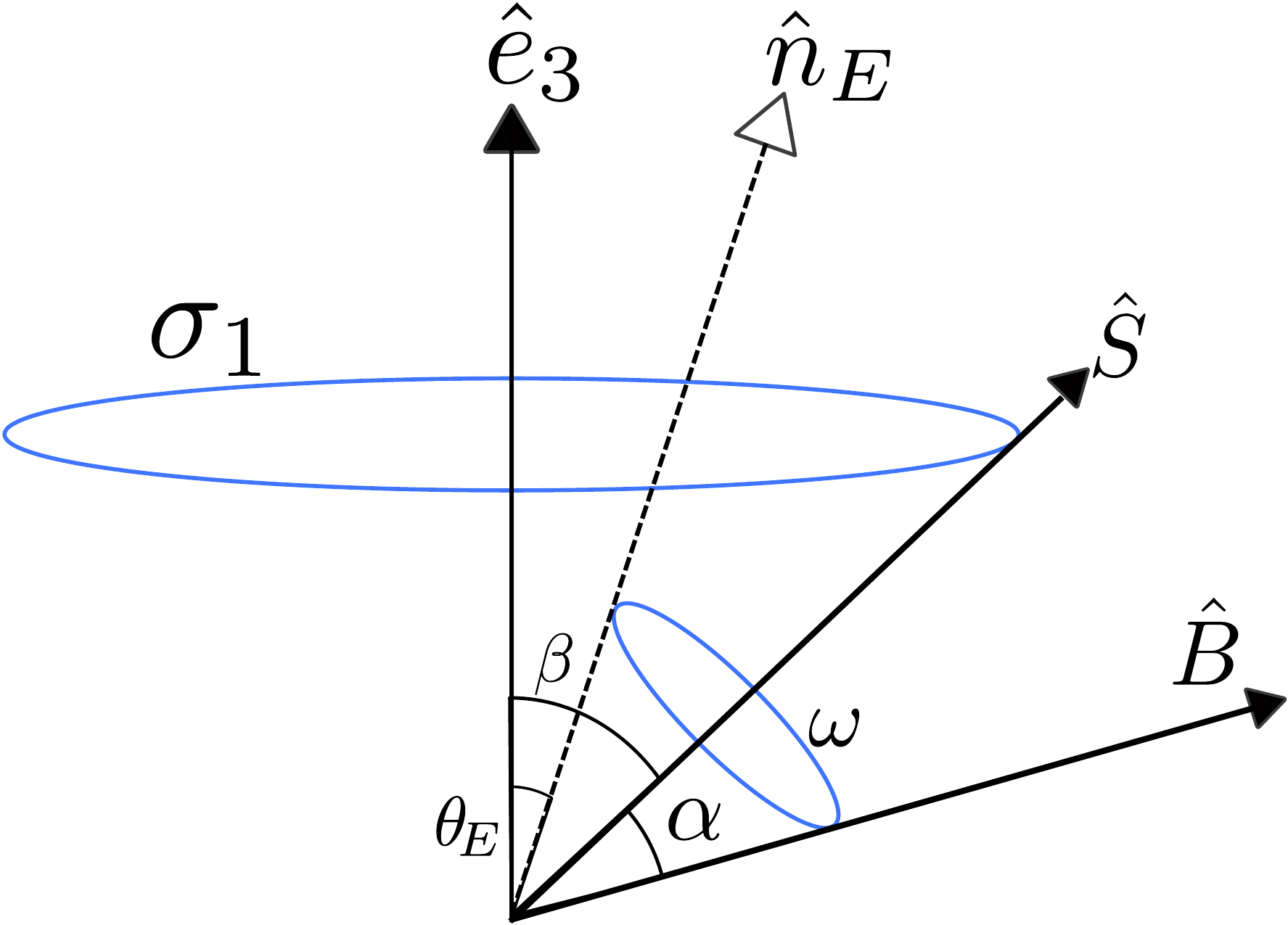}
\caption{If a pulsar that remains spatially fixed, or moves at small orbital angular velocities $\Omega \approx 0$ in the equatorial plane of a Kerr black hole, its spin axis $\hat{S}$ precesses around a precession axis $\hat{e}_3$, at the precession frequency $\sigma_1$, in the clockwise sense. Here $\beta$ denotes the angle between these axes. Further, the direction in which the beam is emitted $\hat{B}$ rotates around this time-varying spin axis $\hat{S}(\tau)$ at the intrinsic spin angular frequency of the pulsar $+\omega$ (in the counter-clockwise sense: a convention), with $\alpha = \angle(\hat{B}, \hat{S})$ remaining constant. Earth lies along $\hat{n}_{\text{E}}$, with $\theta_E = \angle(\hat{n}_{\text{E}}, \hat{e}_3)$, also a constant.}
\label{fig:Precession_Figure}
\end{figure}

Now, if we represent by $\hat{B}$ and $\omega$ the instantaneous direction along which the pulsar's beam of radiation points and the (fixed) angular frequency at which the pulsar spins about its spin-axis, then the beam vector $\hat{B}$ just moves on the surface of a cone with axis $\hat{S}$ and half-opening angle $\alpha$ (not to be confused with the acceleration four-vector), at an angular speed of $\omega$. See Fig. \ref{fig:Precession_Figure} for the geometry of the various vectors involved. The requisite equations of motion in the Euclidean Frenet-Serret frame are the following set of coupled first order differential equations,
\begin{align}
\dot{\hat{S}} &= -\sigma_1\hat{S} \times \hat{e}_3, \label{eq:EoM_S} \\
\dot{\hat{B}} &= \omega \hat{B} \times \hat{S}, \label{eq:EoM_B}
\end{align}
Finally, we can represent the apparent direction along which earth lies $\hat{n}_{\text{E}}$ as (\ref{eq:Null_Tangent_Static}),
\begin{equation}
\hat{n}_{\text{E}} = (\sin\theta_{\text{E}}, 0, \cos\theta_{\text{E}}).
\end{equation}
We are interested in finding the frequency at which $\hat{B}$ points along the apparent direction of earth, $\hat{n}_{\text{E}}$, first in the FS frame and eventually in earth's frame. At every time $\tau$ such that $\hat{B}(\tau) = \hat{n}_{\text{E}}$, a pulse is seen on earth. This corresponds to the frequency at which the deflection vector $\vec{\zeta}$ vanishes,
\begin{equation} \label{eq:DeflectionVector}
\vec{\zeta}(\tau) = \hat{B}(\tau) - \hat{n}_{\text{E}}.
\end{equation}
If the frequency at which $\zeta$ goes to zero is given by $\nu_{\text{FS}}$, then we can use the red-shift formula to find the observed pulse frequency on earth $\nu_{\text{E}}$ as (see for example \citealt{Misner+73}),
\begin{equation}
1 + z = \frac{d\tau}{dt} = \frac{\nu_{\text{E}}}{\nu_{\text{FS}}} = \left[-g_{00}\left(r,\frac{\pi}{2}\right)\right]^{-1/2},
\end{equation}
where $t$ is the coordinate time measured by an asymptotic static observer, like an astronomer on earth.

\subsection{Initial Conditions}
We shall suppose that the apparent direction (direction cosines) of the Earth is given as (\ref{eq:Null_Tangent_Static}), 
\begin{equation} \label{eq:Edir}
\hat{n}_{\text{E}} = \left(\sin\theta_{\text{E}}, 0, \cos\theta_{\text{E}}\right). 
\end{equation}
Let us use the initial condition that at time $\tau = 0$, a pulse is received on earth, i.e. $\hat{B}$ points along the apparent direction of earth,
\begin{equation} \label{eq:B_Initial_Condition}
\hat{B}_0 = \hat{B}(\tau = 0) = \left(\sin\theta_{\text{E}}, 0, \cos\theta_{\text{E}}\right).
\end{equation}
Since $\hat{S}$ moves on a cone of half-angle $\beta$ around $\hat{e}_3$, the most general initial condition that we can write is,
\begin{equation} \label{eq:S_Initial_Condition}
\hat{S}_0 = (\sin\beta\cos\psi, \sin\beta\sin\psi, \cos\beta).
\end{equation}
Now, since $\hat{B}$ moves on a cone of half-angle $\alpha$ around $\hat{S}(\tau)$, we require that the following be satisfied,
\begin{equation}
\hat{B}_0 \! \cdot \! \hat{S}_0 = \cos\alpha.
\end{equation}
Let us note here that $\alpha, \beta, \theta_E$ are fixed by the geometric configuration of the system. $\psi$ however is simply an initial phase for $\hat{S}_0$, which we are free to choose thereby exhausting all freedom in initial conditions. The above equation imposes the following constraint,
\begin{equation} \label{eq:alpha_beta_thetaE_psi}
\cos\alpha = \sin\theta_E\sin\beta\cos\psi + \cos\theta_E\cos\beta.
\end{equation}

\subsection{Solution for the Spin and Beam Vectors} \label{sec:Beam_Solution}
One can simply obtain the solution $\hat{S}(\tau)$ to the spin-equation of motion (\ref{eq:EoM_S}) by using Rodrigues' rotation formula (see for example \citealt{Goldstein80}) as,
\begin{equation} \label{eq:S_pm}
\hat{S} = (\sin\beta\cos{(\psi + \sigma_1\tau)}, \sin\beta\sin{(\psi + \sigma_1\tau)}, \cos\beta).
\end{equation}
It can be checked that $\hat{S}(\tau = 0)$ satisfies the initial conditions given in (\ref{eq:S_Initial_Condition}). The solution for the beam vector that satisfies the relevant equation of motion (\ref{eq:EoM_B}) and the initial condition (\ref{eq:B_Initial_Condition}) is then given as (see Appendix \ref{sec:Beam_Vector_Solution} for an analytic derivation),
\begin{align} \label{eq:B_General}
B_1 \! =& D_1\left[\cos{(\omega_{\text{eff}} \tau)}\cos{\left(\psi \! + \! \sigma_1\tau\right)}\cos\chi \! + \! \sin{(\omega_{\text{eff}} \tau)}\sin{\left(\psi \! + \! \sigma_1\tau\right)}\right]\nonumber \\
& \!\!\!\!\!\! + \! D_2\left[\sin{(\omega_{\text{eff}} \tau)}\cos{\left(\psi \! + \! \sigma_1\tau\right)}\cos\chi \! - \! \cos{(\omega_{\text{eff}} \tau)}\sin{\left(\psi \! + \! \sigma_1\tau\right)}\right] \nonumber \\
& \!\!\!\!\!\! +\! D_3\cos{\left(\psi \! + \! \sigma_1\tau\right)}\sin\chi, \\
B_2 \! =& 
D_1\left[\cos{(\omega_{\text{eff}} \tau)}\sin{\left(\psi \! + \! \sigma_1\tau\right)}\cos\chi \! - \! \sin{(\omega_{\text{eff}} \tau)}\cos{\left(\psi \! + \! \sigma_1\tau\right)}\right] \nonumber \\
& \!\!\!\!\!\! + \! D_2\left[\sin{(\omega_{\text{eff}} \tau)}\sin{\left(\psi \! + \! \sigma_1\tau\right)}\cos\chi \! + \! \cos{(\omega_{\text{eff}} \tau)}\cos{\left(\psi \! + \! \sigma_1\tau\right)}\right], \nonumber \\
& \!\!\!\!\!\! +\! D_3\sin{\left(\psi \! + \! \sigma_1\tau\right)}\sin\chi \nonumber \\
B_3 \! =& 
-\sin\chi\left[D_1\cos{(\omega_{\text{eff}} \tau)} \! + \! D_2\sin{(\omega_{\text{eff}} \tau)}\right]\!+\! D_3\cos\chi, \nonumber
\end{align}
where
\begin{align} \label{eq:D1D2D3}
D_1 &= \sin\theta_E\cos\chi\cos\psi - \cos\theta_E\sin\chi, \\
D_2 &= - \sin\theta_E\sin\psi, \nonumber\\
D_3 &= \cos\theta_E\cos\chi + \sin\theta_E\sin\chi\cos\psi. \nonumber
\end{align}
Also we have introduced,
\begin{align} \label{eq:omega_eff_chi}
\omega_{\text{eff}}^2 &= \omega^2 + \sigma_1^2 + 2\omega\sigma_1\cos{\beta}, \\
\chi &= \sin^{-1}\left(\frac{\omega}{\omega_{\text{eff}}}\sin\beta\right).
\end{align}
Now, let us pick the initial phase for $\hat{S}_0$, without loss of generality, to be $\psi = 0$.  From (\ref{eq:alpha_beta_thetaE_psi}), it is clear that one obtains pulses only for specific geometric configurations, $\theta_E = \pm \alpha + \beta$. Also,
\begin{align} \label{eq:tD1tD2tD3}
D_1 &= \sin{(\theta_E - \chi)}, \\
D_2 &= 0, \nonumber\\
D_3 &= \cos{(\theta_E - \chi)}, \nonumber
\end{align}
Then the deflection vector is given by,
\begin{align} \label{eq:zeta}
\zeta_1\!=&\!
\cos{\left(\sigma_1\tau\right)}\left[\cos{(\omega_{\text{eff}} \tau)}\sin{(\theta_E\!-\!\chi)}\cos\chi \!+\!\cos{(\theta_E\!-\!\chi)}\sin\chi\right] \nonumber \\
+ &\!\sin{\left(\sigma_1\tau\right)}\sin{(\omega_{\text{eff}} \tau)}\sin{(\theta_E\!-\!\chi)} -\sin\theta_E, \\
\zeta_2\!=&\!
-\!\cos{\left(\sigma_1\tau\right)}\sin{(\omega_{\text{eff}} \tau)}\sin{(\theta_E\!-\!\chi)} \nonumber \\
&\!\!\!\!\!\! + \sin{\left(\sigma_1\tau\right)}\left[\cos{(\omega_{\text{eff}} \tau)}\sin{(\theta_E\!-\!\chi)}\cos\chi\!+\!\cos{(\theta_E\!-\!\chi)}\sin\chi\right], \nonumber \\
\zeta_3\!=&\! -\!\cos{(\omega_{\text{eff}} \tau)}\sin{(\theta_E\!-\!\chi)}\sin\chi\!+\!\cos{(\theta_E\!-\!\chi)}\cos\chi - \cos\theta_E. \nonumber
\end{align}
The components of $\vec{\zeta}$ are, in general, \textit{almost} periodic functions of $\tau$ \citep{Besicovitch32, Bohr47}. That is, $\vec{\zeta}$ approaches zero arbitrary closely but $\vec{\zeta}$ does not necessarily vanish periodically, or even vanish exactly at all. 
This is clear immediately if one remembers that $\sigma_1$ and $\omega$ (and therefore $\omega_{\text{eff}}$) are in general incommensurate frequencies. When $\sigma_1, \omega_{\text{eff}}$ are rational multiples of each other, $\vec{\zeta}$ is exactly periodic and one obtains pulses indeed, at the lowest common multiple (LCM) of these frequencies. 

One can immediately anticipate implications for the existence of sub-millisecond pulsars. That is, ordinary garden variety pulsars could appear to pulse at faster (sub-millisecond) rates when in the vicinity of a Kerr black hole, due to gravitomagnetic spin-precession. Further, one could possibly explain quasi-periodic oscillations (QPOs) in pulsar observations, pulsar nulling and multi-peaked pulsar frequency profiles by accounting for the presence of a Kerr black hole in the vicinity of the pulsar. 

Let us remember that the analysis thus far has been conducted entirely at the location of the pulsar, in its comoving FS frame. To obtain the relation between differences in times measured in the pulsar frame $d\tau$ and on earth $dt$, as mentioned before we assume that the earth is at asymptotic infinity, and use the gravitational redshift to write,
\begin{equation}
\frac{dt}{d\tau} =  \left[-g_{tt}\left(r,\frac{\pi}{2}\right)\right]^{-1/2}.
\end{equation}
In particular when when $\vec{\zeta}$ is periodic i.e., when the ratio of $\omega_{\text{eff}}$ and $\sigma_1$ is rational, the zeroes of $\vec{\zeta}$ will occur at the LCM of these two frequencies. If we denote this frequency as $\nu_{\text{FS}}$, then the frequency at which pulses are observed on earth $\nu$ can be obtained as,
\begin{equation} \label{eq:redshift}
\nu_{\text{E}} = \left(1 - \frac{2M}{r}\right)^{1/2}\nu_{\text{FS}}.
\end{equation}

\subsection{Non-Zero Beam Width} \label{sec:Non_Zero_Beam_Width}
Furthermore, since the beam width is not exactly of zero measure, pulses will still be recorded on earth when $\sigma_1, \omega_{\text{eff}}$ are not exact rational multiples of each other. More concretely, let us suppose that the beam has some finite size and model it here as a cone of half-opening angle $\mu$. Then if $\hat{n}_{\text{E}}$ lies anywhere within this cone, i.e. when $\angle(\hat{B},\hat{n}_{\text{E}}) \leq \mu$ or equivalently whenever,
\begin{equation} \label{eq:zeta_mu}
|\vec{\zeta}|^2 \leq 4\sin^2{(\mu/2)},
\end{equation}
one `sees the pulsar.' This is particularly important to take into account since we now have access to a continuous pulse profile whenever the above condition is satisfied, as opposed to a far lesser amount of information corresponding to just the information of when a pulse was observed in our earlier consideration of zero beam width, $\mu=0$. 

In the case of an isolated pulsar, the Fourier spectrum of its pulse profile will effectively contain only one peak, corresponding to its intrinsic spin frequency $\omega$. In stark contrast, Fourier spectra corresponding to observations of pulsars present in strong gravitational fields like near the ergosurface of a Kerr black hole or naked singularity will be multiply peaked and will contain information regarding the relevant frequencies in the problem, $\sigma_1, \omega_{\text{eff}}$, and also regarding other geometry parameters like $\beta, \theta_{\text{E}}$. And from this spectrum, one can potentially extract black hole parameters $a, M$. 

The aim of the present work is just to point out that it is possible from measurements of pulsar spin-precession to extract information regarding black hole parameters but we do not attempt a detailed analysis of the `inverse problem' here, namely what the shape of the pulse profile is as seen in earth's frame etc. However, it is clear that if we define in the pulsar frame, two sets of times $\tau_{\text{on}, i}, \tau_{\text{off}, j}$ as follows,
\begin{align} \label{eq:on_off}
& |\zeta^2|(\tau_{\text{on}, i}) = |\zeta^2|(\tau_{\text{off}, j}) = 4\sin^2{(\mu/2)}, \\
& \frac{d|\zeta^2|(\tau_{\text{on}, i})}{d\tau} < 0,\ \ \frac{d|\zeta^2|(\tau_{\text{off}, j})}{d\tau} > 0, \nonumber
\end{align}
then one sees the pulsar for the duration $\Delta\tau_i = \tau_{\text{off}, i} - \tau_{\text{on}, i}$. In the above, we have used the same index $i$ to denote that these are consecutive on and off times that satisfy (\ref{eq:on_off}). Note that $\Delta\tau_i$ are in general not of the same length. The amount of time for which the pulsar is visible is captured by the set $\Delta\tau_i$ and how frequently it is visible is indicated by the set $\tau_{\text{on}, i}$, and these quantities depend on black hole parameters. Finally, one can obtain the relevant times and time differences in earth's frame via a simple redshift calculation.

\subsection{Kerr Static Pulsars: Approach to the Ergosurface} \label{sec:Static_Pulsars_Ergosurface}
Very slowly moving pulsars can be modelled as test spinning objects that move along Kerr static Killing orbits. In particular, when such pulsars are close to the ergosurface $r \rightarrow r_+$, the precession frequency they experience becomes unboundedly large $\sigma_1 \rightarrow \infty$ and in this limit, we have $\chi \approx 0$ and $\omega_{\text{eff}} \approx \sigma_1$. In this extreme case, the deflection vector just becomes periodic and the pulse frequency of such a pulsar, measured in the FS frame, locks onto the spin-precession frequency, i.e., 
\begin{equation}
\lim_{r\rightarrow r_+}\nu_{\text{FS}} = \frac{\sigma_1}{2\pi}.
\end{equation}
If we introduce a dimensionless parameter $\delta = \frac{r}{2M} - 1$, which measures the radial distance from the ergosurface in the equatorial plane, then in this limit, the redshift goes as $1 + z \approx \delta^{1/2}$, the precession frequency goes as $\sigma_1 \approx \frac{a}{8 M^2}\frac{1}{\delta(1 + \delta)^2}$ and we find that the observed pulse frequency on earth $\nu_{\text{E}}$ behaves as,
\begin{equation}
\nu_{\text{E}} \approx \frac{a}{8\pi M^2}\frac{1}{\delta^{1/2}}.
\end{equation}
If the pulsar is present in this region, one obtains pulses on the Earth far more rapidly than 
an isolated pulsar, i.e., $\nu_{\text{E}} \gg \omega/2\pi$. Specifically, in the limit $\delta \rightarrow 0$, we obtain $\nu \rightarrow \infty$, i.e., the pulses disappear and we see a rather continuous beam from the earth. This is somewhat reminiscent of the chirp one sees in a gravitational wave calculation \citep{Abbott+16}.

Finally, we would like to note that even if for a short proper time a pulsar moves on a circular orbit with a small orbital angular velocity $\Omega \approx 0$ near the ergosurface of a Kerr black hole (or even a naked singularity), and is observed on earth, then since they experience nearly vanishing geodetic precession, measurements from such pulsars could allow us to pick out pure gravitomagnetic effects in the Kerr spacetime. This could potentially lead to an independent estimate of the spin parameter of the central Kerr black hole.

\subsection{Pulsar Precession: Kerr Stationary Observers} \label{sec:PulsarPrecession_Stationary}
We had discussed the equivalence between Kerr stationary and adapted-Kerr static observers in Section \ref{sec:AdaptedKerr} and our analysis of the evolution of the beam vector of a pulsar moving along the integral curve of a Kerr static observer presented in this section thus far can now be immediately extended to study the case of pulsars moving on Kerr stationary Killing orbits. 

In terms of the precession frequency experienced by an adapted-Kerr static observer measured relative to its Frenet-Serret tetrad  $\Omega^\prime_{\text{p}}$ which is given as,
\begin{align}
\Omega_{\text{p}}^\prime =& -\bar{\sigma}_1 e_{\hat{\bar{3}}}, \\
\bar{\sigma}_1 =& \frac{\Omega r^3 + 3 M \Omega r^2(a\Omega - 1) + a M (a\Omega - 1)^2}{r^3\left[1 \! - \! (r^2 + a^2)\Omega^2 \! - \! \frac{2M(a\Omega - 1)^2}{r}\right]}, \nonumber
\end{align}
we introduce,
\begin{align}
\bar{\omega}_{\text{eff}} &= \sqrt{\omega^2 + \bar{\sigma}_1^2 + 2\omega\bar{\sigma}_1\cos{\beta}},\ \ \text{and}\\ 
\bar{\chi} &= \sin^{-1}{\left(\frac{\omega}{\bar{\omega}_{\text{eff}}}\sin{\beta}\right)}, \nonumber
\end{align}
and proceed to directly write out the expressions for the deflection vector $\vec{\bar{\zeta}}$ as,
\begin{align} \label{eq:zeta_Gen}
\bar{\zeta}_1\!=&\!
\cos{\left(\bar{\sigma}_1\tau\right)}\left[\cos{(\bar{\omega}_{\text{eff}} \tau)}\sin{(\bar{\theta}_{\text{E}}\!-\!\bar{\chi})}\cos{\bar{\chi}} \! + \!\cos{(\bar{\theta}_{\text{E}}\!-\!\bar{\chi})}\sin{\bar{\chi}}\right] \nonumber \\
+ &\!\sin{\left(\bar{\sigma}_1\tau\right)}\sin{(\bar{\omega}_{\text{eff}} \tau)}\sin{(\bar{\theta}_{\text{E}}\!-\!\bar{\chi})} - \cos{(\Omega\tau)}\sin{\bar{\theta}_{\text{E}}}, \\
\bar{\zeta}_2\!=&
- \cos{\left(\bar{\sigma}_1\tau\right)}\sin{(\bar{\omega}_{\text{eff}} \tau)}\sin{(\bar{\theta}_{\text{E}}\!-\!\bar{\chi})} \pm \sin{(\Omega\tau)}\sin{\bar{\theta}_{\text{E}}} \nonumber \\
&\!\!\!\!\!\! + \sin{\left(\bar{\sigma}_1\tau\right)}\left[\cos{(\bar{\omega}_{\text{eff}} \tau)}\sin{(\bar{\theta}_{\text{E}}\!-\!\bar{\chi})}\cos{\bar{\chi}}\!+\!\cos{(\bar{\theta}_{\text{E}}\!-\!\bar{\chi})}\sin{\bar{\chi}}\right]  \nonumber \\
\bar{\zeta}_3\!=&
- \cos{(\bar{\omega}_{\text{eff}} \tau)}\sin{(\bar{\theta}_{\text{E}}\!-\!\bar{\chi})}\sin{\bar{\chi}}\! + \!\cos{(\bar{\theta}_{\text{E}}\!-\!\bar{\chi})}\cos{\bar{\chi}} - \cos{\bar{\theta}_{\text{E}}}. \nonumber
\end{align}
where we used the important fact that earth lies along $n_{\text{E}}^{\hat{\bar{j}}}$ in the adapted-Kerr static Killing FS spatial-triad (\ref{eq:Null_Tangent_Stationary}) or equivalently in the Euclidean notation along,
\begin{equation} \label{eq:n_aFS_tau}
\hat{\bar{n}}_{\text{E}}(\tau) = \left(\cos{(\Omega \tau)}\sin{\bar{\theta}_{\text{E}}}, \mp\sin{(\Omega \tau)}\sin{\bar{\theta}_{\text{E}}}, \cos{\bar{\theta}_{\text{E}}}\right).
\end{equation}
In the above (\ref{eq:zeta_Gen}, \ref{eq:n_aFS_tau}) the signs correspond to the signs of $e_{\hat{\bar{3}}}$ relative to the direction of the spin of the black hole $\hat{z}$ (\ref{eq:e3_relative_z}).

The Fourier transform of this beam signal will contain information regarding various combinations of all three frequencies in the problem now, $\Omega, \bar{\sigma}_1, \bar{\omega}_{\text{eff}}$, from which one can extract necessary data to potentially solve the inverse problem and obtain black hole parameters namely $a, M$, as well as the orbital frequency of the pulsar $\Omega$. We do not attempt a detailed analysis of the trends of the Fourier transforms here but in Section \ref{sec:Results}, we argue that the precession frequencies in various realistic scenarios become comparable to the intrinsic spin period of the pulsar and hence these effects must be taken into account when analysing pulsar profiles of pulsars present near rotating compact objects. The redshift relations for these observers can be found in \cite{Herrera-Aguilar_Nucamendi15}, for example.

Lastly, for pulsars moving on circular orbits close to the event horizon of a Kerr black hole, i.e. for $r \rightarrow r_{\text{H}}$ we have, $\bar{\sigma}_1 \rightarrow \infty$ and $\Omega \rightarrow \Omega_{\text{H}}$, where we have defined the horizon frequency as,
\begin{equation}
\Omega_{\text{H}} \equiv \lim_{r\rightarrow r_{\text{H}}} \Omega_+ = \lim_{r\rightarrow r_{\text{H}}} \Omega_- = \frac{a}{2 M r_{\text{H}}}.
\end{equation}
In this limit, $\bar{\chi} \approx 0$ and $\bar{\omega}_{\text{eff}} \approx \sigma_1$, and in this extreme case, the deflection vector just becomes, 
\begin{align} \label{eq:zeta_Event_Horizon}
\bar{\zeta}_1 \approx& \left[1 - \cos{\left(\Omega_{\text{H}}\tau\right)}\right]\sin{\bar{\theta}_{\text{E}}}, \\
\bar{\zeta}_2 \approx& \pm \sin{\left(\Omega_{\text{H}}\tau\right)}\sin{\bar{\theta}_{\text{E}}}, \nonumber \\
\bar{\zeta}_3 \approx& 0. \nonumber
\end{align}
That is, for pulsars very close to the event horizon of a Kerr black hole, one obtains pulses on earth every $\nu_{\text{FS}} \approx \Omega_{\text{H}}/2\pi$.

\section{Analysis \& Results} \label{sec:Results}
\begin{figure}
\centering
\includegraphics[scale=.65]{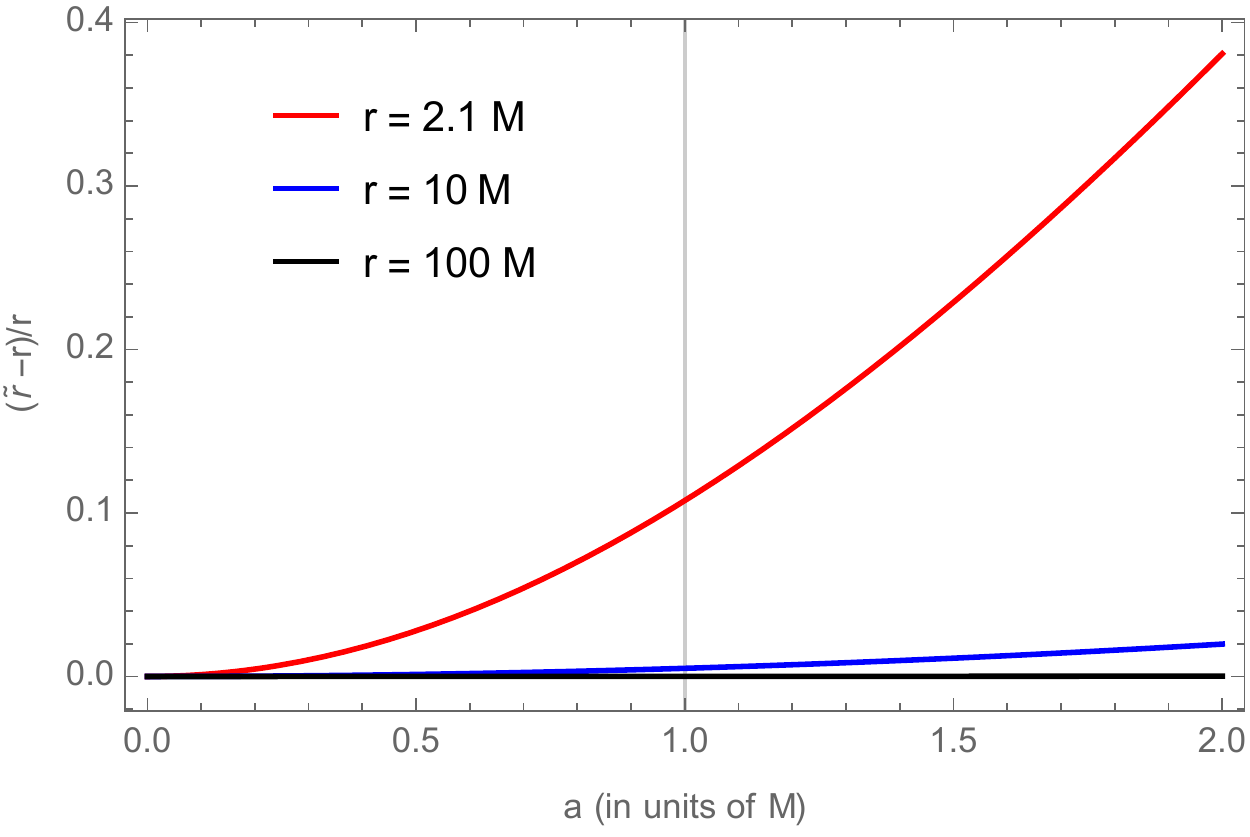}
\caption{Here we highlight the difference in the useful Boyer-Lindquist (BL) coordinate $r$ and the physical Kerr-Schild coordinate $\tilde{r}$ in the equatorial plane of the Kerr spacetime, by displaying the variation of $\tilde{r}/r - 1$ with change in $a$ and $r$. Therefore, this roughly measures the distortion from spherical symmetry, and we show here that these coordinates agree well for larger distances $r$ and smaller central spins $a$. Also, since this is a fractional quantity, evidently these qualitative features are representative of black holes ($0 \leq a \leq M$) and naked singularities ($M < a$), independently of their masses. In particular, if we consider a pulsar at $r = 2.1 M$ near a BH with spin $a = .1 M$ and one with $a = M$, the corresponding physical distances would be $\tilde{r} \approx M$ and $\tilde{r} \approx 1.1 \times 2.1 M$ respectively. Clearly, for larger central masses, this difference in measurement of distances can be significant ($1 M_\odot = 1.5$ km). Note that $r_+ = 2M$ is the location of the ergosurface (BL), and is independent of the central spin.}
\label{fig:BL_vs_KS_a}
\end{figure}

In the remainder of this article, we will switch from Geometrized units ($G \! = \! c \! = \! 1$) to physical units, and keep only two significant digits when reporting values of physical quantities. The conversions for distances, angular frequencies, and accelerations are given as: $1 M_\odot = 1.5$ km, $1 M_\odot^{-1} = 1.3 \times 10^6$ rad/s and $1 M_\odot^{-1} = 6.0 \times 10^{10}$ km/$s^2$ respectively. Also, we will freely use both the convenient Boyer-Lindquist $r$ and the more physical Cartesian Kerr-Schild $\tilde{r}$ radial coordinates (see eq. \ref{eq:KS_and_BL}). Therefore, we find it useful to mention here that the fractional change in the distances measured in these coordinate systems, $\tilde{r}/r - 1$, decreases with distance from and increases with spin of the central Kerr object, as is evident from Fig.~\ref{fig:BL_vs_KS_a}.

\begin{figure*}
\centering
\subfigure[$\omega = 2\pi$ rad/s, $a = .1 M, r=50 M$, $f_{\text{p}} = 5.9, f_{\text{o}} = 5.9$]
{\includegraphics[scale=.69]{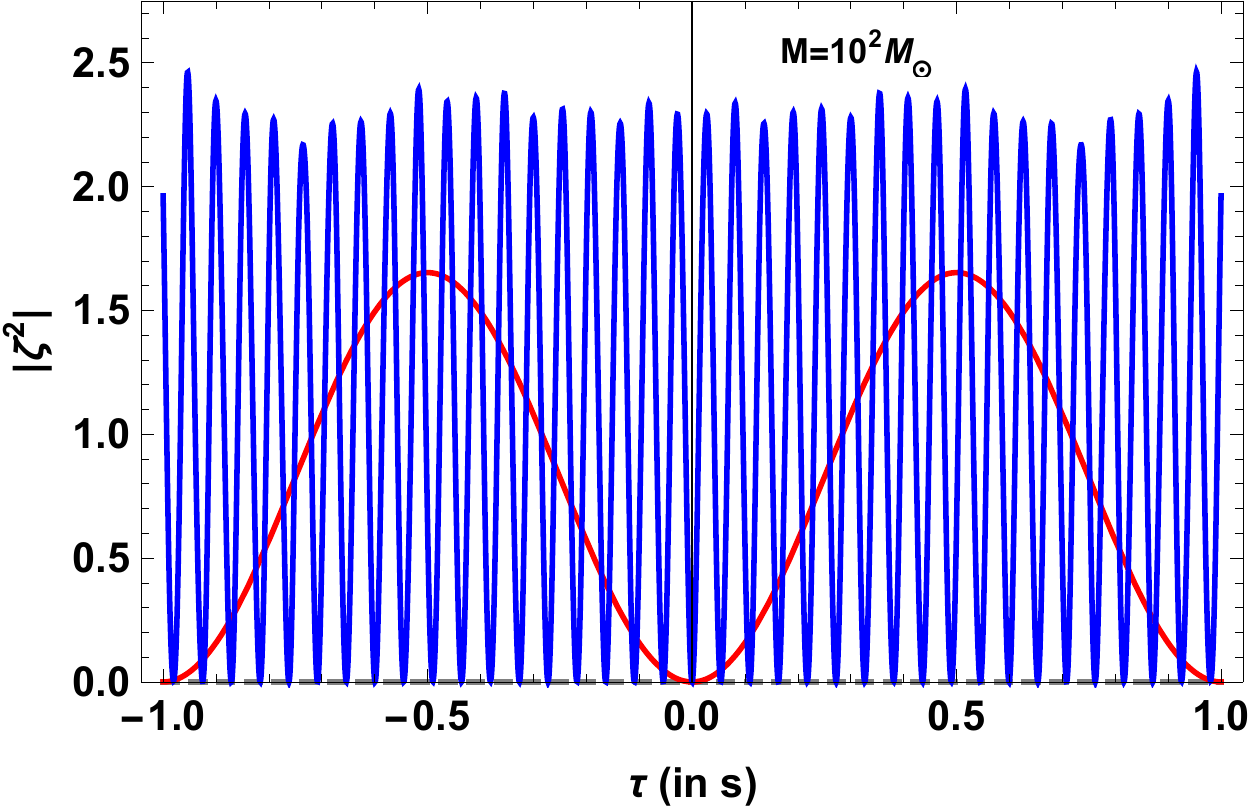}}
\subfigure[$\omega = 2\pi$ rad/s, $a = .1 M, r=50 M$, $f_{\text{p}} = 5.9 \times 10^{-3}, f_{\text{o}} = 5.9 \times 10^{-3}$]
{\includegraphics[scale=.69]{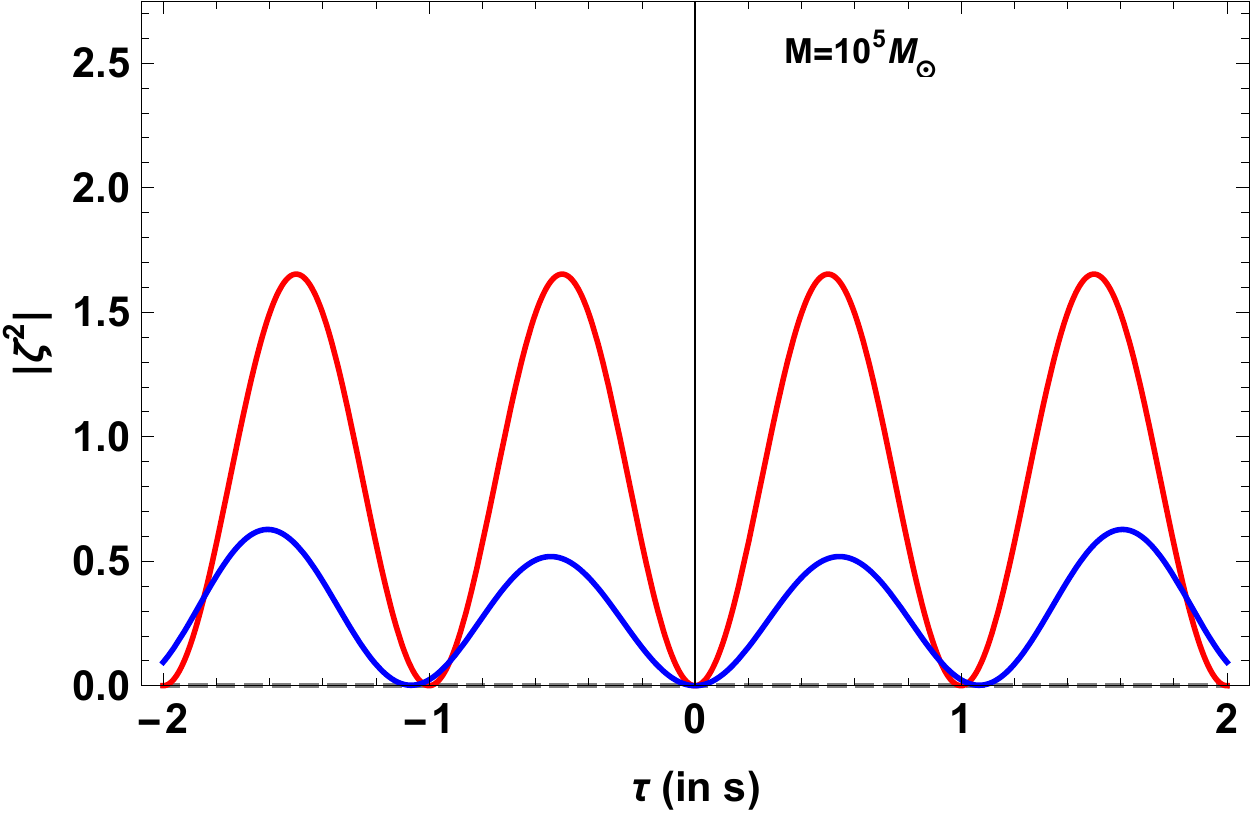}}
\subfigure[$a = .5 M, \omega = 2\pi$ rad/s, $r=10^2 M$, $f_{\text{p}} = 2.1, f_{\text{o}} = 2.1$]
{\includegraphics[scale=.69]{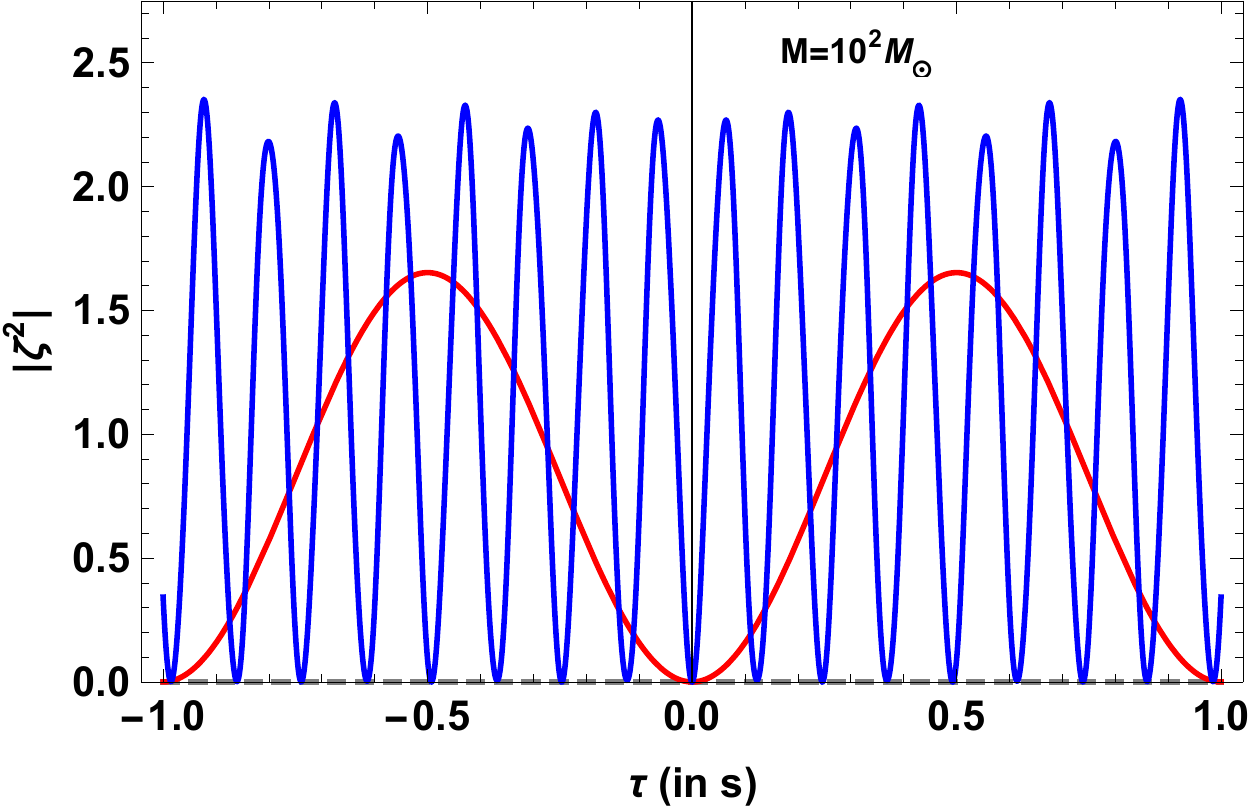}}
\subfigure[$a = .9 M, \omega = 2\pi$ rad/s, $r=100 M$, $f_{\text{p}} = 2.1 \times 10^{-3}, f_{\text{o}} = 2.1 \times 10^{-3}$]
{\includegraphics[scale=.69]{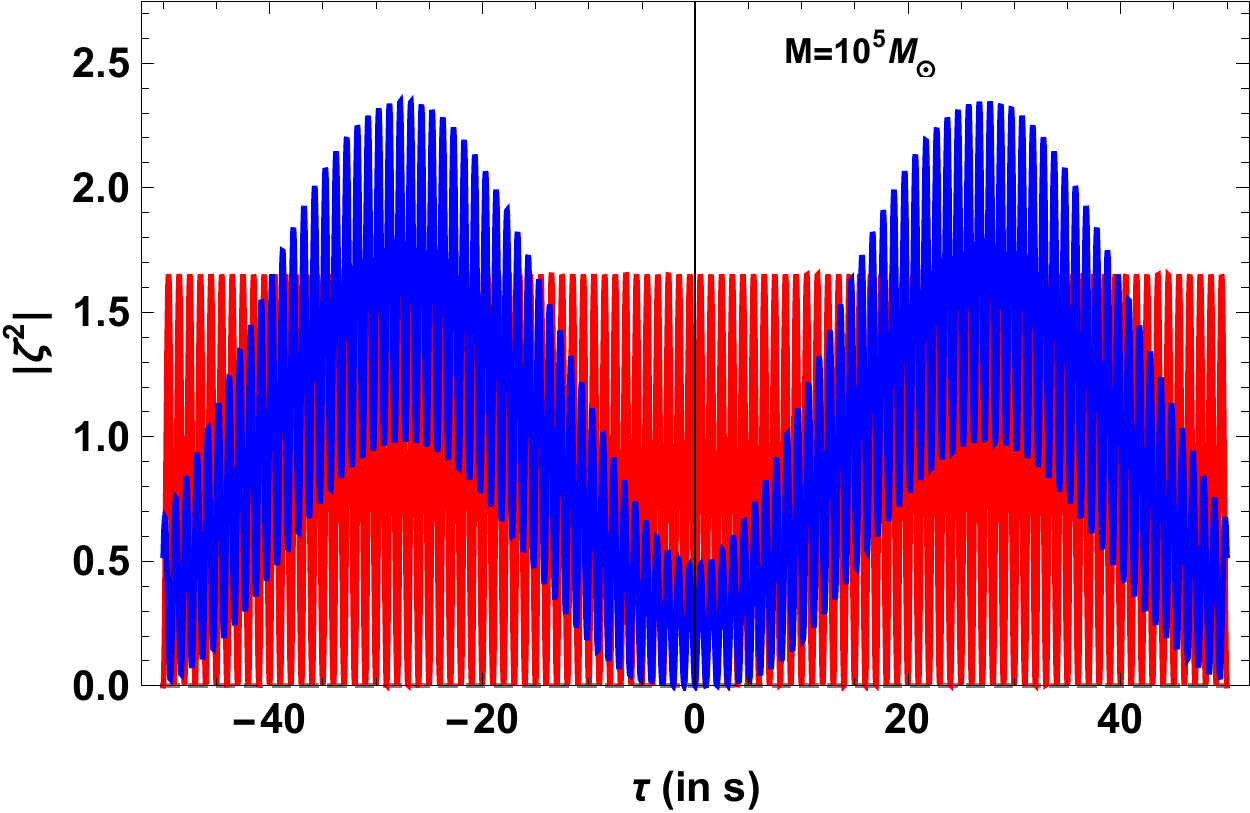}}
\subfigure[$a = .9 M, \omega = 200\pi$ rad/s, $r=10 M$, $f_{\text{p}} = .65, f_{\text{o}} = .64$]
{\includegraphics[scale=.69]{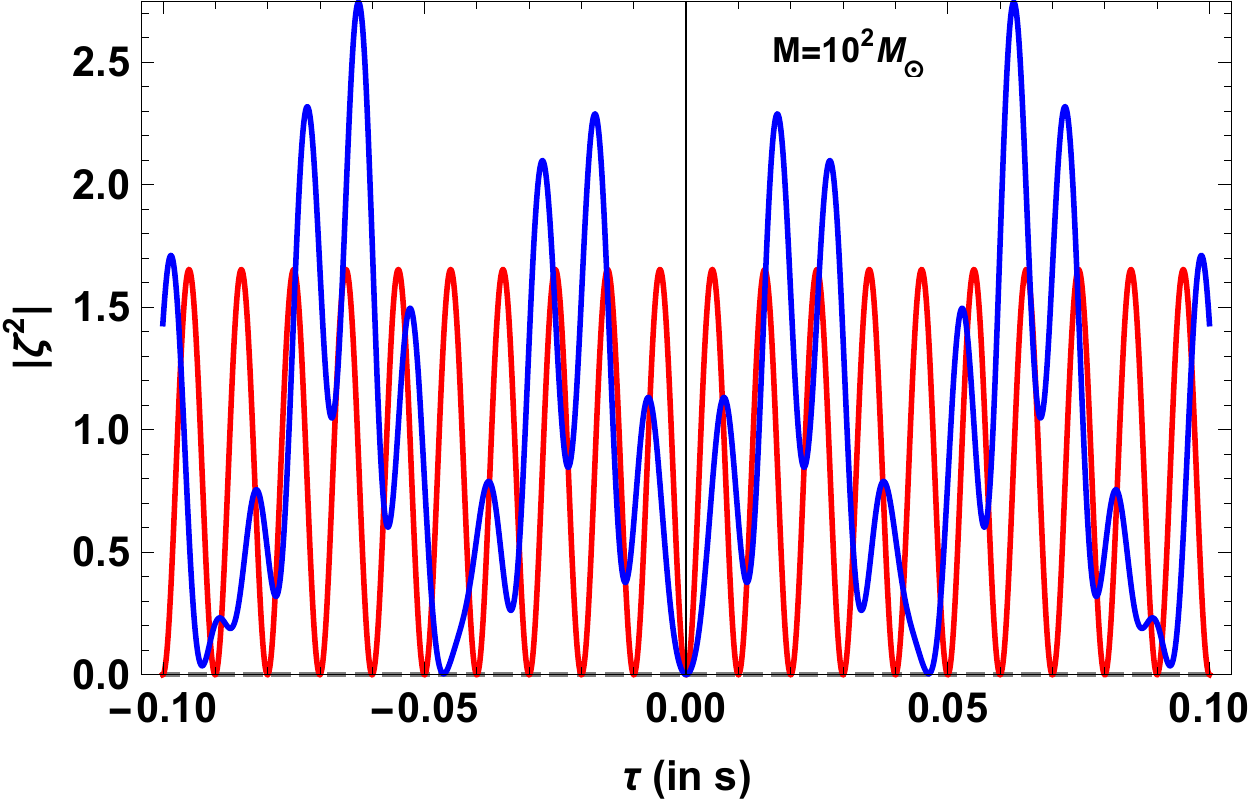}}
\subfigure[$a = .5 M, \omega = 200\pi$ rad/s, $r=10 M$, $f_{\text{p}} = 6.5 \times 10^{-4}, f_{\text{o}} = 6.4 \times 10^{-4}$]
{\includegraphics[scale=.69]{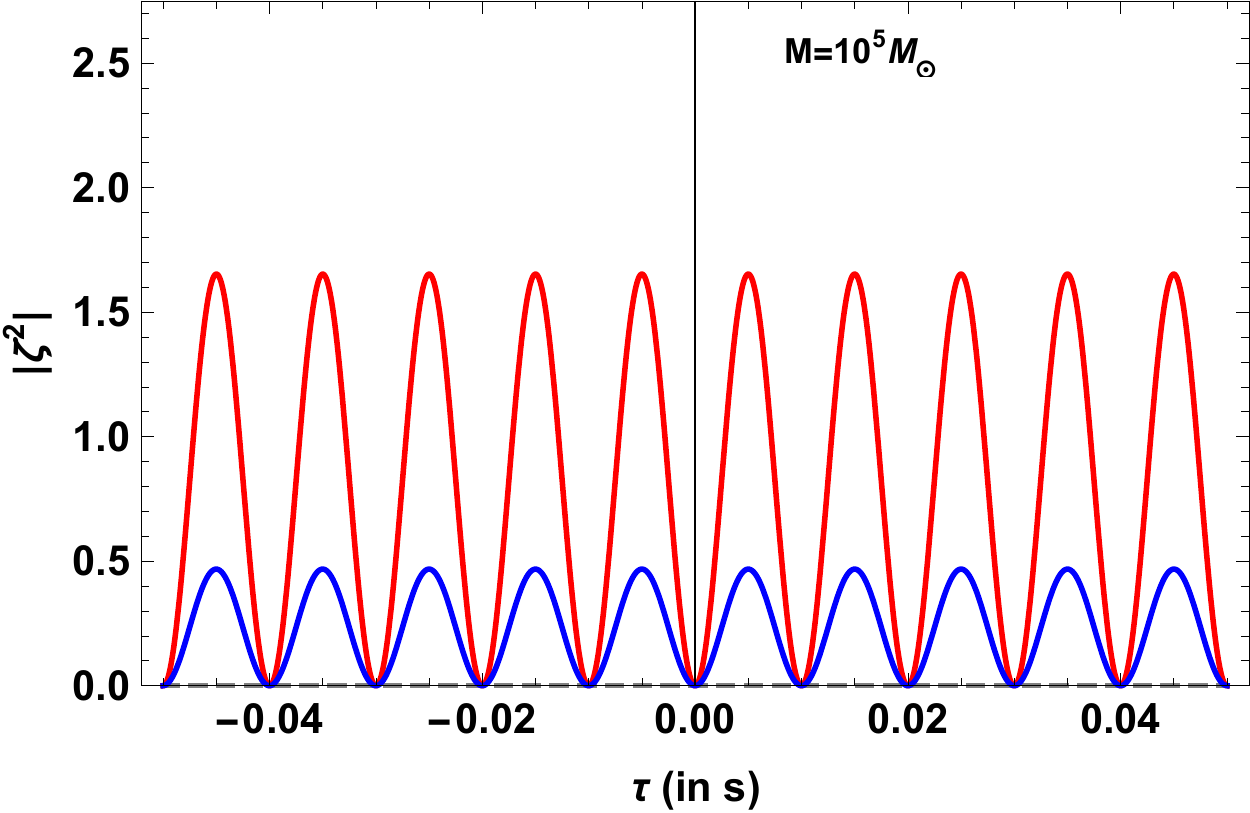}}
\caption{We demonstrate here changes in the time profile of the squared-modulus of the deflection vector $|\zeta^2|$ of a pulsar present near Kerr BHs, of masses $10^2 M_\odot$ (IMBHs) and $10^5 M_\odot$ (SMBHs), due to gravitomagnetism. In red we represent an isolated spatially-fixed pulsar $\left(\Omega = 0, \sigma_1 = 0\right)$, and in blue we have a pulsar moving on an equatorial circular geodesic $\left(\Omega = \Omega_{\text{K}+}\right)$ around a Kerr black hole. The organisation in this figure is as follows: the top and middle rows are for pulsars with spin frequencies $\omega = 2\pi$ rad/s, and the bottom row is for ms-pulsars with $\omega = 200\pi$ rad/s. We have used placeholder values of $\beta = 30^\circ, \theta_{\text{E}} = 50^\circ$ in all panels. We have used assorted values for the black hole spin parameter $a$ and the orbit radius $r$ for pulsars in the Kerr spacetime, indicated below each panel; we have also included there the values of the associated precession and orbital angular velocity ratios $f_{\text{p}}$ and $f_{\text{o}}$, which effectively determine the Fourier spectrum of $|\zeta^2(\tau)|$. Note that these plots are associated with the rest frame of the pulsar. Nonetheless, it is clear that even after incorporating redshift considerations gravitomagentic spin-precession in the vicinity of a Kerr black hole can cause significant deviations in the observed time profile of such a pulsar from that of an isolated pulsar. Now, even though the precession frequencies $\bar{\sigma}_1$ (see eq. \ref{eq:Kepler_Precession}) experienced by pulsars on such Keplerian orbits are independent of the spin-parameter of the central Kerr compact object $a$, their orbital angular velocities $\Omega_{\text{K}+}$ (see eq. \ref{eq:Kepler_Frequencies}) and the physical size of their orbits $\tilde{r}$ (Kerr-Schild radial coordinate) are not. Therefore, from observed profiles of such pulsars, one can obtain independent estimates for both BH parameters. It is useful to remember that $\tilde{r}$ can be obtained from the Boyer-Lindquist radius values given here using eq. (\ref{eq:KS_and_BL}) as $\tilde{r} = \sqrt{r^2 + a^2} \times 1.5 \times (M/M_\odot)\ $km, where $r$ and $a$ are in units of $M$. So, the size of the orbit in panel (c) is $\tilde{r} = 1.5 \times 10^4$ km.}
\label{fig:timeplotters}
\end{figure*}

\begin{figure*}
\centering
\subfigure[$M = 10^2 M_\odot, \Omega = 0, \omega = 2\pi\ $rad/s]
{\includegraphics[scale=.46]{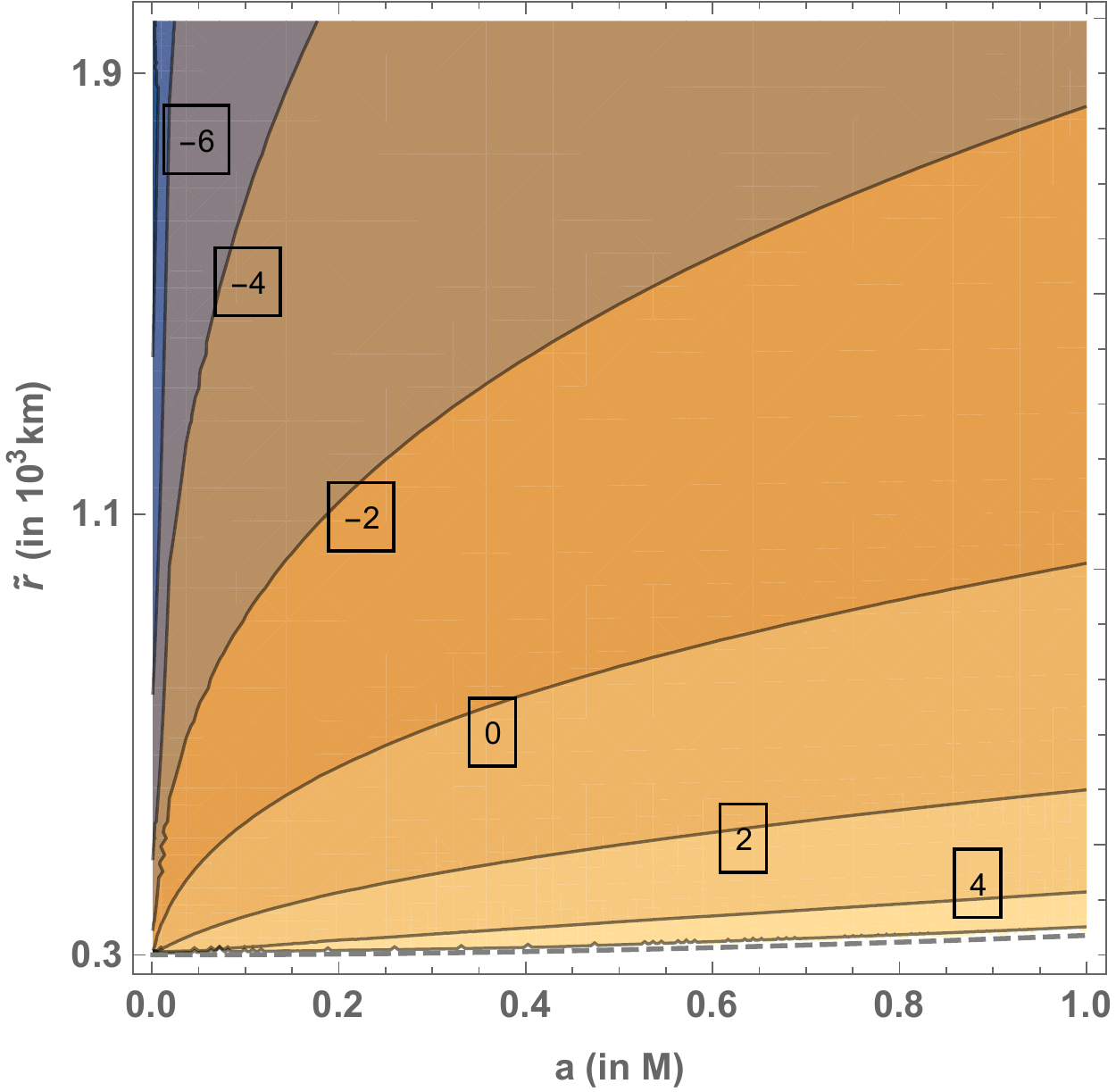}}
\subfigure[$M = 10^2 M_\odot, \Omega = \Omega_{\text{Z}}, \omega = 2\pi\ $rad/s]
{\includegraphics[scale=.46]{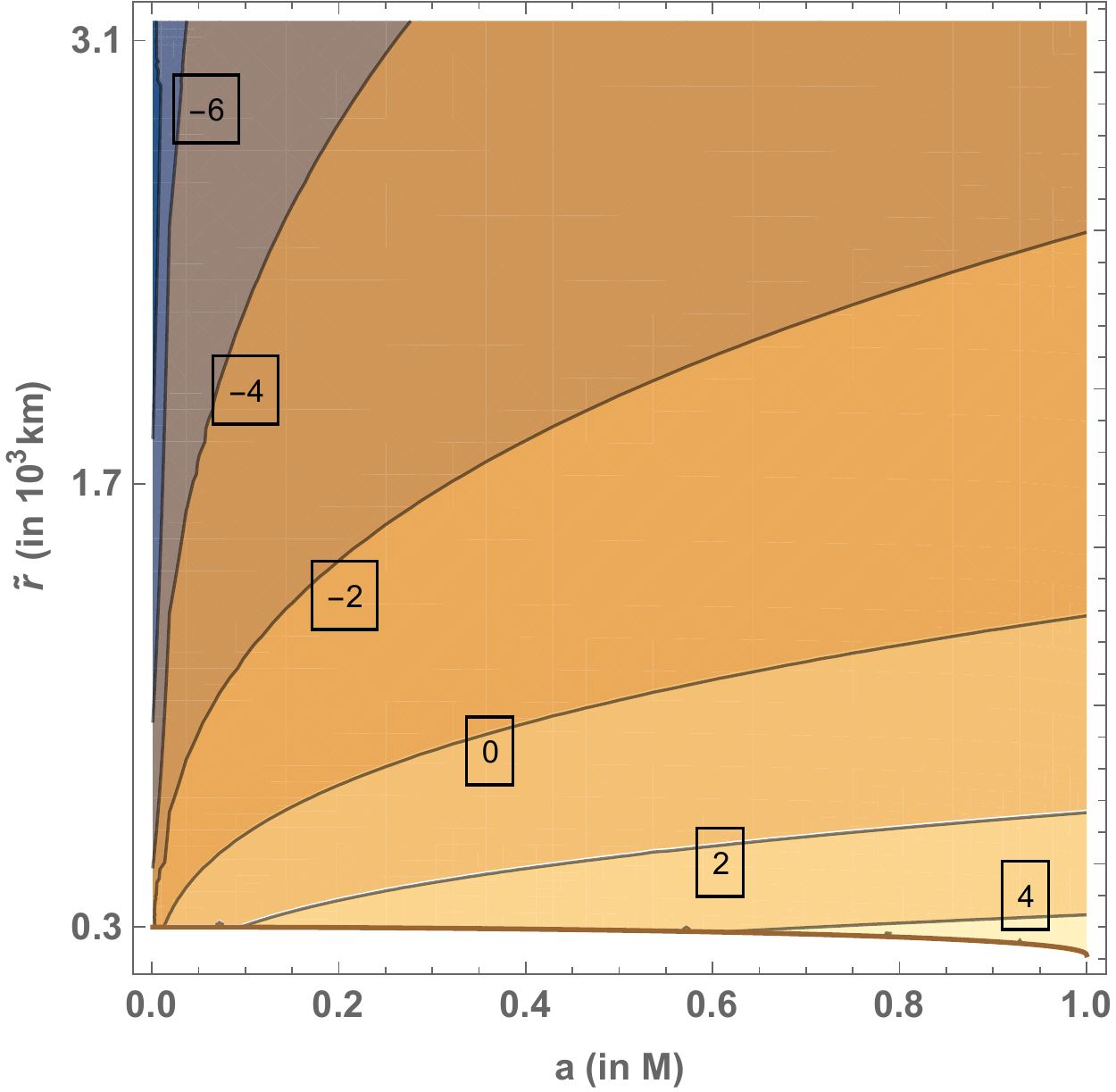}}
\subfigure[$M = 10^2 M_\odot, \Omega = \Omega_{\text{K}+}, \omega = 2\pi\ $rad/s]
{\includegraphics[scale=.46]{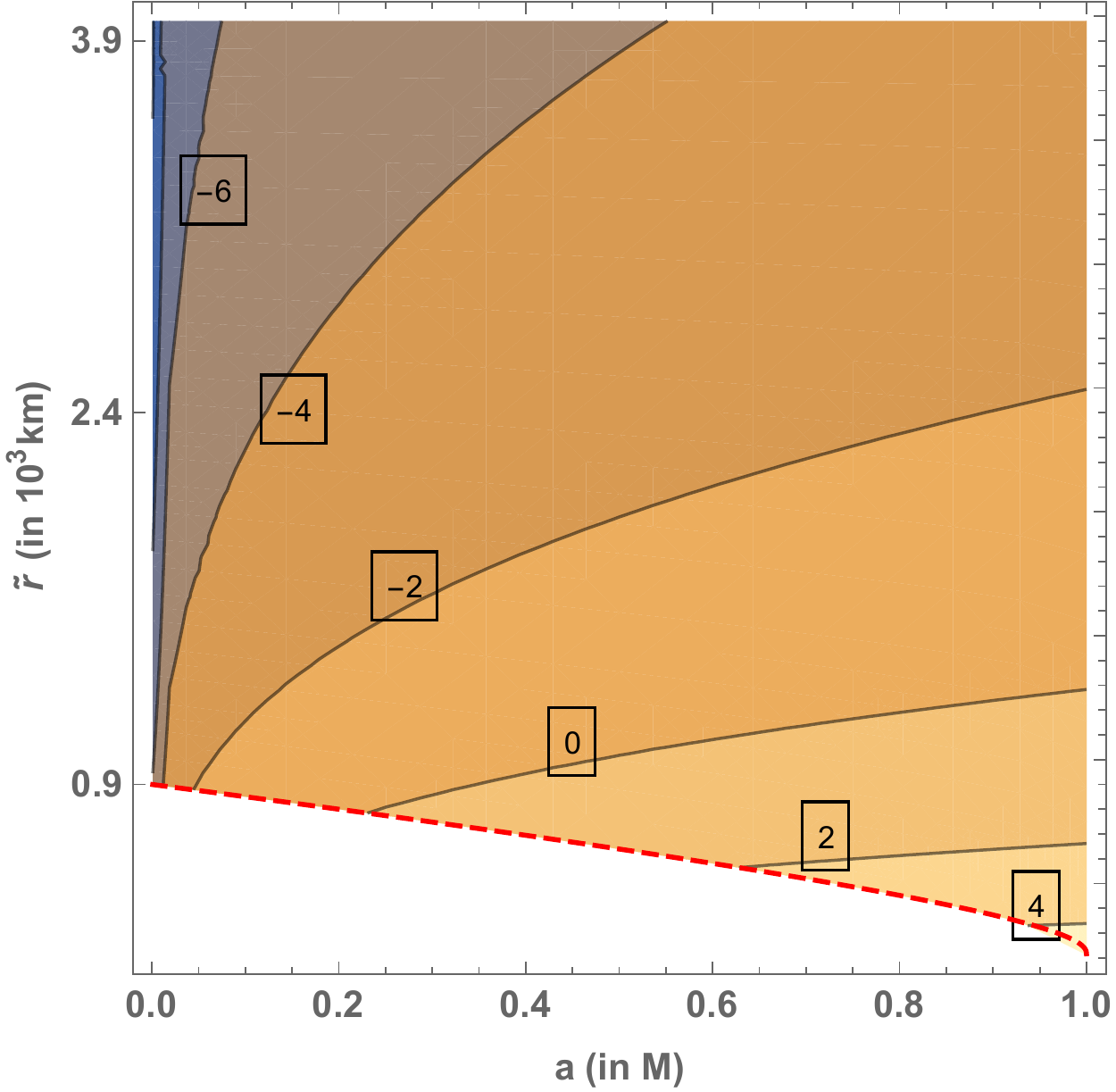}}
\subfigure[$M = 10^2 M_\odot, \Omega = 0, \omega = 200\pi\ $rad/s]
{\includegraphics[scale=.46]{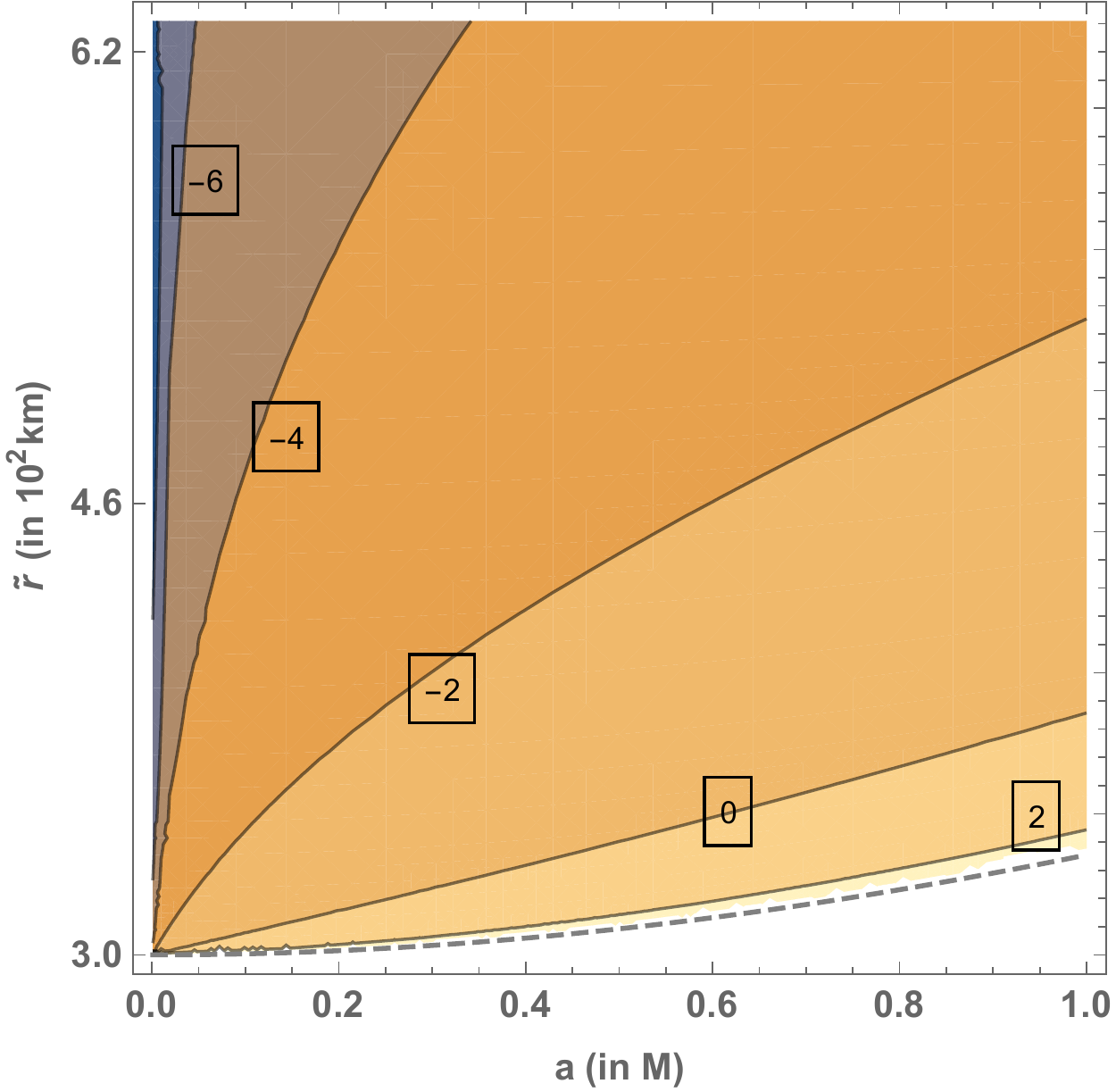}}
\subfigure[$M = 10^2 M_\odot, \Omega = \Omega_{\text{Z}}, \omega = 200\pi\ $rad/s]
{\includegraphics[scale=.46]{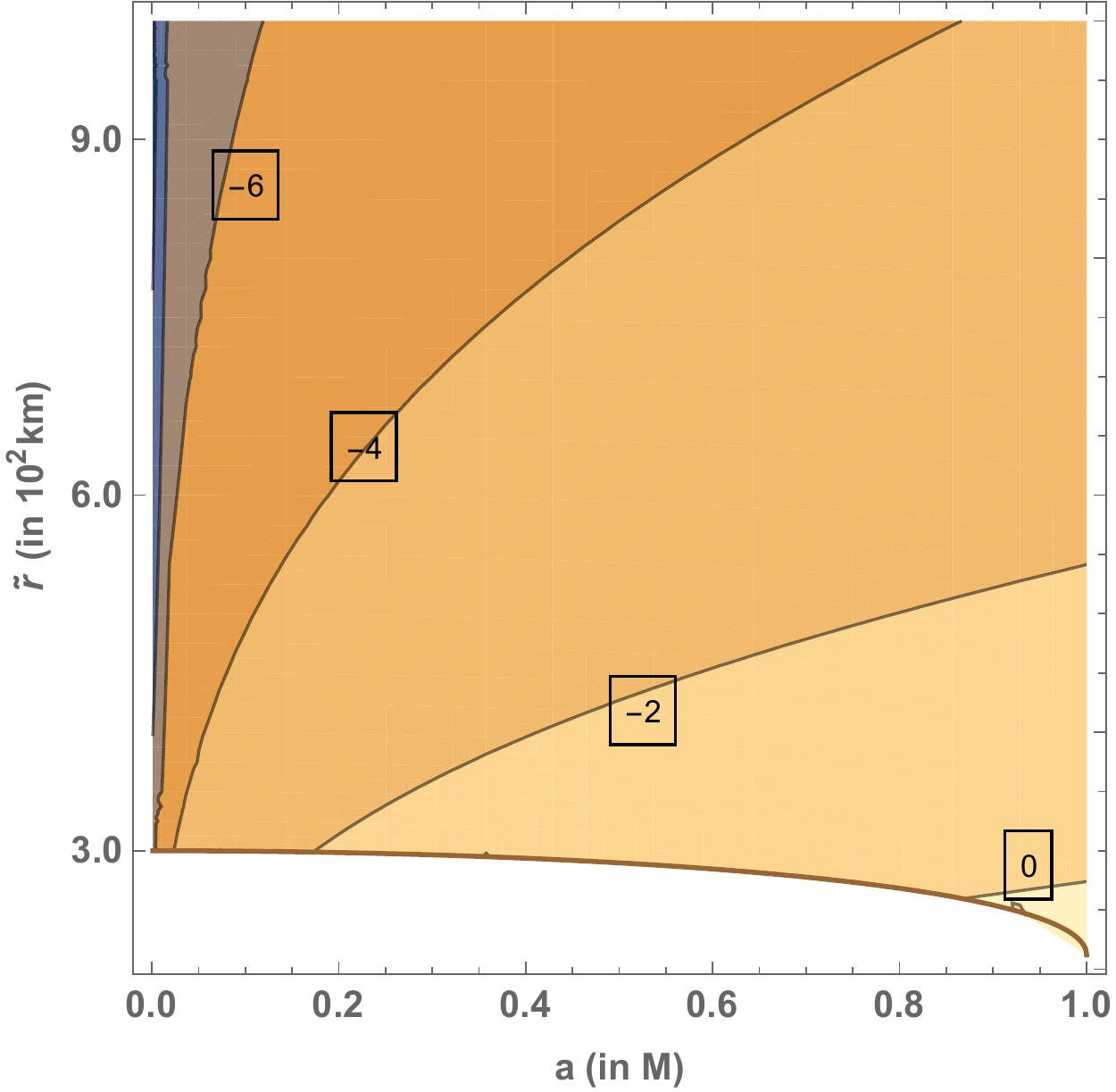}}
\subfigure[$M = 10^2 M_\odot, \Omega = \Omega_{\text{K}+}, \omega = 200\pi\ $rad/s]
{\includegraphics[scale=.46]{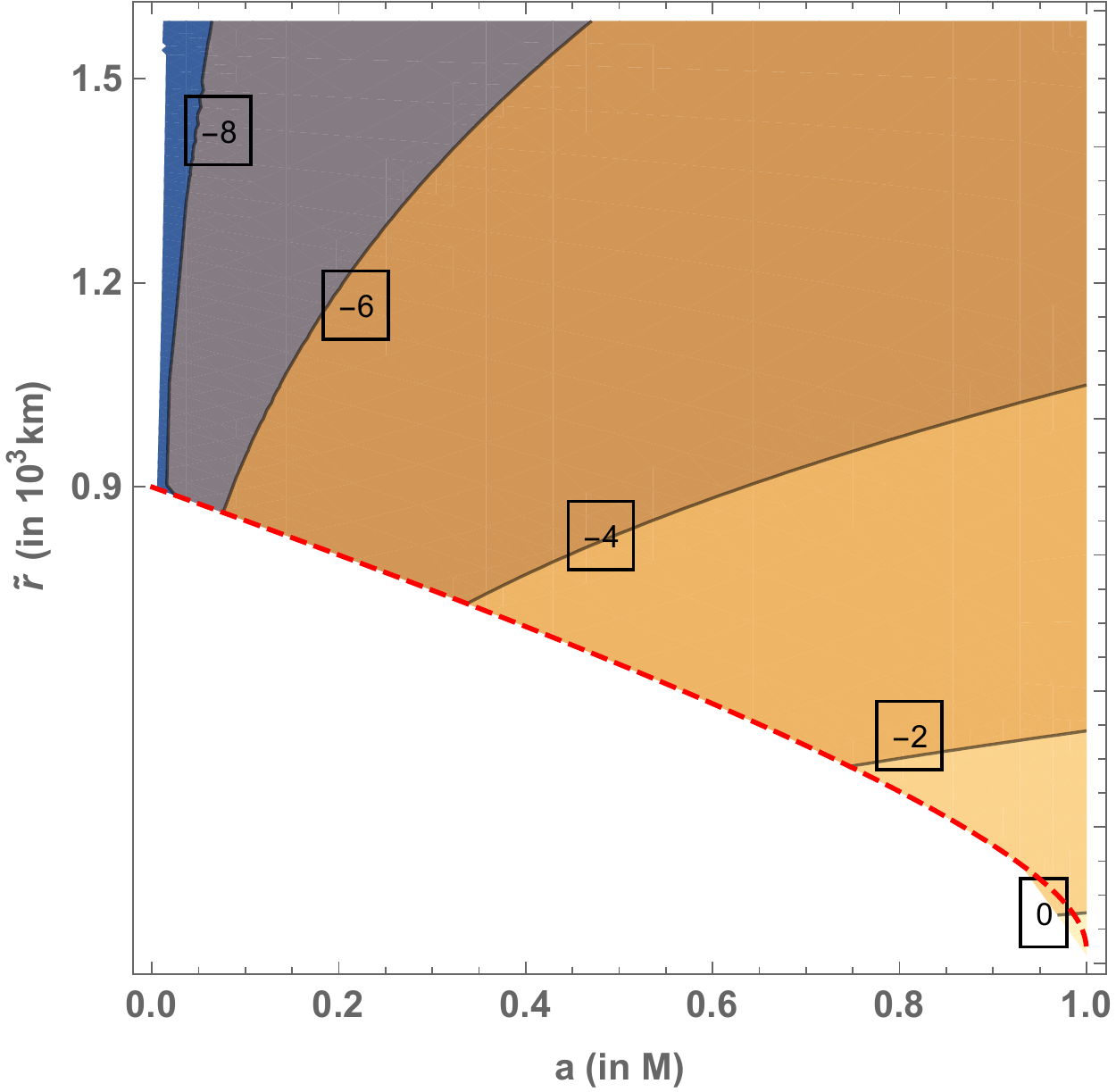}}
\subfigure[$M = 10^5 M_\odot, \Omega = 0, \omega = 2\pi\ $rad/s]
{\includegraphics[scale=.46]{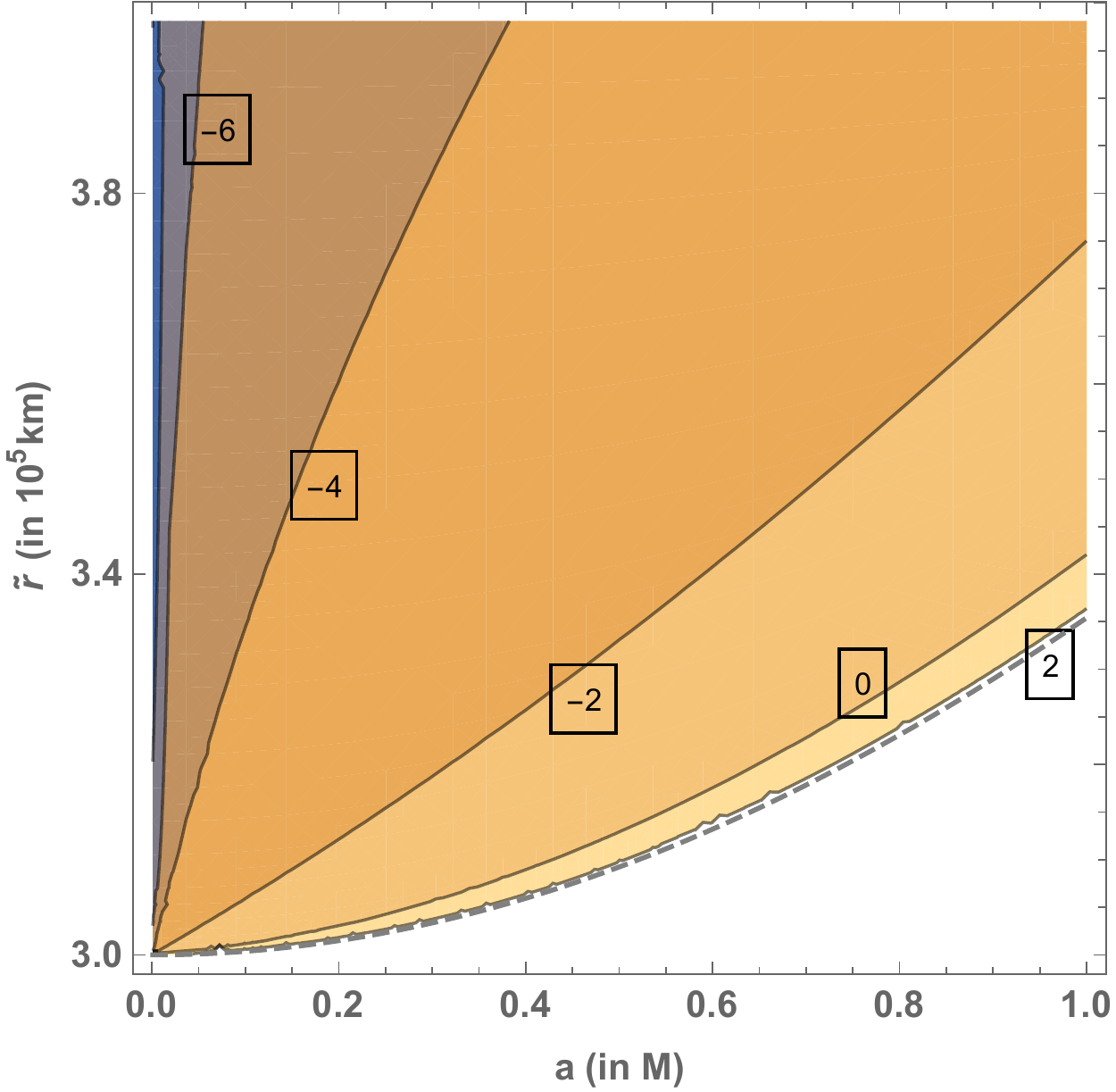}}
\subfigure[$M = 10^5 M_\odot, \Omega = \Omega_{\text{Z}}, \omega = 2\pi\ $rad/s]
{\includegraphics[scale=.46]{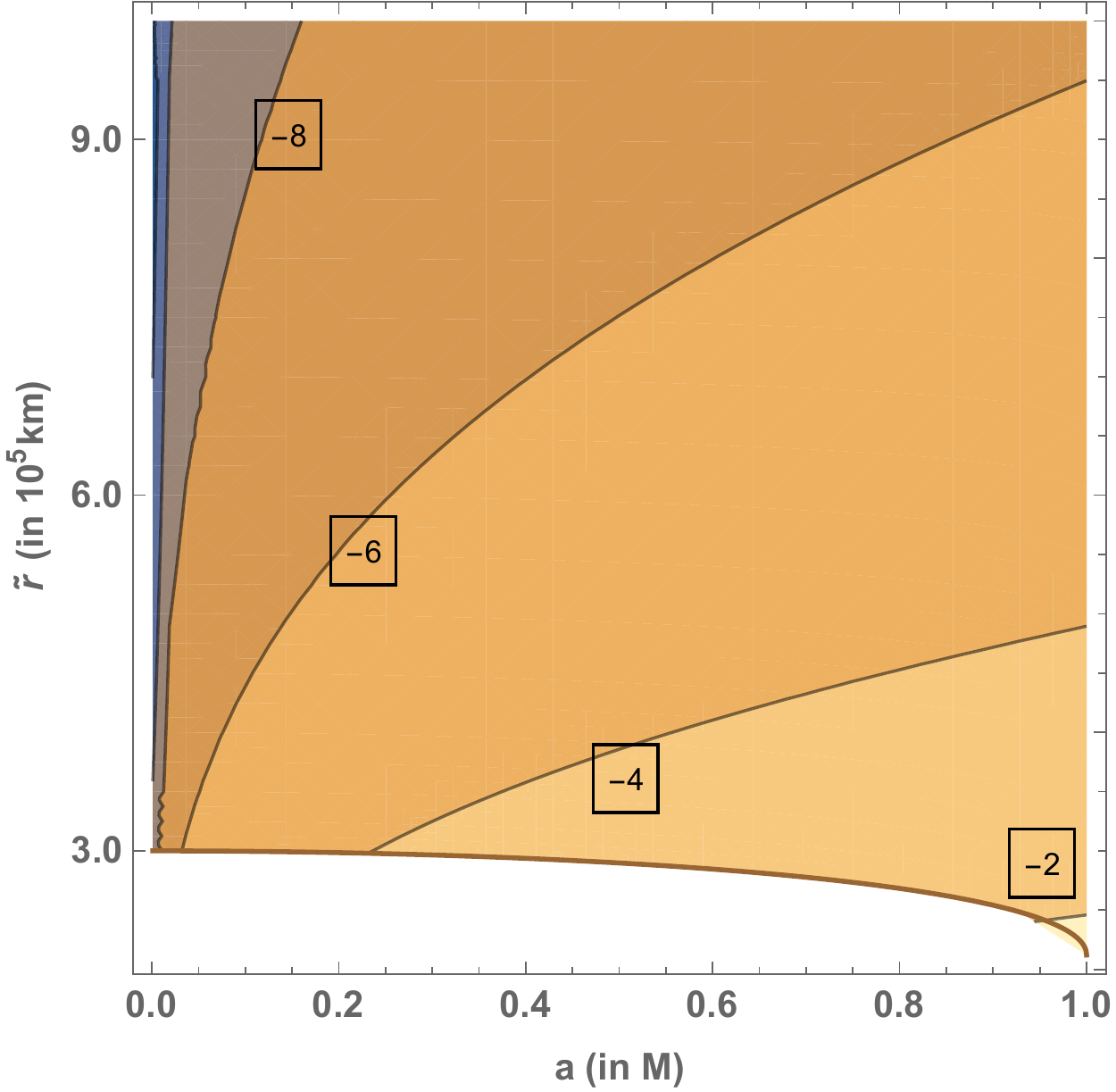}}
\subfigure[$M = 10^5 M_\odot, \Omega = \Omega_{\text{K}+}, \omega = 2\pi\ $rad/s]
{\includegraphics[scale=.46]{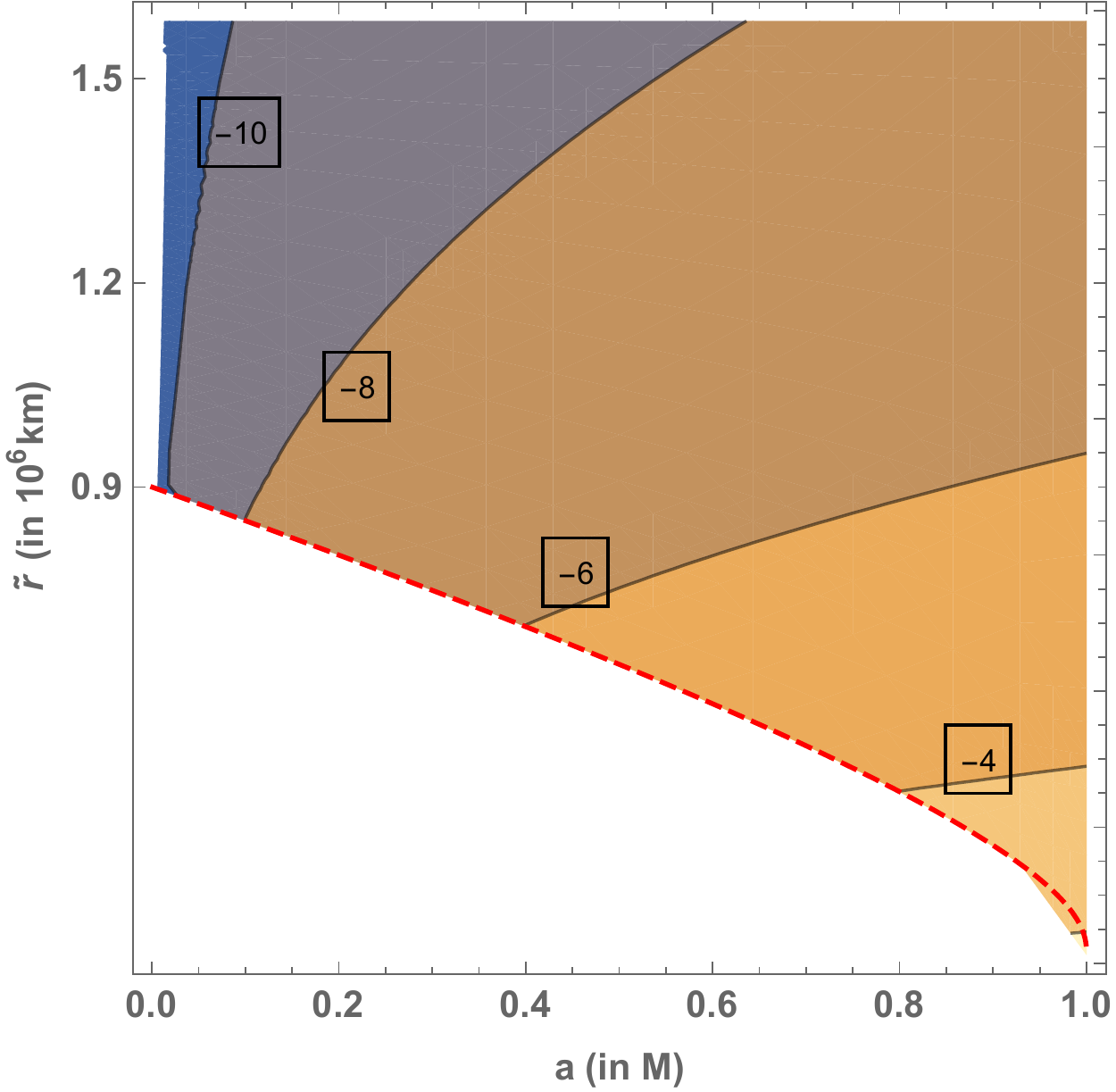}}
\caption{The aim of this figure is to serve as an indicator for how the strength of the precession frequency varies with BH mass and spin, and position and orbital angular velocity of the pulsar. Towards this end, we show here the contour plot of $\log{f_{\text{p}}} = \log{\left(|\bar{\sigma}_1|/\omega\right)}$ in the $a \! - \! \tilde{r}$ parameter space. On the $x$-axis, $a$ is represented in units of $M$ and on the $y$-axis, the physical Kerr-Schild size $\tilde{r}$ (see eq. \ref{eq:KS_and_BL}) of the pulsar orbit is given in km. The masses of the BHs and the intrinsic spin angular frequencies of the pulsars are given below panels. We consider here BHs of masses $10^2 M_\odot$ (IMBHs) and $10^5 M_\odot$ (SMBHs), the associated gravitational radii ($r_{\text{g}} = M$) of which are $1.5 \times 10^2\ $km and $1.5 \times 10^5\ $km respectively. In the left column, pulsars move at very low orbital angular frequencies $\Omega \approx 0$ (static observers). In the centre column, pulsars move at the ZAMO orbital angular frequency, $\Omega = \Omega_{\text{Z}}$, and in the right column, pulsars move on stable co-rotating Kepler orbits, i.e. $\Omega = \Omega_{\text{K}+}$. It is useful to remember that static observers, ZAMOs and stable co-rotating Kepler observers are allowed outside the ergoregion (dashed-gray), horizon (brown) and the co-rotating ISCO (dashed-red) respectively. Now, when $\log{f_{\text{p}}} \gtrsim 0$, effects of gravitomagnetic spin-precession appear on the dynamical time-scale of the pulsar. When this quantity is much smaller, we require long-time observations to not only extract the precession frequency (and therefore BH parameters), but to even \textit{see} the pulsar again (see for example panel (d) of Fig.~\ref{fig:timeplotters}). From the current figure, it is evident that gravitomagnetic spin-precession effects are significant, and therefore must necessarily be incorporated in pulsar timing analyses, for slowly-spinning (sub-ms) pulsars present sufficiently close to IMBHs Further, studying the spin-precession properties of such systems would yield excellent constraints on BH parameters.}
\label{fig:Ratios_sigma_1_omega_eff}
\end{figure*}

Firstly, let us note that for an isolated pulsar ($\sigma_1 = 0$) at rest ($\Omega = 0$), we can write the deflection vector from (\ref{eq:zeta}) as, 
\begin{align}
\zeta_1=& (\cos{(\omega \tau)} - 1)\sin{(\theta_E - \beta)}\cos\beta, \nonumber \\
\zeta_2 =& -\sin{(\omega \tau)}\sin{(\theta_E - \beta)}, \\
\zeta_3=& (1 - \cos{(\omega \tau)})\sin{(\theta_E - \beta)}\sin\beta. \nonumber
\end{align}
Clearly, the Fourier spectrum of the squared modulus of the deflection vector $|\zeta^2(\tau)|$ for such a pulsar has a single peak corresponding to its intrinsic spin angular frequency $\omega$. On the other hand, the Fourier spectrum of $|\zeta^2(\tau)|$ for a pulsar near a Kerr black hole is multiply peaked, with these peaks correspond to the magnitudes of its orbital angular velocity $\Omega$, its precession frequency $\bar{\sigma}_1$ and the effective vector sum of the intrinsic spin angular frequency ($\omega$) and the precession frequency, denoted by $\omega_{\text{eff}}$. Of course, this spectrum contains other peaks corresponding to certain sums and differences of these three frequencies. However, for qualitative insight into how gravitomagnetic spin-precession (spin-spin coupling) and orbital motion (spin-orbit coupling) affect pulsar timing, it is useful to consider the following ratios, 
\begin{equation}
f_{\text{p}} = \frac{|\bar{\sigma}_1|}{\omega},\ f_{\text{o}} = \frac{\Omega}{\omega},
\end{equation}
and we can capture the role of $\omega_{\text{eff}}$ on pulsar timing measurements, using $f_{\text{p}}$ and $\beta$, via,
\begin{equation}
\frac{\omega_{\text{eff}}}{\omega} = \sqrt{1 + f_{\text{p}}^2 \pm 2 f_{\text{p}} \cos\beta}.
\end{equation}
Now, since the spin-precession \eqref{eq:Precession_Frequency_aFS_Eq} and orbital angular frequencies 
\eqref{eq:Omegapm} typically decrease with an increase in size of pulsar orbit $r$, it makes sense to analyse pulsar-BH binaries depending on their sizes. Therefore, we can broadly divide these systems into two categories: (a) when a pulsar is present sufficiently close to the BH such that the relevant ratios $f_{\text{p}}(r)$ and $f_{\text{o}}(r)$ are comparable to, or much larger than unity, then gravitomagnetic precession effects modify the pulse profile on the time-scale of the pulsar's intrinsic spin period; and (b) when a pulsar is sufficiently far away such that these ratios are much smaller than unity, one finds longer-time-scale variations in the pulse profile (for example, as secular shifts in the times of arrival of pulses). To best demonstrate these points, we will now consider Figs.~\ref{fig:timeplotters} and \ref{fig:Ratios_sigma_1_omega_eff}, and Tab.~\ref{table:Exploring_Omega_kappa_sigma}. In Fig.~\ref{fig:timeplotters}, we consider the all-important case of pulsars moving on stable equatorial circular geodesics around IMBHs of mass $10^2 M_\odot$ and SMBHs of mass $10^5 M_\odot$ (similar to the BHs of the Seyferts in the study of \citealt{Greene_Ho07}), and analyse how gravitomagnetic spin-precession effects modify the times of arrival of pulses. In Fig.~\ref{fig:Ratios_sigma_1_omega_eff}, the focus will be on the strength of the spin-precession frequency (in physical units) relative to the intrinsic spin-frequency of typical pulsars, to understand which effect dominates the beam evolution, with (a) increase in distance between the pulsar and BH, (b) change in the orbital angular velocity of the pulsar, and (c) with change in mass and spin of the BH. We conclude the present analysis with an exhaustive catalogue of the possible orbital angular velocities, accelerations and precession frequencies experienced by pulsars present at varying distances near intermediate-mass Kerr BHs of varying spins $a = .1 M, .5 M, .9 M$ but of same mass $10^2 M_\odot$.

Now, we show in Fig. \ref{fig:timeplotters} the time plots of the squared-modulus of the deflection vector $|\zeta^2(\tau)|$ for pulsars moving on (stable) equatorial circular Kepler orbits with $\Omega = \Omega_{\text{K}+}$ near intermediate-mass ($10^2 M_\odot$) and supermassive ($10^5 M_\odot$) Kerr BHs (in blue), contrasted against $|\zeta^2(\tau)|$ for isolated pulsars (in red), for different values of the spin for the BH $a$, spin-frequencies of the pulsar $\omega$ (corresponding to normal pulsars which have $\omega = 2\pi\ $rad/s, and ms-pulsars for which $\omega = 200\pi\ $rad/s) moving on assorted values of orbit radii $r$. As discussed in Section \ref{sec:Non_Zero_Beam_Width}, when $|\zeta^2(\tau)| \approx 0$, one obtains a pulse on earth. Therefore, the period at which this quantity vanishes corresponds to period of the pulsar as seen on earth, once redshift is accounted for. From this figure, it is evident that as the BH mass increases, the precession frequency drops, and this is directly manifest in the $|\zeta^2|$ profile. A similar statement holds also for increase in distance between pulsar and black hole. Now, when $f_{\text{p}}, f_{\text{o}} \gtrsim 1$ - see for example panels (a) and (c) - gravitomagnetic spin-precession (significantly) modifies the observed pulse period on the dynamical time-scale itself (which is determined by the spin-period of the pulsar $2\pi/\omega$). Further, even when spin-precession effects are not as strong (panels b, d, e), we still see significant modifications in the pulse period albeit on time-scales larger than the spin-period of the pulsar. These findings indicate that when modelling pulse profiles of pulsars near BHs, one must account for gravitomagnetic spin-precession effects, even if the pulsar is not `very close' to the BH. Finally, when spin-precession effects are very small ($f_{\text{p}}, f_{\text{o}} \ll 1$), as in panel (f), then pulses appear to arrive at $2\pi/\omega$. However, the gravitomagnetic spin-precession effects due to the companion BH are imprinted onto the shape of the pulse. What these results imply are that if one were to find pulsars moving on astrophysically relevant (equatorial stable circular Kepler) orbits in a reasonably large radial-band around a BH, one can obtain an independent estimate of BH parameters by looking for effects of gravitomagnetic spin-precession. For an idea of how far away a pulsar must be from a BH to neglect spin-precession effects, see Fig.~\ref{fig:Ratios_sigma_1_omega_eff} and also glance at Tab.~\ref{table:Exploring_Omega_kappa_sigma}.


Now, in Fig.~\ref{fig:Ratios_sigma_1_omega_eff}, we present a systematic study of how $\log{f_{\text{p}}}$ varies with changes in the mass $M$ and spin $a$ of the BH, the distance of the pulsar from it $\tilde{r}$, the orbital angular frequency of the pulsar $\Omega$ and the intrinsic spin frequency of the pulsar $\omega$. The contour $\log{f_{\text{p}}} \approx 0$ divides the $a - \tilde{r}$ parameter space into the `strong-precession effects' region ($f_{\text{p}} \! \gtrsim \! 1$) and the weak-precession effects region ($1 \! > \! f_{\text{p}} \! \gtrsim \! 10^{-4}$), as discussed above. Now, for pulsars with small spins around IMBHs (top row), the strong- and weak-precession region can be identified roughly (in Kerr-Schild coordinates) as being upto about $\approx \! 1.3 \times 10^3$ km and $\approx \! 4.9 \times 10^3$ km respectively. For pulsars with high spins around BHs of similar masses (middle row), these regions correspond to distances of $\approx \! 3. \! \times \! 10^3$ and $\approx \! 1.1 \! \times \! 10^3$ km respectively. Finally, for pulsars with small spins around SMBHs (bottom row), spin-precession effects are severely suppressed since they scale inversely with mass of the central BH; one can still expect to see weak imprints of spin-precession upto a region of about $\approx \! 5.0 \! \times 10^5$ km. The picture that emerges is when considering the interaction of spin gravitating objects like pulsars present close ($\lesssim \! 100 M$) to intermediate-mass BHs ($10^2 \! - \! 10^5 M_\odot$), one cannot neglect the effect that gravitomagentic spin-precession has on their observed pulse profiles. Additionally, these systems would serve as excellent probes of properties of BH spacetimes.

Finally, in Tab.~\ref{table:Exploring_Omega_kappa_sigma}, we show the magnitudes, in physical units, of the accelerations $\bar{\kappa}$ and spin-precession frequencies $\bar{\sigma}_1$ experienced by pulsars on equatorial circular time-like orbits with varying sizes $\tilde{r}$ and angular frequencies $\Omega$. Although the calculations presented here apply generally to BHs (and naked singularities; see \citealt{Chakraborty+17b}) of masses $\gtrsim \! 10^2 M_\odot$, as an example, here we display the values corresponding to central masses of $10^2 M_\odot$. We choose this mass for the central objects in order to ensure that the test spinning object approximation for the pulsar $m_p \! \ll M \! < r$ holds well, as discussed in Section \ref{sec:PulsarPrecession_Static}. The angular frequency values $(\Omega, \bar{\sigma}_1)$, which are the relevant entities that modify observed pulse periods, can be easily compared against the intrinsic spin angular frequency of typical pulsars which lie in the range $\omega \! \approx \! 1 \! - \! 10^2$ rad/s. Clearly, gravitomagnetic spin-precession would be easily detectable from the modifications in the Fourier transform of the pulse profile when $f_{\text{p}}, f_{\text{o}} \! \gtrsim \! 10^{-4}$. For larger mass black holes, the accelerations, orbital angular frequencies and precession frequencies would be suppressed by a factor of $100 M_\odot/M$. However, the radius of the orbit also scales as $M/100 M_\odot$, making pulsars on such orbits much easier to detect. Also, it can be seen from this table that pulsars that experience accelerations ($q \neq q_{\text{K}\pm}$) typically experience larger spin-precession frequencies. Therefore, observations corresponding to pulsars that experience a temporary phase of acceleration could be helpful to obtain black hole parameters from the pulsar's Fourier spectrum. 

\begin{center}
\begin{table*}
\caption{
We report here the absolute values of the accelerations $\bar{\kappa}$ (in km/s$^2$) and the spin-precession frequencies $|\bar{\sigma}_1|$ (in rad/) experienced by pulsars moving on equatorial circular orbits with physical (Kerr-Schild) radii $\tilde{r}$ (in km) in the equatorial plane of a Kerr spacetime with \textit{arbitrary} orbital angular velocities $\Omega$ (in rad/s) around black holes of mass $M = 100 M_\odot$ with spin parameters $a = .1 M, .5 M, .9 M$. We also include these values for a Kerr naked singularity of the same mass and spin parameter $a = 1.01 M$. To compare, the intrinsic spin frequency of a pulsar $\omega$ lies between $1-10^2\ $rad/s. We consider pulsars with varying Boyer-Lindquist orbit radii $r = 2.1 M, 10 M, 10^2 M, 10^3 M, 10^4 M$. We choose these values for $r$ since the ergosurface in the equatorial plane is at $r_+ = 2 M$. At each of these radii, we vary the angular speeds of the pulsars between their maximum and minimum allowed values $\Omega_- < \Omega < \Omega_+$ and parametrize them with $q$ as $\Omega = q\Omega_+ + (1-q)\Omega_-$. We choose $q = q_{\text{K}+}, q_{\text{K}-}, q_{\text{static}}, .1, .3, .5, .7, .9$, where we have defined $q_{\text{K}\pm}$ and $q_{\text{static}}$ as satisfying $\Omega(q_{\text{K}\pm}) = \Omega_{\text{K}\pm}$ and $\Omega(q_{\text{static}}) = 0$, corresponding to the equatorial non-accelerating co-rotating and counter-rotating circular Kepler observers and the Kerr static Killing observers respectively. Dashes in entries below denote that Kepler observers are disallowed for those values of $a, r,$ and $q$ (since $r < r_{\text{ISCO}\pm}$). Note also that whenever allowed, Kepler observers have zero acceleration $\bar{\kappa} = 0$ and static observers have zero orbital angular velocity $\Omega = 0$. The number in a parenthesis $(n)$ denotes the order of magnitude of the entry, i.e. 3.2 (2) = 3.2 $\times 10^2$, and serves as a rough indicator for the ratios $\log{f_{\text{o}}}$ and $\log{f_{\text{p}}}$, for second-period pulsars. For ms-pulsars, $n-2$ is the appropriate estimate for these ratios naturally.}
\label{table:Exploring_Omega_kappa_sigma}
\begin{tabular}[t]{|c|c| |c|c|c|c|c| |c|c|c|c|c|}
\hline
& $a$ & \multicolumn{5}{c||}{$\bm{.1 M}$ \textbf{(Black Hole)}} & \multicolumn{5}{c|}{$\bm{.5 M}$ \textbf{(Black Hole)}} \\
\cline{2-12}
& $r$ & 2.1 & 1.0 (1) & 1.0 (2) & 1.0 (3) & 1.0 (4) & 2.1 & 1.0 (1) & 1.0 (2) & 1.0 (3) & 1.0 (4) \\
\cline{2-12}
&$\tilde{r}$ & 3.2 (2) & 1.5 (3) & 1.5 (4) & 1.5 (5) & 1.5 (6) & 3.2 (2) & 1.5 (3) & 1.5 (4) & 1.5 (5) & 1.5 (6)\\
\hline
$q = q_{\text{K}+}$ 
& $\Omega_{\text{K}+}$ 
& - & 4.1 (2) & 1.3 (1) & 4.1 (-1) & 1.3 (-2)
& - & 4.0 (2) & 1.3 (1) & 4.1 (-1) & 1.3 (-2) \\
& $\bar{\kappa}$ 
& - & 0 & 0 & 0 & 0
& - & 0 & 0 & 0 & 0\\
& $|\bar{\sigma}_1|$
& - & 4.1 (2) & 1.3 (1) & 4.1 (-1) & 1.3 (-2)
& - & 4.1 (2) & 1.3 (1) & 4.1 (-1) & 1.3 (-2)\\
\hline
$q = q_{\text{K}-}$ 
& $\Omega_{\text{K}-}$ 
& - & -4.1 (2) & -1.3 (1) & -4.1 (-1) & -1.3 (-2)
& - & -4.2 (2) & -1.3 (1) & -4.1 (-1) & -1.3 (-2)\\
& $\bar{\kappa}$
& - & 0 & 0 & 0 & 0
& - & 0 & 0 & 0 & 0 \\
& $|\bar{\sigma}_1|$ 
& - & 4.1 (2) & 1.3 (1) & 4.1 (-1) & 1.3 (-2)
& - & 4.1 (2) & 1.3 (1) & 4.1 (-1) & 1.3 (-2)\\
\hline
$q = q_{\text{static}}$ 
& $\Omega$ 
& 0 & 0 & 0 & 0 & 0
& 0 & 0 & 0 & 0 & 0\\
& $\bar{\kappa}$ 
& 6.4 (8) & 6.7 (6) & 6.1 (4) & 6.0 (2) & 6.0
& 9.2 (8) & 6.7 (6) & 6.1 (4) & 6.0 (2) & 6.0\\
& $|\bar{\sigma}_1|$ 
& 2.9 (3) & 1.6 & 1.3 (-3) & 1.3 (-6) & 1.3 (-9)
& 1.5 (4) & 8.1 & 6.6 (-3) & 6.5 (-6) & 6.5 (-9)\\
\hline
$q = .1$ 
& $\Omega$ 
& -8.2 (2) & -9.3 (2) & -1.0 (2) & -1.0 (1) & -1.0
& -1.8 (2) & -9.2 (2) & -1.0 (2) & -1.0 (1) & -1.0\\
& $\bar{\kappa}$ 
& 1.7 (9) & 7.6 (7) & 1.0 (7) & 1.1 (6) & 1.1 (5)
& 1.3 (9) & 7.3 (7) & 1.0 (7) & 1.1 (6) & 1.1 (5)\\
& $|\bar{\sigma}_1|$ 
& 2.8 (4) & 2.2 (3) & 2.8 (2) & 2.9 (1) & 2.9
& 2.3 (4) & 2.2 (3) & 2.8 (2) & 2.9 (1) & 2.9\\
\hline
$q = .3$ 
& $\Omega$ 
& -2.7 (2) & -4.6 (2) & -5.1 (1) & -5.2 & -5.2 (-1)
& 5.4 (2) & -4.5 (2) & -5.1 (1) & -5.2 & -5.2 (-1)\\ 
& $\bar{\kappa}$ 
& 7.3 (8) & 2.1 (6) & 1.1 (6) & 1.1 (5) & 1.1 (4)
& 5.2 (8) & 1.4 (6) & 1.1 (6) & 1.1 (5) & 1.1 (4)\\
& $|\bar{\sigma}_1|$ 
& 6.2 (3) & 4.8 (2) & 6.1 (1) & 6.2 & 6.2 (-1)
& 5.8 (3) & 4.6 (2) & 6.1 (1) & 6.2 & 6.2 (-1)\\
\hline
$q = .5$ 
& $\Omega_{\text{Z}}$ 
& 2.8 (2) & 2.6 & 2.6 (-3) & 2.6 (-6) & 2.6 (-9)
& 1.3 (3) & 1.3 (1) & 1.3 (-2) & 1.3 (-5) & 1.3 (-8) \\
& $\bar{\kappa}$ 
& 6.1 (8) & 6.7 (6) & 6.1 (4) & 6.0 (2) & 6.0
& 3.8 (8) & 6.7 (6) & 6.1 (4) & 6.0 (2) & 6.0\\
& $|\bar{\sigma}_1|$ 
& 4.2 (2) & 3.9 & 3.9 (-3) & 3.9 (-6) & 3.9 (-9) 
& 1.9 (3) & 1.9 (1) & 1.9 (-2) & 1.9 (-5) & 1.9 (-8)\\
\hline
$q = .7$ 
& $\Omega$ 
& 8.3 (2) & 4.7 (2) & 5.1 (1) & 5.2 & 5.2 (-1)
& 2.0 (3) & 4.8 (2) & 5.1 (1) & 5.2 & 5.2 (-1)\\
& $\bar{\kappa}$ 
& 6.9 (8) & 2.4 (6) & 1.1 (6) & 1.1 (5) & 1.1 (4)
& 3.6 (8) & 3.1 (6) & 1.1 (6) & 1.1 (5) & 1.1 (4)\\ 
& $|\bar{\sigma}_1|$ 
& 5.0 (3) & 4.9 (2) & 6.1 (1) & 6.2 & 6.2 (-1)
& 4.5 (2) & 5.1 (2) & 6.1 (1) & 6.2 & 6.2 (-1)\\
\hline
$q = .9$ 
& $\Omega$ 
& 1.4 (3) & 9.3 (2) & 1.0 (2) & 1.0 (1) & 1.0
& 2.7 (3) & 9.4 (2) & 1.0 (2) & 1.0 (1) & 1.0\\
& $\bar{\kappa}$ 
& 1.5 (9) & 7.8 (7) & 1.0 (7) & 1.1 (6) & 1.1 (5)
& 5.2 (8) & 8.1 (7) & 1.0 (7) & 1.1 (6) & 1.1 (5)\\
& $|\bar{\sigma}_1|$ 
& 2.4 (4) & 2.3 (3) & 2.8 (2) & 2.9 (1) & 2.9
& 5.7 (3) & 2.3 (3) & 2.8 (2) & 2.9 (1) & 2.9\\
\hline
\end{tabular}

\begin{tabular}[t]{|c|c| |c|c|c|c|c| |c|c|c|c|c|}
\hline
& $a$ & \multicolumn{5}{c||}{$\bm{.9 M}$  \textbf{(Black Hole)}} & \multicolumn{5}{c|}{$\bm{1.1 M}$ \textbf{(Naked Singularity)}} \\
\cline{2-12}
& $r$ & 2.1 & 1.0 (1) & 1.0 (2) & 1.0 (3) & 1.0 (4) & 2.1 & 1.0 (1) & 1.0 (2) & 1.0 (3) & 1.0 (4) \\
\cline{2-12}
&$\tilde{r}$ & 3.4 (2) & 1.5 (3) & 1.5 (4) & 1.5 (5) & 1.5 (6) & 3.6 (2) & 1.5 (3) & 1.5 (4) & 1.5 (5) & 1.5 (6)\\
\hline
$q = q_{\text{K}+}$ 
& $\Omega_{\text{K}+}$ 
& - & 4.0 (2) & 1.3 (1) & 4.1 (-1) & 1.3 (-2)
& 3.2 (3) & 4.0 (2) & 1.3 (1) & 4.1 (-1) & 1.3 (-2)\\
& $\bar{\kappa}$ 
& - & 0 & 0 & 0 & 0
& 0 & 0 & 0 & 0 & 0\\
& $|\bar{\sigma}_1|$
& - & 4.1 (2) & 1.3 (1) & 4.1 (-1) & 1.3 (-2)
& 4.3 (3) & 4.1 (2) & 1.3 (1) & 4.1 (-1) & 1.3 (-2)\\
\hline
$q = q_{\text{K}-}$ 
& $\Omega_{\text{K}-}$ 
& - & -4.2 (2) & -1.3 (1) & -4.1 (-1) & -1.3 (-2)
& - & -4.2 (2) & -1.3 (1) & -4.1 (-1) & -1.3 (-2)\\
& $\bar{\kappa}$
& - & 0 & 0 & 0 & 0
& - & 0 & 0 & 0 & 0 \\
& $|\bar{\sigma}_1|$ 
& - & 4.1 (2) & 1.3 (1) & 4.1 (-1) & 1.3 (-2)
& - & 4.1 (2) & 1.3 (1) & 4.1 (-1) & 1.3 (-2)\\
\hline
$q = q_{\text{static}}$ 
& $\Omega$ 
& 0 & 0 & 0 & 0 & 0
& 0 & 0 & 0 & 0 & 0\\
& $\bar{\kappa}$ 
& 1.4 (9) & 6.7 (6) & 6.1 (4) & 6.0 (2) & 6.0
& 1.5 (9) & 6.8 (6) & 6.1 (4) & 6.0 (2) & 6.0\\
& $|\bar{\sigma}_1|$ 
& 2.7 (4) & 1.5 (1) & 1.2 (-2) & 1.2 (-5) & 1.2 (-8)
& 3.0 (4) & 1.6 (1) & 1.3 (-2) & 1.3 (-5) & 1.3 (-8)\\
\hline
$q = .1$ 
& $\Omega$ 
& 1.1 (2) & -9.0 (2) & -1.0 (2) & -1.0 (1) & -1.0
& 1.5 (2) & -9.0 (2) & -1.0 (2) & -1.0 (1) & -1.0\\
& $\bar{\kappa}$ 
& 1.1 (9) & 6.9 (7) & 1.0 (7) & 1.1 (6) & 1.1 (5)
& 1.0 (9) & 6.8 (7) & 1.0 (7) & 1.1 (6) & 1.1 (5)\\
& $|\bar{\sigma}_1|$ 
& 2.0 (4) & 2.1 (3) & 2.8 (2) & 2.9 (1) & 2.9
& 1.9 (4) & 2.1 (3) & 2.8 (2) & 2.9 (1) & 2.9\\
\hline
$q = .3$ 
& $\Omega$ 
& 9.8 (2) & -4.4 (2) & -5.1 (1) & -5.2 & -5.2 (-1)
& 1.1 (3) & -4.4 (2) & -5.1 (1) & -5.2 & -5.2 (-1)\\ 
& $\bar{\kappa}$ 
& 3.7 (8) & 6.6 (5) & 1.1 (6) & 1.1 (5) & 1.1 (4)
& 3.5 (8) & 4.7 (5) & 1.1 (6) & 1.1 (5) & 1.1 (4)\\
& $|\bar{\sigma}_1|$ 
& 5.4 (3) & 4.3 (2) & 6.1 (1) & 6.2 & 6.2 (-1)
& 5.4 (3) & 4.3 (2) & 6.1 (1) & 6.2 & 6.2 (-1)\\
\hline
$q = .5$ 
& $\Omega_{\text{Z}}$ 
& 1.9 (3) & 2.3 (1) & 2.3 (-2) & 2.3 (-5) & 2.3 (-8)
& 2.0 (3) & 2.6 (1) & 2.6 (-2) & 2.6 (-5) & 2.6 (-8) \\
& $\bar{\kappa}$ 
& 2.2 (8) & 6.7 (6) & 6.1 (4) & 6.0 (2) & 6.0
& 1.9 (8) & 6.7 (6) & 6.1 (4) & 6.0 (2) & 6.0\\
& $|\bar{\sigma}_1|$ 
& 3.0 (3) & 3.5 (1) & 3.5 (-2) & 3.5 (-5) & 3.5 (-8) 
& 3.2 (3) & 3.9 (1) & 3.9 (-2) & 3.9 (-5) & 3.9 (-8)\\
\hline
$q = .7$ 
& $\Omega$ 
& 2.7 (3) & 4.9 (2) & 5.1 (1) & 5.2 & 5.2 (-1)
& 2.9 (3) & 4.9 (2) & 5.2 (1) & 5.2 & 5.2 (-1)\\
& $\bar{\kappa}$ 
& 1.1 (8) & 3.7 (6) & 1.1 (6) & 1.1 (5) & 1.1 (4)
& 7.1 (7) & 3.9 (6) & 1.1 (6) & 1.1 (5) & 1.1 (4)\\ 
& $|\bar{\sigma}_1|$ 
& 2.8 (3) & 5.3 (2) & 6.1 (1) & 6.2 & 6.2 (-1)
& 3.3 (3) & 5.3 (2) & 6.1 (1) & 6.2 & 6.2 (-1)\\
\hline
$q = .9$ 
& $\Omega$ 
& 3.6 (3) & 9.5 (2) & 1.0 (2) & 1.0 (1) & 1.0
& 3.8 (3) & 9.5 (2) & 1.0 (2) & 1.0 (1) & 1.0\\
& $\bar{\kappa}$ 
& 1.6 (8) & 8.3 (7) & 1.0 (7) & 1.1 (6) & 1.1 (5)
& 2.8 (8) & 8.4 (7) & 1.0 (7) & 1.1 (6) & 1.1 (5)\\
& $|\bar{\sigma}_1|$ 
& 7.4 (3) & 2.4 (3) & 2.8 (2) & 2.9 (1) & 2.9
& 9.6 (3) & 2.4 (3) & 2.8 (2) & 2.9 (1) & 2.9\\
\hline
\end{tabular}
\end{table*}
\end{center}

\section{Discussion \& Astrophysical Implications} \label{sec:Astrophysical_Implications}


It is thought that globular clusters \citep{Lutzgendorf+13, Perera+17} and ultra-compact dwarf galaxies \citep{Mapelli07, Mieske+13} house IMBHs $(10^2 \! - \! 10^4 M_\odot)$ at their centres. In fact, some large galaxies like M83 may even host two such BHs \citep{Thatte+00}. Now, globular clusters are known also to host a large population of pulsars \citep{Lorimer08}; these are mostly ms-pulsars \citep{Manchester+91, Ransom+05, Hui+10, Pan+16} and so, the formation of a ms-pulsar-IMBH binary becomes quite plausible in such globular clusters \citep{Devecchi+07, Clausen+14, Verbunt_Freire14}. Further, if the estimate of the frequency of occurrence of such binaries per cluster $\sim \!\! 1 \! - \! 10$ is accurate, then a few tens of such binaries could reside in the globular clusters in the Local Group galaxies, meaning that pulse emission from such ms-pulsars could be detected by SKA and FAST. Normal pulsars have also been found in globular clusters \citep{Biggs+94, Freire+08}; for example, of the $\sim \! \! 150$ detected globular cluster pulsars in our Galaxy%
\footnote{see for example the comprehensive catalogue of Galactic globular clusters compiled by P. Freire at \url{http://www.naic.edu/~pfreire/GCpsr.html}}, %
$\lesssim \! 10$ of them are of this variety. Therefore, a case can be made for the discovery of a normal-pulsar-IMBH binary in a globular cluster. Searches for pulsars in dwarf galaxies are yet to yield positive results as of \cite{Rubio-Herrera_Maccarone12}.

On the other hand, SMBHs $(10^5 \!\! - \!\! 10^9 M_\odot)$ are expected to be present abundantly at the centres of nearly all large galaxies \citep{Kormendy_Richstone95, Kormendy_Ho13}. Of these, although the existence of SMBHs with masses below $10^6 M_\odot$ has not been firmly established \citep{Peterson_Horne+04}, there is some evidence in their favour \citep{Greene_Ho07, Xiao+11}. The higher end of the mass-spectrum of astrophysical BHs (ultra-massive BHs) is thought to be around $\sim \!\! 10^{10} M_\odot$ (for example, the central BH in NGC 1277; \citealt{vandenBosch+12}). Large populations of about $\sim \!\! 10^5$ active normal pulsars and $\sim \!\! 10^4$ ms-pulsars are estimated to reside in our Galaxy \cite{Lorimer08}. 
Of these pulsars, about $\sim \! 10^3$ are expected to exist in the central region of our Galaxy \citep{Pfahl_Loeb04, Wharton+12, Zhang+14}. Due to indications that such high number densities of pulsars near the centre of galactic nuclear SMBHs are typical, binaries are expected to form, either through sequences of stellar interactions in the case of large spheroidal galaxies or due to capture by the central BH of a small elliptical or spiral galaxy (see for example the excellent discussion in \citealt{Li+19}), and it is not unreasonable to expect to see inspiral events \citep{Merritt+11, Clausen+14}.

Astrophysical systems involving pulsars inspiralling into massive BHs are typically divided into two categories depending on the mass of the BH, as intermediate-mass-ratio inspiral (IMRI; $10^3 \! - \! 10^4 M_\odot$) systems and extreme-mass-ratio-inspiral (EMRI; $10^5 \! - \! 10^6 M_\odot$) systems, and form major classes of gravitational wave sources for LISA \citep{Amaro-Seoane+07}.  If we consider, in particular, the scenario of a pulsar that is slowly inspiralling into a BH, then its orbit can be well approximated as being a quasi-circular orbit \citep{Miller+05, Shibata_Taniguchi11}; for example, when spin-curvature coupling is negligible and the pulsar orbit lies outside the ISCO. Also, when IMRIs/EMRIs enter the relativistic regime, pulsar orbits are greatly circularized by gravitational wave emission (see for example \citealt{Li+19} and references therein for a discussion on this). Now, such orbits are approximately the world-lines of stationary observers (and hence our results apply to such systems), with their four-velocities given by $u^\mu_{\text{slow-infall}} \propto (1,\epsilon,0,\Omega)$, with $\epsilon \ll \Omega$. Now, when the pulsar is within about $\approx\! 100 M$, spin-precession effects due to gravitomagnetism would be measurable from pulsar profiles, as we have shown here, either due to change in the observed period of pulses or in the systemic change in the morphology of the pulse shape (see \citealt{Lorimer08} for a discussion on how morphology can affect timing). Therefore, with high-precision pulsar timing observations courtesy of extremely sensitive astronomy missions like SKA and FAST, these astrophysical scenarios can be investigated independently, complimentary to gravitational wave observations.

Further, we have shown (see Section 4.1, Section 4.2 and Section 4.5) that a pulsar that is moving on circular orbits near a BH, and under the influence of gravitomagnetic spin-precession, is detectable on earth if and only if the following geometric condition is satisfied, 
\begin{equation} \label{eq:Geometric_Condition_Precession}
\theta_{\text{E}} = \pm \alpha + \beta,
\end{equation}
where in the above, $\theta_{\text{E}}$ is the angle between the direction of the earth and the precession-axis of the pulsar, $\beta$ is the angle between the spin-axis and the precession-axis and $\alpha$ is the angle between the spin and radiation axes. The precession-axis in the current paper (for pulsars moving on equatorial orbits) lies along the $z$-axis, i.e. it is parallel to the spin-axis of the central BH. Also, when the pulsar doesn't experience spin-precession ($\beta = 0$), the condition for pulses to be observed that must be met is,
\begin{equation} \label{eq:Geometric_Condition_Isolated}
\theta_{\text{E}} = \alpha.
\end{equation}
Now, let us consider the evolution of a single pulsar-BH system. When the pulsar is relatively far away from the BH, if one initially obtained pulses on earth (condition \ref{eq:Geometric_Condition_Isolated} is met), and the orientation of the spin-axis of the pulsar w.r.t. the $z$-axis is not small ($\beta \not\approx 0$), then the geometric condition for when the pulsar begins to experience non-trivial spin-precession \eqref{eq:Geometric_Condition_Precession} would not be met. In such cases, the pulsar would eventually vanish. On the other hand, if $\beta \approx 0$, then the pulsar would be visible for the entire duration of its inspiral, opening up wonderful possibilities. Further, pulsars in binaries with BHs that did not initially satisfy \eqref{eq:Geometric_Condition_Isolated}, and therefore went undetected, could be caught by our detectors close to the BH if they started to approximately satisfy \eqref{eq:Geometric_Condition_Precession}. In this regard, we think it relevant to mention here that there have been numerous studies that highlight the role that pulsars could play in the discovery (and   the subsequent analysis of the properties) of new IMBHs near Sgr$A^\star$ (see for example \citealt{Kocsis+12, Ray+13, Konar+16}), and our results here further strengthen the case for pulsars as probes of BH spacetimes.

Another implication of our findings here is that a normal or ms-pulsar with intrinsic spin-angular frequency in the range $\omega \! = \! 1 \! - \! 10^2$ rad/s could appear to pulsate at much faster frequencies when present near a massive BH, with the consequence that it could even masquerade as a genuine sub-millisecond pulsar (see for example panels a and c of Fig.~\ref{fig:timeplotters}). Moreover, the pulsar mass-shed or break-up spin frequency is typically around $\nu_{b} \simeq 1200$~Hz (see for example \citealt{Bhattacharyya+16}), depending on the equation of state, and an ordinary pulsar present near a massive BH could appear to pulsate at frequencies larger than even $\nu_b$. Therefore, neglecting gravitomagnetic spin-precession effects near BHs could possible lead to incorrect conclusions regarding the internal structure of neutron stars. Furthermore, gravitomagnetic effects present one possible explanation for pulsar nulling and for pulsar-related quasiperiodic oscillations, since spin-precession could cause pulsars to appear at frequencies smaller than the intrinsic spin-period (see panels d and e of Fig.~\ref{fig:timeplotters}). If this hypothesis renders a partial explanation for QPOs (maybe for a specific class of them), this could even lead to the discovery of black hole binary partners for some pulsars.

The Kerr spacetime, depending on the relation between the specific angular momentum $a$ and mass $M$ of the central object, describes either a BH or a naked singularity (NS), or possibly even more exotic hypothetical objects like superspinars \citep{Gimon_Horava09}. Disregarding for now the latter case, when $a \leq M$, the spacetime contains a BH, and a NS otherwise. We note that all of the analysis presented in this paper applies to Kerr NSs equally well \citep{Chakraborty+17b}. Further, since the precession frequency (and consequently the pulse frequency) depends on the spin-parameter $a$, one could identify what the nature of the companion of a pulsar is, namely whether or not it possesses an event horizon. In \cite{Chakraborty+17a} and \cite{Chakraborty+17b}, the spin-precession frequency along Killing orbits in the Kerr spacetime was revisited and various trends were outlined. It was noted there that precession frequencies experienced by such observers can become quite large in regions of strong gravitational fields. First, spinning objects that remain fixed spatially (static Killing orbits) close to a Kerr ergosurface would experience drastically large spin-precession frequencies. Now, as was discussed in \cite{Visser08} for example, since the topology of the ergosurface is drastically different for Kerr BHs (spheroidal) and NSs (toroidal), one could potentially distinguish these objects from pulsar measurements, due to the effect that gravitomagnetic spin-precession has on them \citep{Chakraborty+17a}. Similarly, sharp rises in the spin-precession frequencies were found to occur for observers moving on equatorial circular orbits (stationary Killing orbits) either around Kerr BHs near their event horizons or around Kerr naked singularities when near the ring singularity itself. Since the physical sizes of the horizon ($\tilde{r}_{\text{H}} = \sqrt{2 r_{\text{H}}}$; see eq. \ref{eq:Ergo_Horizon}) and the ring singularity ($\tilde{r}_{\text{sing}} = a$) can be easily distinguished between, one could use pulsar timing measurements to identify the nature of the compact object. Additionally, the decay of the spin-precession frequency as one moves off of the equatorial plane ($\theta = \pi/2$) to much smaller, reasonable values is much more drastic near a naked singularity, as compared to near a BH, roughly due to the presence of the horizon for $\theta \neq \pi/2$ \citep{Chakraborty+17b}. And so, small deviations in the pulsar's orbit from the equatorial plane could also be immensely useful in distinguishing the two compact objects. Therefore, in practice, the detection of a pulsar near a supermassive collapsed object with a frequency much higher than the maximum observed pulsar frequency (716 Hz; \citealt{Hessels+06}) could strongly suggest the existence of an ergoregion, event horizon or a ring singularity depending on the state of the motion of the pulsar. Further, one could also potentially test the no hair theorem for black holes (see \citealt{Misner+73}) if, for example, spin-precession measurements of `Killing pulsars' indicate deviations from the norm in the structure of the associated horizon or ergosphere.

While the event horizon, if present, could be detected by the above method, additional measurements, for example, the ratio of the pulsar radial distance to the central collapsed object mass, could be used to estimate the spin of this collapsed object, and hence to identify its nature (BH or NS). This can be done by comparing the measured pulsar spin-precession rate with the theoretical computation of this rate as a function of the pulsar radial distance in the unit of the collapsed object mass and the collapsed object spin. The required estimation of the distance can be obtained either from the gravitational radiation chirp characteristics or from the known mass of the supermassive collapsed object and the orbital period of the pulsar. The orbital period can be inferred from a periodic variation of the pulsar's intensity, which can be associated with its orbital motion. For example, such an intensity variation (or even a periodic disappearence of the pulses) can happen due to the light bending effect, as the pulsar moves behind the central collapsed object and comes in front of it periodically. The spectrum (or the intensities in a few radio bands) may also periodically vary due to the orbital motion related Doppler effect.

Let us consider now more modest tests involving, for simplicity, a pulsar that moves with very small orbital angular velocities ($\Omega \approx 0$) around a Kerr BH or naked singularity (static Killing orbits). For such pulsars, the overall periodicity of the deflection vector $|\zeta^2(\tau)|$, and therefore of the observed pulse profile, is determined by the play-off between the precession frequency $\Omega_{\text{p}}$ and the intrinsic spin frequency of the pulsar $\omega$. Roughly, if $\Omega_{\text{p}} > \omega$, then pulses appear at rates faster than the intrinsic spin frequency. On the other hand, if this relation is reversed, one misses out pulses. Now, from Tab. 1, one can safely conclude that spin-precession effects grow faster as a pulsar slowly moves towards the ergoregion of a Kerr naked singularity, as opposed to that of a Kerr BH. Similar statements can be made for pulsars moving on equatorial circular orbits, and so over the course of an inspiral of a pulsar into a BH or naked singularity, if one could extract and characterize the evolution of the spin-precession frequency ($\Omega_{\text{p}}(r)$), then one could distinguish the two.

\section{Conclusions}
\cite{Sakina_Chiba79} studied the precession of the spin-vectors of test spinning objects present in the vicinity of a Schwarzchild BH, due to geoedetic effects, and \cite{Iyer_Vishveshwara93} calculated the total precession around Kerr BHs by including gravitomagnetic effects. In both papers, the overall deflection of the spin-vector after an entire orbit was obtained and as far as we can tell, this paper gives a new application of \textit{instantaneous} spin-precession effects, as opposed to the cumulative spin-precession effects that have generally been studied. 

If an isolated pulsar spins around its axis at an angular frequency of $\omega$, then one obtains pulses on earth every $\Delta t = \omega/2\pi$ and the Fourier spectrum of this pulse profile contains a single unique peak. However, we find here that if the same pulsar was actually present near a Kerr BH, then its Fourier spectra would exhibit multiple peaks corresponding to its orbital angular velocity $\Omega$, its precession frequency $\Omega_{\text{p}}$, the vector sum of $\vec{\omega}$ and $\vec{\Omega}_{\text{p}}$ denoted by $\omega_{\text{eff}} = \sqrt{\omega^2 + \Omega_{\text{p}}^2 + 2\omega\Omega_{\text{p}}\cos\beta}$, and various combinations of the sums and differences of these three frequencies. In particular, only when $\Omega_{\text{p}}$ and $\omega_{\text{eff}}$ are commensurate, one obtains exactly periodic pulses on earth. Although here we worked mostly in the pulsar's frame, we indicated that it was possible to obtain the change in the rate of pulses obtained on earth (see the discussion in Section \ref{sec:Non_Zero_Beam_Width}). 

Our computations for the astrophysically important cases corresponding to pulsars moving around IMBHs and lower-end SMBHs on equatorial circular orbits show that gravitomagnetic spin-precession leads to significant modifications in the associated observed pulse profile, either leading to substantial modifications in either pulse-arrival times or in pulse-shape. Therefore, these effects need to be accounted for properly to interpret the timing of pulsar signals when the source closely orbits a IMBH that may exist in globular cluster cores, for example. Furthermore, models for pulsar timing observations that include these effects will therefore provide accurate tests of BH spacetimes and parameters.

Standard pulsar searches use Fourier techniques \citep{Lorimer_Kramer05} to search for \textit{a priori} unknown periodic signals. Since our analysis predicts precisely the change in the number of peaks in the Fourier profile of the power spectrum of a pulsar, as discussed in Section \ref{sec:PulsarPrecession_Stationary} above, looking for gravitomagnetic spin-precession effects via pulsar timing would likely be easily implementable. However, typically one assumes in such searches that the apparent pulse period remains constant throughout the observation \citep{Lorimer08}, which when violated, as will likely be the case during inspirals, could lead to a loss in signal-to-noise ratio (SNR) of the signal power in the Fourier-domain. In such instances, one could use the so-called `acceleration searches' \citep{Middleditch_Kristian84}, which have been proven to improve SNR in such instances. This technique, on the other hand, assumes that the pulsar has a constant acceleration during the observation, which works excellently well for the pulsar orbits we have considered here. In fact in Tab. \ref{table:Exploring_Omega_kappa_sigma} we have even catalogued the values of the accelerations experienced by pulsars on slowly inspiralling orbits around BHs for ready reference.

We note that an analytic derivation of the evolution of the beam vector for a pulsar that experiences gravitomagnetism is detailed in Appendix \ref{sec:Beam_Vector_Solution} to make clear that the extension to include pulsars moving along non-equatorial Killing orbits is straightforward. It is worth mentioning that it would be useful to extend the analysis presented here to include even more astrophysically interesting scenarios of pulsars moving on arbitrary time-like orbits, building on the results of \cite{Bini+17}, where the precession frequency experienced by such observers was recently obtained. Recently, \cite{Rana_Mangalam19} obtained closed-form analytic solutions for the properties of the motion (evolution of orbit with time etc.) of test objects moving on non-equatorial eccentric bound trajectories around a Kerr black hole; this work could drastically simplify the calculational aspect of obtaining spin-precession frequencies along arbitrary trajectories, and potentially help in generating pulse profile templates for pulsars around IMBHs and SMBHs for SKA and FAST to use (one would have to assume a beam structure model, though; see for instance Fig.~5 of \citealt{Lorimer08}).

Also, as mentioned before, here we have avoided solving the full Mathisson-Papapetrou equations by neglecting the effect of spin-curvature coupling on the motion of the pulsar, which is an excellent approximation for pulsars that either (a) spin slowly, or (b) are present near SMBHs ($M_{\text{BH}} \geq 10^6 M_\odot$), or (c) are sufficiently far away ($r \gtrsim 50 M_{\text{BH}}$) from massive BHs ($M_{\text{BH}} \geq 10^2 M_\odot$). Further, as was discussed in a couple of excellent papers recently \citep{Singh+14, Li+19}, the pulse profile of ms-pulsars present in the equatorial plane of a massive Kerr black hole (and sufficiently close to it) can be significantly altered due to this effect and so, it would be safe to conclude that our calculations work best for normal pulsars around IMBHs/SMBHs and for ms-pulsars present near SMBHs or when sufficiently far away from IMBHs.

Finally, it is worth noting that we have presented proof that radiation from Killing pulsars does in fact reach earth when a certain simple geometric condition is met (see Section 3.3 and Appendix C). This is equivalent to a ray tracing analysis.

{\it Acknowledgements}.-- The authors thank the referee for driving us to better represent and discuss our results, and their astrophysical implications. Prashant Kocherlakota thanks Rohan Poojary (CMI, India), Ronak Soni (Stanford University, USA), Abhimanyu Sashobanan (TIFR, India) and Sourav Chatterjee (TIFR, India) for useful comments and discussions. Chandrachur Chakraborty gratefully acknowledges support from the National Natural Science Foundation of China (NSFC), Grant No. 11750110410. Alak Ray acknowledges the support of Raja Ramanna Fellowship at Homi Bhabha Centre for Science Education (TIFR). 

\begin{appendix}
\section{Conversions between Frames} \label{sec:Conversions_Between_Frames}
We discuss here the conversions between the various frames that we have introduced. $e_{\hat{a}}, e_{\hat{\bar{b}}}$ denote elements of the Frenet-Serret basis associated with static Killing observers in the Kerr and adapted-Kerr spacetimes respectively, and $e_\mu, e_{\bar{\nu}}$ are elements of the Boyer-Lindquist and adapted-Boyer-Lindquist coordinate basis respectively. 

We already know the Jacobian $J$ for the coordinate transformation that transforms elements of the BL coordinate basis to their counterparts in the adapted-BL coordinate bases, $e_\mu \rightarrow e_{\bar{\nu}}$ (\ref{eq:Jacobian_BL_aBL}),
\begin{equation}
e_{\bar{\nu}} = \left(J^{-1}\right)_{\bar{\nu}}^{\ \mu} e_\mu.
\end{equation}
Let us now define $P_1$ to be the projection matrix that projects vectors defined in the BL coordinate basis onto the FS tetrad associated with Kerr static observers and $P_2$ to be the projection matrix that projects vectors defined in the adapted-BL coordinate basis onto the FS tetrad associated with adapted-Kerr static observers, that is,
\begin{align}
e_{\hat{a}} =& \left(P_1\right)_{\hat{a}}^{\ \mu}e_\mu, \\
e_{\hat{\bar{b}}} =& \left(P_2\right)_{\hat{\bar{b}}}^{\ \bar{\nu}}e_{\bar{\nu}}. \nonumber
\end{align}
Then, the entries of $P_2$ can be read off from (\ref{eq:FS_AdaptedKerr_Static}) to be,
\begin{equation*} \label{eq:Projection_Matrix_BL_FS}
\left(P_2\right)_{\hat{\bar{b}}}^{\ \bar{\nu}} = 
\begin{bmatrix}
\frac{1}{\sqrt{-g_{\bar{0}\bar{0}}}} & 0 & 0 & 0 \\
0 & \frac{g_{\bar{0}\bar{0},1}}{2\bar{\kappa} g_{\bar{0}\bar{0}}g_{11}} & \frac{g_{\bar{0}\bar{0},2}}{2\bar{\kappa}g_{\bar{0}\bar{0}}g_{22}} & 0 \\
-\frac{g_{\bar{0}\bar{3}} }{\sqrt{g_{\bar{0}\bar{0}}\Delta_{03}}} & 0 & 0 & \sqrt{\frac{g_{\bar{0}\bar{0}}}{\Delta_{03}}} \\
0 &  \frac{g_{\bar{0}\bar{0},2}}{2\bar{\kappa} g_{\bar{0}\bar{0}}\sqrt{g_{11}g_{22}}} & -\frac{g_{\bar{0}\bar{0},1}}{2\bar{\kappa} g_{\bar{0}\bar{0}}\sqrt{g_{11}g_{22}}} & 0
\end{bmatrix} .
\end{equation*}
Similarly, $P_1$ is obtained simply by replacing the barred metric components in the above by the unbarred Kerr metric components. Then the conversion between the FS tetrads associated with static Killing observers in the Kerr and adapted-Kerr metrics is given as,
\begin{equation}
e_{\hat{\bar{b}}} = \left(P_2\right)_{\hat{\bar{b}}}^{\ \bar{\nu}}\left(J^{-1}\right)_{\bar{\nu}}^{\ \mu}\left(P_1^{-1}\right)_\mu^{\ \hat{a}}e_{\hat{a}},
\end{equation}
Let us denote the conversion matrix in the above succinctly as $Q$, i.e.,
\begin{equation}
Q_{\hat{\bar{b}}}^{\ \hat{a}} = \left(P_2\right)_{\hat{\bar{b}}}^{\ \bar{\nu}}\left(J^{-1}\right)_{\bar{\nu}}^{\ \mu}\left(P_1^{-1}\right)_\mu^{\ \hat{a}}.
\end{equation}

\section{Reversal of the Acceleration and the Precession Frequency} \label{sec:Reversals}
Kerr stationary Killing observers (or equivalently adapted-Kerr static Killing observers) have four-velocities $u^\prime$ (\ref{eq:StationaryObserver}) and four-accelerations denoted by $\alpha^\prime = \nabla_{u^\prime}u^\prime$. This in the equatorial plane is given as,
\begin{equation}
\alpha^\prime = \left[-\frac{\sqrt{\Delta}(a^2 M - r^3)}{r^3 g_{\bar{0}\bar{0}}}\right](\Omega-\Omega_{\text{K}+})(\Omega-\Omega_{\text{K}-})\frac{\partial_r}{\sqrt{g_{rr}}}.
\end{equation}
The term in the square braces is always negative for $0 \leq a \leq M$ and $r_{\text{H}} < r$ (outside the horizon). Now, remembering that $\Omega_{\text{K}+} \geq 0, \Omega_{\text{K}-} \leq 0$ and adopting the usual convention $\bar{\kappa} \geq 0$, by comparing the expression for the acceleration from the above expression with $\alpha^\prime = \bar{\kappa} e_{\hat{\bar{1}}}$, it can be verified that $e_{\hat{\bar{1}}}$ should be defined as,
\begin{align}
e_{\hat{\bar{1}}} =
\begin{cases}
-\frac{\partial_1}{\sqrt{g_{11}}}, & \text{for}\ \Omega_ - < \Omega \leq \Omega_{\text{K}-}, \\
\frac{\partial_1}{\sqrt{g_{11}}}, & \text{for}\ \Omega_{\text{K}-} < \Omega \leq \Omega_{\text{K}+}, \\
-\frac{\partial_1}{\sqrt{g_{11}}}, & \text{for}\ \Omega_ {\text{K}+} < \Omega < \Omega_+.
\end{cases}
\end{align}
That is, the acceleration experienced by these observers changes in direction across $\Omega = \Omega_{\text{K}\pm}$. Physically, as was discussed in \cite{Nayak_Vishveshwara96}, this is related to the change in the sense of the centrifugal forces experienced by these observers, and a more general analysis for arbitrary axially symmetric stationary spacetimes is presented in \cite{Nayak_Vishveshwara98}. Also, for a discussion on how to define $e_{\hat{\bar{1}}}$ for non-accelerating Kepler observers, see Section  IV.A.3 of \cite{Iyer_Vishveshwara93}.

As noted in Section \ref{sec:Precession_Freq_Statio}, with the introduction of $\epsilon_{\bar{3}}$ as (see Section 3 of \citealt{Bini+99b}),
\begin{equation}
\epsilon_{\bar{3}} = \frac{-g_{\bar{0}\bar{3}}\partial_{\bar{0}} + g_{\bar{0}\bar{0}}\partial_{\bar{3}}}{\sqrt{g_{\bar{0}\bar{0}}\Delta_{03}}},
\end{equation}
we can rewrite the right-handed FS tetrad for these observers (\ref{eq:FS_AdaptedKerr_Static}) as,
\begin{align}
\left\{e_{\hat{\bar{1}}}, e_{\hat{\bar{2}}}, e_{\hat{\bar{3}}}\right\} =
\begin{cases}
\left\{-\frac{\partial_1}{\sqrt{g_{11}}}, \epsilon_{\bar{3}}, \frac{\partial_2}{g_{22}}\right\}, & \text{for}\ \Omega_ - < \Omega \leq \Omega_{\text{K}-}, \\
\left\{\frac{\partial_1}{\sqrt{g_{11}}}, \epsilon_{\bar{3}}, -\frac{\partial_2}{g_{22}}\right\}, & \text{for}\ \Omega_{\text{K}-} < \Omega \leq \Omega_{\text{K}+}, \\
\left\{-\frac{\partial_1}{\sqrt{g_{11}}}, \epsilon_{\bar{3}}, \frac{\partial_2}{g_{22}}\right\}, & \text{for}\ \Omega_ {\text{K}+} < \Omega < \Omega_+.
\end{cases}
\end{align}
At this juncture, let us remember that both $\Omega_{\text{K}\pm}$ are allowed orbital angular frequencies only when $r_{\text{ISCO}-} \leq r$. More fully,
\begin{align}
\Omega_{\text{K}-}, \Omega_{\text{K}+} \in (\Omega_-, \Omega_+),\ \ & \text{for}\ r_{\text{ISCO}-} \leq r, \\
\text{only}\ \Omega_{\text{K}+} \in (\Omega_-, \Omega_+),\ \  & \text{for}\
 r_{\text{ISCO}+} \leq r < r_{\text{ISCO}-}, \nonumber \\
\Omega_{\text{K}-}, \Omega_{\text{K}+} \notin (\Omega_-, \Omega_+),\ \  & \text{for}\ r < r_{\text{ISCO}+}. \nonumber
\end{align}
Since we know that the acceleration experienced by static observers is always along $+ \partial_1$, and that sign changes occur at $\Omega_{\text{K}\pm}$, we can find the direction in which the acceleration experienced by an arbitrary Killing observer in the Kerr spacetime points along. 

\begin{figure}
\centering
\subfigure[~$a = .1 M$, Sense of the Acceleration]
{\includegraphics[scale=.65]{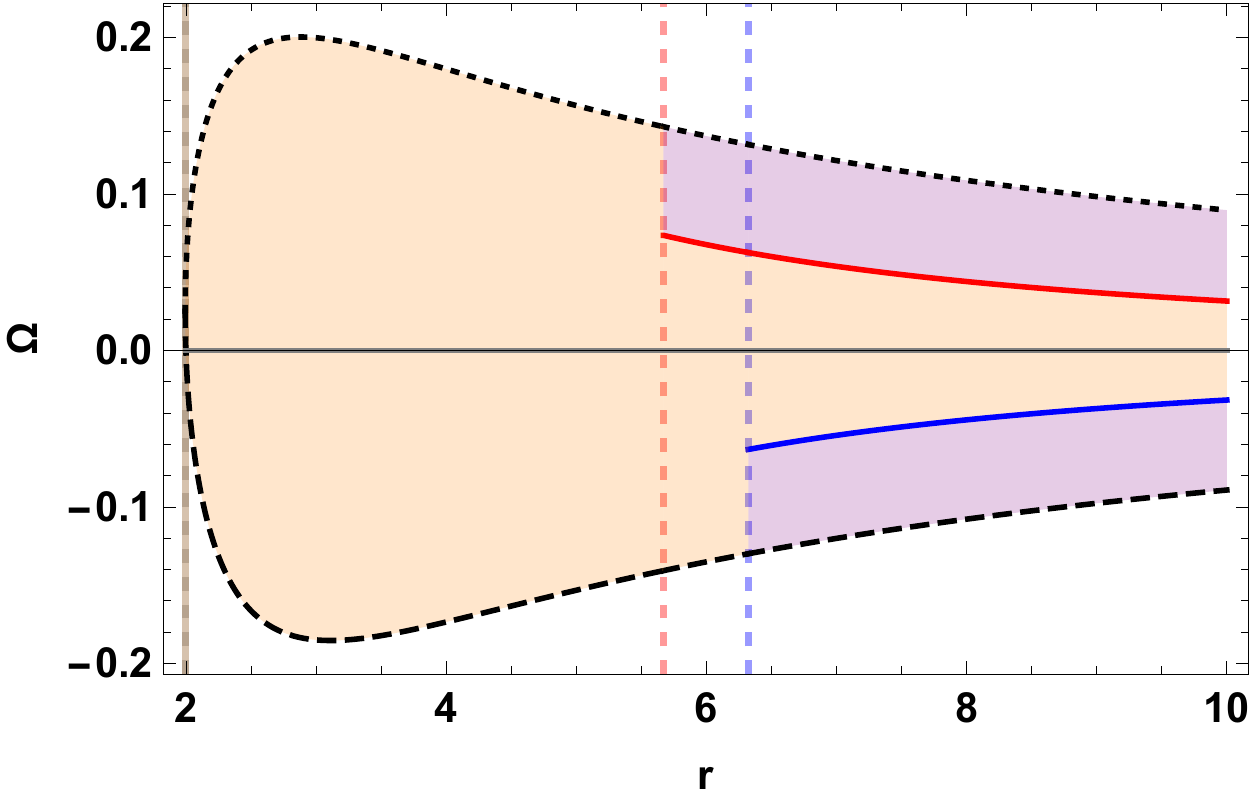}}
\hspace{0.1cm}
\subfigure[~$a = .5 M$, Sense of the Acceleration]
{\includegraphics[scale=.65]{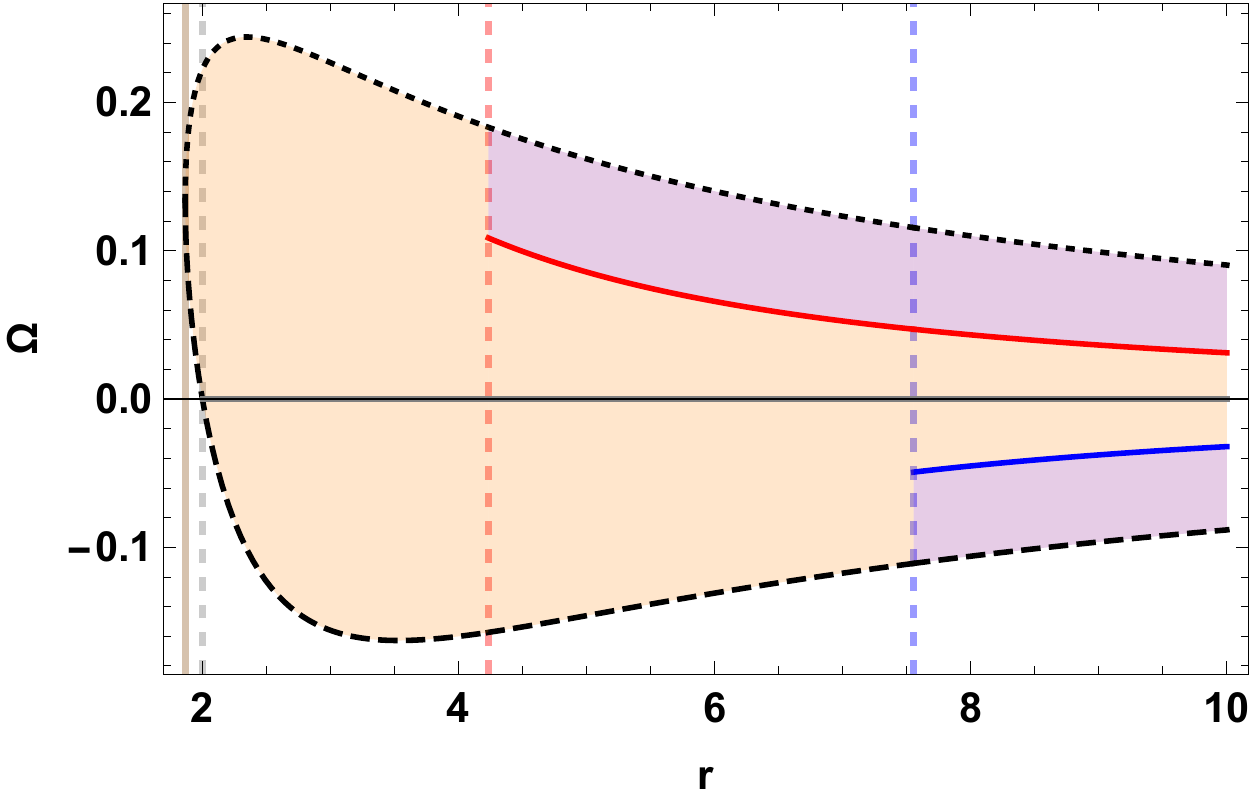}}
\hspace{0.1cm}
\subfigure[~$a = .9 M$, Sense of the Acceleration]
{\includegraphics[scale=.65]{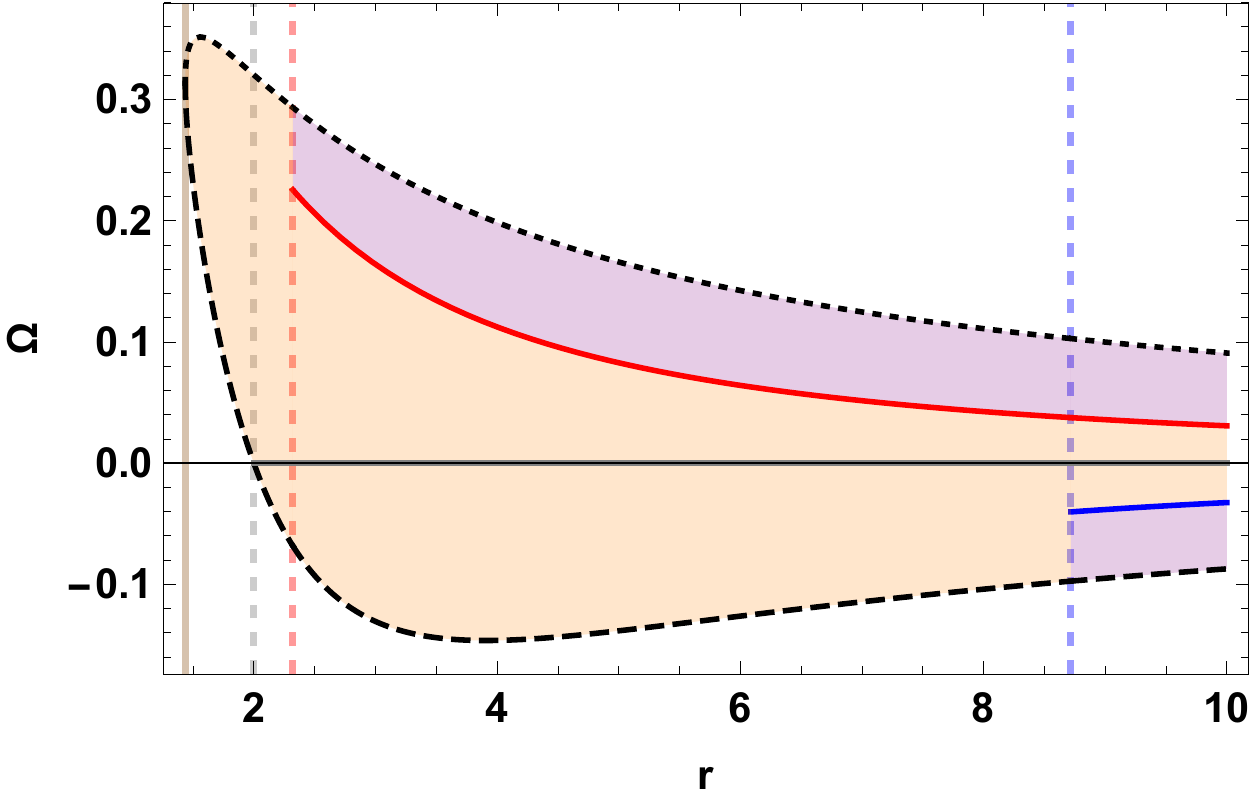}}
\caption{We plot here the variation in $\Omega_{\pm}, \Omega_{\text{K}\pm}$ with $r$ in dotted-black (topmost line), dashed-black, red and blue respectively. The vertical brown, dashed-gray, dashed-red and dashed-blue lines represent the location of the horizon $r_{\text{H}}$, the ergoradius in the equatorial plane $r_+(\pi/2)$ and the ISCOs of co-rotating and counter-rotating Kepler observers $r_{\text{ISCO}\pm}$ respectively. We plot over the range $r_{\text{H}} < r \leq 10 M$ and use units of $M^{-1}$ on the y-axis and of $M$ on the x-axis, for black holes of mass $M$ and spin parameters $a=.1 M, .5 M, .9 M$ in panels (a-c) respectively. The shading represents the sign of the acceleration experienced by adapted-Kerr static Killing observers. Orange and purple represent regions where the acceleration vector points along $\pm\partial_1$  and the appropriate definitions for the right-handed Frenet-Serret tetrads in these regions are $\left\{\pm\frac{\partial_1}{\sqrt{g_{11}}}, \epsilon_{\bar{3}}, \mp\frac{\partial_2}{g_{22}}\right\}$ respectively. To compare, the direction in which the black hole spin points in is given by $\hat{z} = -\frac{\partial_2}{\sqrt{g_{22}}}$.}
\label{fig:Acceleration_Handedness}
\end{figure}

For a pictorial representation of the above discussion, we plot in Fig. \ref{fig:Acceleration_Handedness} the variation in $\Omega_{\pm}, \Omega_{\text{K}\pm}$ with $r$, for black holes with spin parameters $a=.1 M, .5 M, .9 M$. Orange and purple represent regions where the acceleration vector points along $\pm\partial_1$ and the appropriate definitions for the right-handed Frenet-Serret tetrads in these regions are $\left\{\pm\frac{\partial_1}{\sqrt{g_{11}}}, \epsilon_{\bar{3}}, \mp\frac{\partial_2}{g_{22}}\right\}$ respectively.

\begin{figure}
\centering
\subfigure[~$a = .1 M$, Sense of the Precession Frequency]
{\includegraphics[scale=.65]{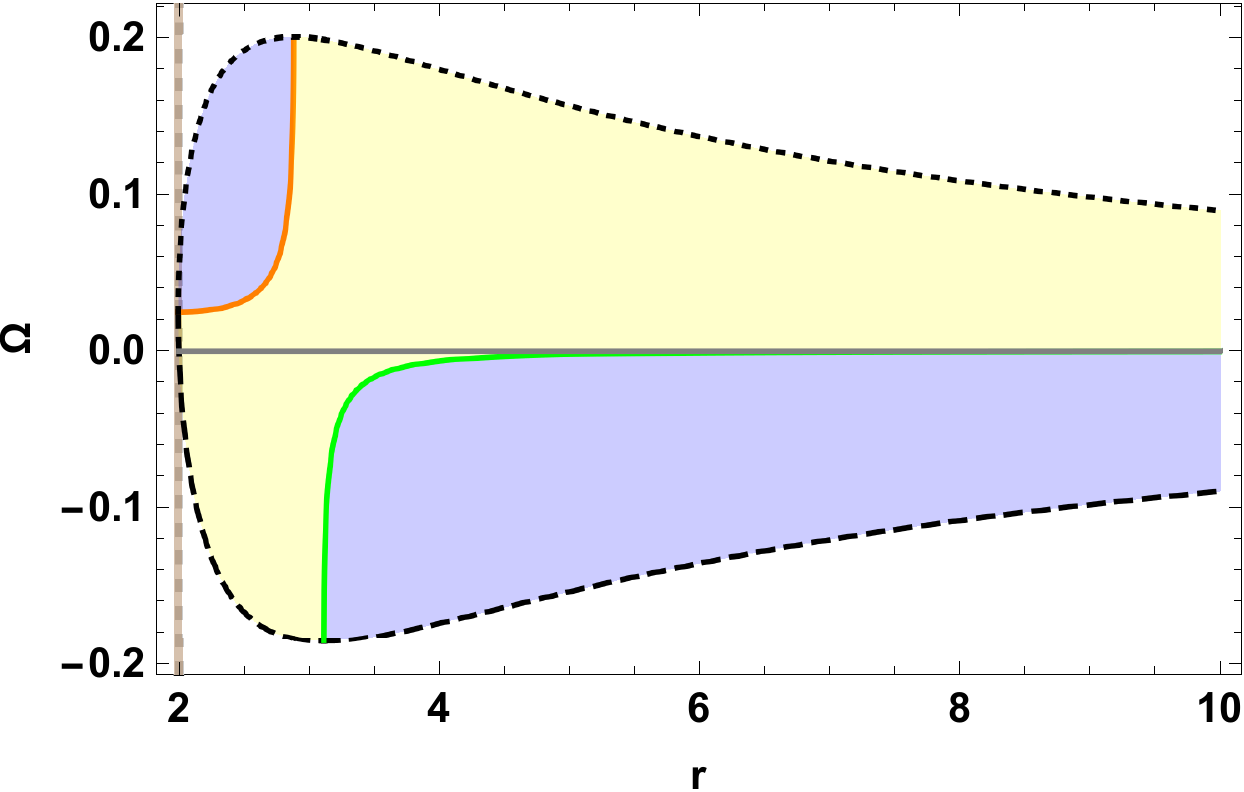}}
\hspace{0.1cm}
\subfigure[~$a = .5 M$, Sense of the Precession Frequency]
{\includegraphics[scale=.65]{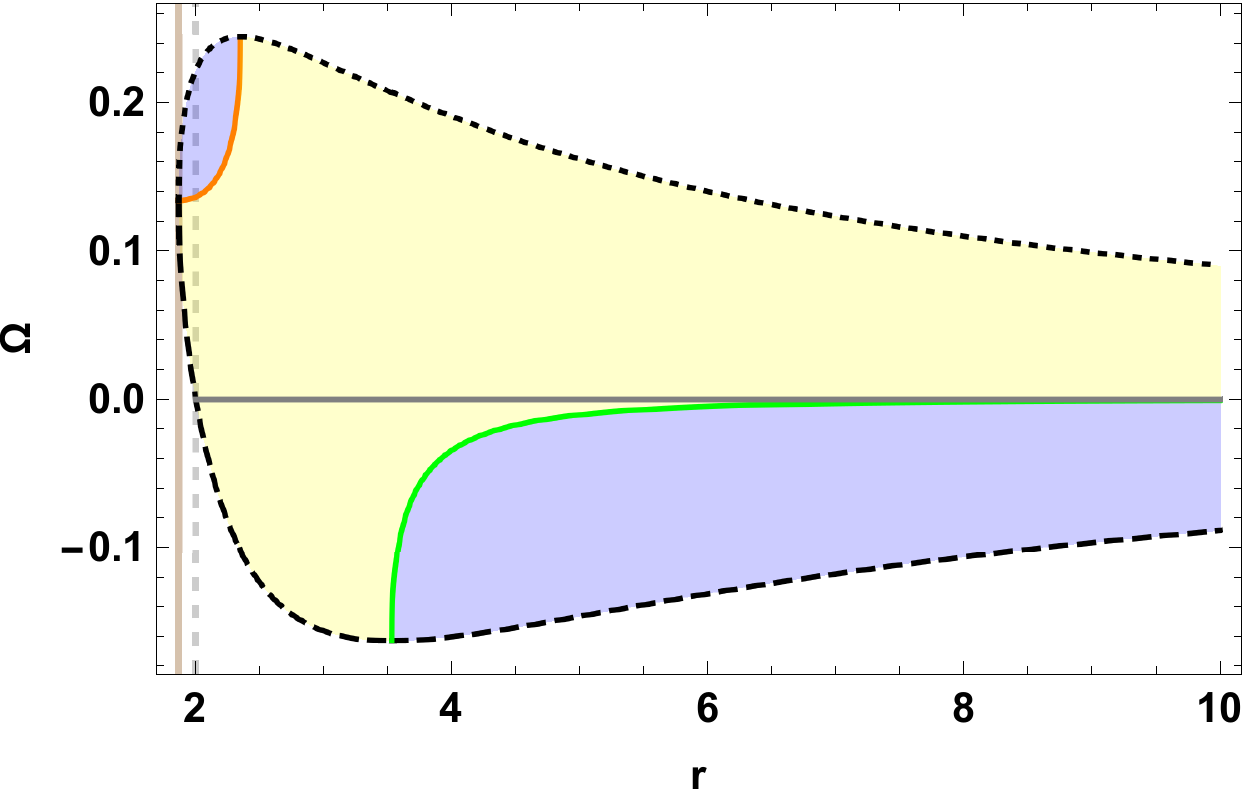}}
\hspace{0.1cm}
\subfigure[~$a = .9 M$, Sense of the Precession Frequency]
{\includegraphics[scale=.65]{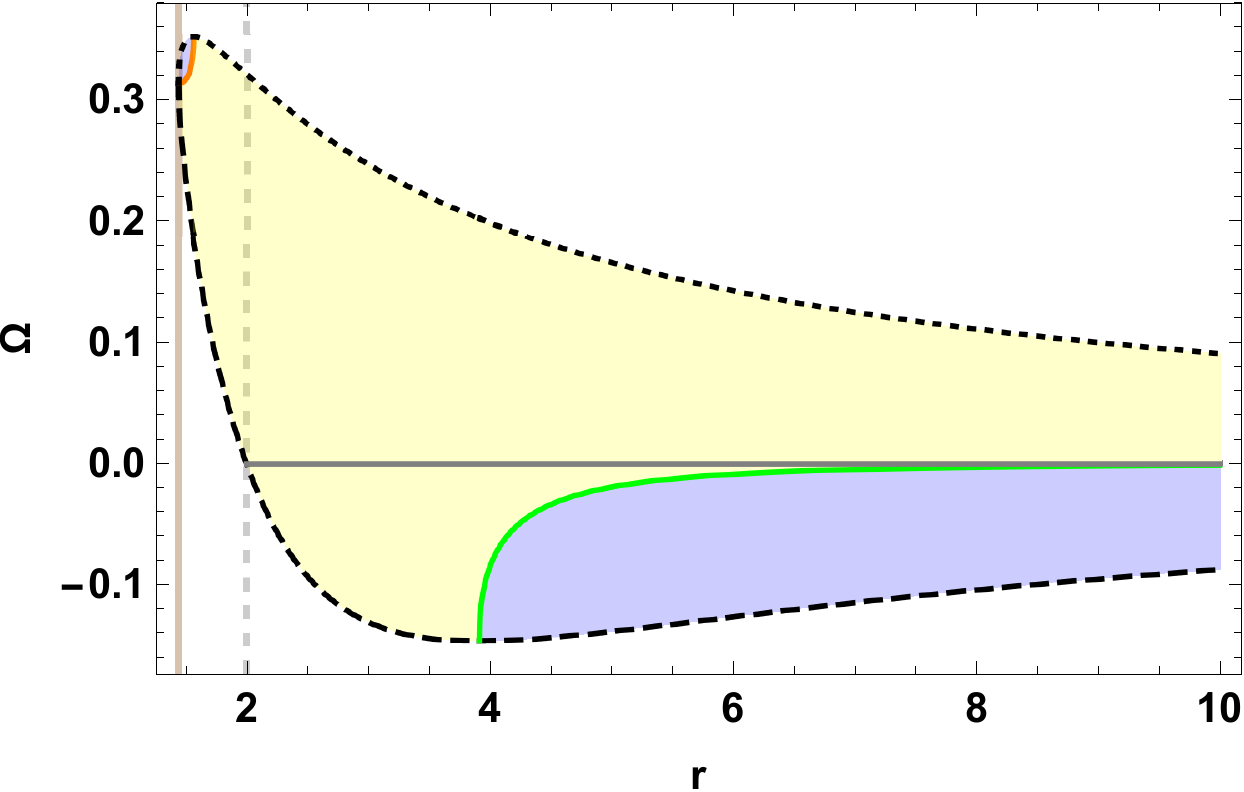}}
\caption{We plot here the variation in $\Omega_{\pm}, \Omega_1, \Omega_2$ with $r$ in dotted-black (topmost line), dashed-black, orange and green respectively. The vertical brown and dashed-gray lines represent the location of the horizon $r_{\text{H}}$ and ergoradius in the equatorial plane $r_+(\pi/2)$ respectively. We display the orange and green lines only when they satisfy $\Omega_- < \Omega_1, \Omega_2 < \Omega_+$, over the range $r_{\text{H}} < r \leq 10 M$. We plot in units of $M^{-1}$ on the $y$-axis and in units of $M$ on the $x$-axis, for black holes with spin parameters $a=.1 M, .5 M, .9 M$ in panels (a-c) respectively. We display the regions in the Kerr spacetime where observers moving on equatorial circular geodesics experience positive and negative precession frequencies relative to the $e_{\hat{\bar{3}}}$ leg of their respective right-handed FS tetrads, in blue and yellow respectively. That is, the precession frequencies in these regions are given as $\Omega^\prime = \pm|\bar{\sigma}_1|e_{\hat{\bar{3}}}$ respectively. Further, these plots imply that the precession frequency associated with stationary observers of fixed orbital radius radius $r$ changes in sense at most only once with change in $\Omega$, i.e. either when $\Omega = \Omega_1$ or $\Omega = \Omega_2$. It is also interesting to note that there exist regions (in $r$) where the precession frequency does not change in sense.}
\label{fig:Precession_Frequency_Sense}
\end{figure}

With respect to the right-handed FS triad defined above, the precession frequency is given as,
\begin{align}
\Omega_{\text{p}}^\prime =& -\bar{\sigma}_1 e_{\hat{\bar{3}}}, \\
\bar{\sigma}_1 =& -\frac{\Omega r^3 + 3 M \Omega r^2(a\Omega - 1) + a M (a\Omega - 1)^2}{r^3g_{\bar{0}\bar{0}}}. \nonumber
\end{align}
Clearly then the precession frequency changes signs (relative to the right-handed FS triad) at the zeroes of the numerator, i.e. when,
\begin{equation} \label{eq:Precession_Vanish}
\Omega r^3 + 3 M \Omega r^2(a\Omega - 1) + a M (a\Omega - 1)^2 = 0,
\end{equation}
and the orbits where the reversal of the precession frequency occurs do not, in general, coincide with the Kepler orbits where the centrifugal force reverses. If we denote the roots of (\ref{eq:Precession_Vanish}) by $\Omega_1, \Omega_2$, then
\begin{align}
\Omega_1 =& \frac{2 a M}{2 a^2 M - r^2 (r-3 M) + \sqrt{r^6 - 6 M r^5 + 9 M^2 r^4 - 4 a^2 M r^3}}, \nonumber \\
\Omega_2 =& \frac{1}{\Omega_1(a^2 + 3 r^2)}.
\end{align}
Let us define the $q$-values corresponding to $\Omega_1, \Omega_2$ as,
\begin{equation}
q_1 = \frac{\Omega_1 - \Omega_-}{\Omega_+ - \Omega_-}, \ \ q_2 = \frac{\Omega_2 - \Omega_-}{\Omega_+ - \Omega_-}.
\end{equation}
It can be verified that that $q_2 < q_{\text{static}} < q_{\text{Z}} < q_1$. That is, the precession frequencies experienced by static observers and the ZAMO always have the same sense, relative to their right-handed Frenet-Serret triads. Moreover, as can be seen from Fig. \ref{fig:qStatic_qKep+-}, $q_{\text{K}-} < q_{\text{static}} < q_{\text{Z}} < q_{\text{K}+}$ and so the $e_{\hat{\bar{3}}}$ legs of the static observers and the ZAMOs also point along the same direction $\hat{z} \equiv -\partial_2/\sqrt{g_{22}}$, where we have introduced $\hat{z}$ to denote the unit vector that points along the $z$-axis in the Boyer-Lindquist (and also adapted-Boyer-Lindquist) chart and is the direction in which the black hole spin points. Therefore, the precession experienced by static observers and the ZAMOs has the same sense relative to the BH spin orientation as well.

In Fig. \ref{fig:Precession_Frequency_Sense}, we show the variation of $\Omega_1, \Omega_2$ with $r$, when $\Omega_- < \Omega_1, \Omega_2 < \Omega_+$, for black holes with mass $M$ and spin parameters $a = .1 M, .5 M, .9 M$. Since the precession frequency changes sense at $\Omega = \Omega_1, \Omega_2$, this figure implies that the precession frequency associated with stationary observers at a fixed orbital radius radius $r$ changes at most only once, either when $\Omega = \Omega_1$ or $\Omega = \Omega_2$. Furthermore, there also exists a region where the precession frequency does not change in sense. In this figure, we display the regions in the Kerr spacetime where observers moving on equatorial circular geodesics experience positive and negative precession frequencies relative to the $e_{\hat{\bar{3}}}$ leg of their respective right-handed FS tetrads, in blue and yellow respectively. That is, the precession frequencies in these regions are given as $\Omega^\prime = \pm|\bar{\sigma}_1|e_{\hat{\bar{3}}}$ respectively.

\section{The Connecting Vector for Earth \& Stationary Killing Observers} \label{sec:Connecting_Vector_aFS}
The motion of earth in the Kerr spacetime can naturally be modelled by that of an asymptotic static observer. As discussed in Section \ref{sec:SpinPrecession_Killing}, the direction to earth (when causally connected) in the Frenet-Serret spatial triad of an equatorial Kerr static observer is a constant vector. This is associated with the notion of the (null) connecting vector of the Kerr static Killing congruence. Now, if we denote the direction to earth in the Kerr static observer's FS spatial triad as $n^{\hat{i}}$, then we can write it most generally as,
\begin{equation} \label{eq:n_Kerr_Static_FS}
n^{\hat{i}} = \left(\sin\theta_{\text{E}}\cos{\phi_{\text{E}}}, \sin\theta_{\text{E}}\sin{\phi_{\text{E}}}, \cos\theta_{\text{E}}\right).
\end{equation}
Now, we will discuss here how to find the direction towards earth in the frame of a pulsar that moves on a circular orbit in the equatorial plane of a Kerr black hole. Such an observer can be treated as a static Killing observer in the adapted-Kerr metric. In this metric however, earth becomes a stationary Killing observer, and the aim here then is to find the connecting vector between a static and a stationary Killing observer. 

Let the tangent to the null geodesic connecting a pulsar (that can be treated as moving on an adapted-Kerr static Killing orbit) and an asymptotic adapted-Kerr static Killing observer be given as $n^{\hat{\bar{j}}}$, in the associated spatial FS triad. Then, in analogy with (\ref{eq:n_Kerr_Static_FS}), we can write,
\begin{equation} \label{eq:n_aKerr_Static_FS}
n^{\hat{\bar{j}}}(\bar{\phi}_{\text{E}}) = \left(\sin{\bar{\theta}_{\text{E}}}\cos{\bar{\phi}_{\text{E}}}, \sin\bar{\theta}_{\text{E}}\sin{\bar{\phi}_{\text{E}}}, \cos{\bar{\theta}_{\text{E}}}\right).
\end{equation}
Since a Kerr static Killing observer becomes an adapted-Kerr stationary Killing observer, in a coordinate time $dt$, in the adapted-Boyer Lindquist coordinates, it moves by an amount $-\Omega~dt$ along in the $\bar{\phi}$ direction. The components of the connecting vector to an asymptotic adapted-Kerr static Killing observer that is present at an infinitesimal adapted-Boyer-Lindquist $\bar{\phi}$-coordinate shift of $-\Omega~dt$ can be found by a rotation about the $z$-axis in the adapted-Kerr metric by $-\Omega~dt$. Instead, we can simply rotate the above vector (\ref{eq:n_aKerr_Static_FS}) in the FS frame of the adapted-Kerr static Killing observer by the appropriate rotation matrix $R[\pm e_{\hat{\bar{3}}}, -\Omega~d\tau]$. The signs turn up because,
\begin{align} \label{eq:Signs_e3_z}
\hat{z} = -\frac{\partial_2}{\sqrt{g_{22}}} =
\begin{cases}
-  e_{\hat{\bar{3}}}, & \text{for}\ \Omega_ - < \Omega \leq \Omega_{\text{K}-}, \\
e_{\hat{\bar{3}}}, & \text{for}\ \Omega_{\text{K}-} < \Omega \leq \Omega_{\text{K}+}, \\
-  e_{\hat{\bar{3}}}, & \text{for}\ \Omega_ {\text{K}+} < \Omega < \Omega_+.
\end{cases}
\end{align}
Here we have assumed that the pulsar orbit radius $r$ lies outside the counter-rotating ISCO, i.e. $r_{\text{ISCO}-} \leq r$ so that both Kepler frequencies $\Omega_{\text{K}\pm}$ are allowed, $\Omega_- < \Omega_{\text{K}\pm} < \Omega_+$. The analysis can be extended to accommodate observers present on orbital radii $r$ in the range $r_{\text{H}} < r < r_{\text{ISCO}-}$ by following the discussion in Appendix \ref{sec:Reversals}. Now, we can write
\begin{align}
& n^{\hat{\bar{j}}}(\bar{\phi}_{\text{E}} - \Omega d\tau) = R[\pm e_{\hat{\bar{3}}}, -\Omega d\tau]n^{\hat{\bar{j}}}(\bar{\phi}_{\text{E}}) = \\
&\ \ \ \ \left(\sin{\bar{\theta}_{\text{E}}}\cos{\left(\bar{\phi}_{\text{E}} \mp \Omega d\tau\right)}, \sin\bar{\theta}_{\text{E}}\sin{\left(\bar{\phi}_{\text{E}} \mp \Omega d\tau\right)}, \cos{\bar{\theta}_{\text{E}}}\right), \nonumber
\end{align}
where in the above, one can use the appropriate redshift formulae to relate time differences $dt$ and $d\tau$. Furthermore, since all infinitesimal rotations are around the same constant axis, we can write the direction to earth in FS spatial triad associated with the adapted-Kerr static Killing observer as being given by,
\begin{equation} 
n_{\text{E}}^{\hat{\bar{j}}}(\tau) = \left( \cos{(\Omega \tau)}\sin{\bar{\theta}_{\text{E}}}, \mp\sin{(\Omega \tau)}\sin{\bar{\theta}_E}, \cos{\bar{\theta}_{\text{E}}}\right),
\end{equation}
where for convenience we have set $\bar{\phi}_{\text{E}} = 0$ and the signs correspond to the signs of $e_{\hat{\bar{3}}}$ in (\ref{eq:Signs_e3_z}). This expression is consistent with relevant statements in Section  3 of \cite{Bini+99b}. It is useful to note that at $\tau = 0$, 
\begin{equation} 
n_{\text{E}}^{\hat{\bar{j}}}(\tau = 0) = \left(\sin{\bar{\theta}_{\text{E}}}, 0, \cos{\bar{\theta}_{\text{E}}}\right),
\end{equation}
and so, the initial condition for the beam vector remains identical to the case when the pulsar is modelled as a Kerr static Killing observer.

\section{Solution for the Beam Vector} \label{sec:Beam_Vector_Solution}
We show here the analytical solution to solve the beam evolution equation,
\begin{equation}
\dot{\hat{B}} = \omega \hat{B} \times \hat{S},
\end{equation}
where $\hat{S}$ in the above is obtained from (\ref{eq:S_pm}). It is useful to include this simple calculation to indicate that the extension to pulsars moving on non-equatorial Killing orbits in the Kerr spacetime is straightforward, and involves just additional (constant) rotation matrices. Also, this can serve as a starting point for a more general analysis for pulsar's moving on arbitrary time-like orbits.

We use the Euclidean notation that was employed in Section \ref{sec:PulsarPrecession_Static}. We first move to a rotating frame $\{\hat{e}_{i}^\prime, i \! = \! 1, 2, 3\}$ in which $\hat{S}(\tau)$ becomes time-independent i.e., $\hat{S}^\prime(\tau) = (\sin\beta, 0 , \cos\beta)$. This achieved by a rotation around $\hat{e}_{3}$ by $-(\psi + \sigma_1\tau)$, for which we use the rotation matrix $U_1 = R[\hat{e}_{3}, -(\psi + \sigma_1\tau)]$ to write,
\begin{equation} \label{eq:U1}
U_1\hat{S}(t) = (\sin\beta, 0, \cos\beta). 
\end{equation}
Remembering that if $\vec{A} = \vec{B} \times \vec{C}$, then under rotations we have also, $U_1 \vec{A} = U_1\vec{B} \times U_1\vec{C}$. 
We now define $\hat{B}^\prime = U_1\hat{B}, \hat{S}^\prime = U_1\hat{S}$, and write in the rotating frame,
\begin{equation} \label{eq:B_EoM_RotFrame}
U_1\dot{\hat{B}} = \omega \hat{B}^\prime \times \hat{S}^\prime.
\end{equation}
To rewrite the above equation completely in terms of the rotating frame, we write out $\dot{U_1}$ as,
\begin{equation}
\dot{U}_1 = 
-\sigma_1
\begin{bmatrix}
0 & -1 & 0 \\
1 & 0 & 0 \\
0 & 0 & 0
\end{bmatrix}U_1,
\end{equation}
to obtain the following,
\begin{align}
\dot{\hat{B}}^\prime &= \dot{U}_1\hat{B} + \omega \hat{B}^\prime \times \hat{S}^\prime \nonumber \\
&= \dot{U}_1 U_1^T U_1\hat{B} + \omega \hat{B}^\prime \times \hat{S}^\prime \nonumber \\
&= -\sigma_1
\begin{bmatrix}
0 & -1 & 0 \\
1 & 0 & 0 \\
0 & 0 & 0
\end{bmatrix}
\hat{B}^\prime + \omega \hat{B}^\prime \times \hat{S}^\prime \nonumber \\
&= \hat{B}^\prime \times (\sigma_1\hat{e}_3^\prime) + \omega \hat{B}^\prime \times \hat{S}^\prime \nonumber \\
&= \hat{B}^\prime \times \left(\sigma_1\hat{e}_3^\prime+ \omega(\sin\beta  \hat{e}_1^\prime + \cos\beta  \hat{e}_3^\prime)\right) \nonumber \\
&= \omega_{\text{eff}} \hat{B}^\prime \times \hat{S}_{\text{eff}}^\prime,
\end{align}
where we have introduced $\omega_{\text{eff}}$ so $\hat{S}^\prime_{\text{eff}}$ has unit norm, i.e.,
\begin{align} \label{eq:omegaEff_def}
\omega_{\text{eff}}^2 &= \omega^2 + \sigma_1^2 + 2\omega\sigma_1\cos\beta, \\
\hat{S}_{\text{eff}}^\prime &= \left(\frac{\omega\sin\beta}{\omega_{\text{eff}}}, 0, \frac{\sigma_1 + \omega\cos\beta}{\omega_{\text{eff}}}\right).
\end{align}
And we recognize that $\omega_{\text{eff}}$ is simply the norm of the vector difference of the intrinsic spin angular frequency and the precession frequency vectors, $\omega \hat{S}$ and $-\sigma_1\hat{e}_3$. 

Now that $\hat{S}_{\text{eff}}^\prime$ is already a time-independent vector, we can immediately apply Rodrigues' rotation formula to obtain the beam vector in this frame. However, we make a second coordinate transformation and simply read off the beam vector in the new frame. We define $\chi$ from writing $\hat{S}_{\text{eff}}^\prime = (\sin\chi, 0, \cos\chi)$, that is, 
\begin{equation} \label{eq:chi_def}
\chi = \sin^{-1}\left(\frac{\omega}{\omega_{\text{eff}}}\sin\beta\right).
\end{equation}
We can then send $\hat{S}^\prime_{\text{eff}}$ to $\hat{e}^{\prime\prime}_3$ via $\hat{e}^{\prime\prime}_3 = R\left[\hat{e}^\prime_2, -\chi\right]\hat{S}^\prime_{\text{eff}} = U_2\hat{S}_{\text{eff}}^\prime$. Then if we introduce $\hat{B}^{\prime\prime} = U_2\hat{B}^\prime$, we can write
\begin{equation}
U_2\dot{\hat{B}}^\prime = U_2\left(\omega_{\text{eff}} \hat{B}^\prime \times \hat{S}_{\text{eff}}^\prime\right),
\end{equation}
and since we are simply performing a time-independent rotation transformation, $\frac{dU_2}{d\tau} = 0$ and ,
\begin{equation}
\dot{\hat{B}}^{\prime\prime} = \omega_{\text{eff}} \hat{B}^{\prime\prime} \times \hat{e}_{3}^{\prime\prime}.
\end{equation}
By an application of Rodrigues' rotation formula, it can be seen immediately that the solution is simply given as,
\begin{align} \label{eq:Bprimeprime}
B_1^{\prime\prime} &= D_1\cos{(\omega_{\text{eff}} \tau)} + D_2\sin{(\omega_{\text{eff}} \tau)}, \\
B_2^{\prime\prime} &= D_2\cos{(\omega_{\text{eff}} \tau)} - D_1\sin{(\omega_{\text{eff}} \tau)}, \nonumber \\
B_3^{\prime\prime} &= D_3. \nonumber
\end{align}
Note that the integration constants $D_1, D_2, D_3$ are not free since we want to consider a specific initial condition for $\hat{B}$ given in (\ref{eq:B_Initial_Condition}). We will first obtain $\hat{B}$ by performing in series the inverse transformations on $\hat{B}^{\prime\prime}$ and then proceed to set the initial conditions. The requisite transformations to obtain $\hat{B}$ are given as,
\begin{equation}
\hat{B} = U_1^T U_2^T \hat{B}^{\prime\prime},
\end{equation}
to obtain in the Frenet-Serret frame,
\begin{small}
\begin{align} \label{eq:B_General}
B_1 \! =& D_1\left[\cos{(\omega_{\text{eff}} \tau)}\cos{\left(\psi \! + \! \sigma_1\tau\right)}\cos\chi \! + \! \sin{(\omega_{\text{eff}} \tau)}\sin{\left(\psi \! + \! \sigma_1\tau\right)}\right]\nonumber \\
& \!\!\!\!\!\! + \! D_2\left[\sin{(\omega_{\text{eff}} \tau)}\cos{\left(\psi \! + \! \sigma_1\tau\right)}\cos\chi \! - \! \cos{(\omega_{\text{eff}} \tau)}\sin{\left(\psi \! + \! \sigma_1\tau\right)}\right] \nonumber \\
& \!\!\!\!\!\! +\! D_3\cos{\left(\psi \! + \! \sigma_1\tau\right)}\sin\chi, \\
B_2 \! =& 
D_1\left[\cos{(\omega_{\text{eff}} \tau)}\sin{\left(\psi \! + \! \sigma_1\tau\right)}\cos\chi \! - \! \sin{(\omega_{\text{eff}} \tau)}\cos{\left(\psi \! + \! \sigma_1\tau\right)}\right] \nonumber \\
& \!\!\!\!\!\! + \! D_2\left[\sin{(\omega_{\text{eff}} \tau)}\sin{\left(\psi \! +\! \sigma_1\tau\right)}\cos\chi \! + \! \cos{(\omega_{\text{eff}} \tau)}\cos{\left(\psi \! + \! \sigma_1\tau\right)}\right], \nonumber \\
& \!\!\!\!\!\! +\! D_3\sin{\left(\psi \! + \! \sigma_1\tau\right)}\sin\chi \nonumber \\
B_3 \! =& 
-\sin\chi\left[D_1\cos{(\omega_{\text{eff}} \tau)} \! + \! D_2\sin{(\omega_{\text{eff}} \tau)}\right]\!+\! D_3\cos\chi. \nonumber
\end{align}
\end{small}
From the condition that the beam vector above (\ref{eq:B_General}) satisfy the initial condition (\ref{eq:B_Initial_Condition}), we obtain
\begin{align} \label{eq:D1D2D3}
D_1 &= \sin\theta_E\cos\chi\cos\psi - \cos\theta_E\sin\chi, \\
D_2 &= - \sin\theta_E\sin\psi, \nonumber\\
D_3 &= \cos\theta_E\cos\chi + \sin\theta_E\sin\chi\cos\psi. \nonumber
\end{align}
Note that the initial phase of the spin-vector $\hat{S}$ denoted here by $\psi$ is still free to choose. Once this is chosen, we obtain the full solution for the evolution of the beam vector of a pulsar that remains fixed in space near a Kerr black hole or naked singularity.

Without loss of generality, let us pick the initial phase for $\hat{S}_0$ to be $\psi = 0$.  From eq. (\ref{eq:alpha_beta_thetaE_psi}), it is clear that one obtains pulses only for specific geometric configurations, $\theta_E = \pm \alpha + \beta$. Also,
\begin{align} \label{eq:D1D2D3_psi0}
D_1 &= \sin{(\theta_E - \chi)}, \\
D_2 &= 0, \nonumber\\
D_3 &= \cos{(\theta_E - \chi)}. \nonumber
\end{align}
And the beam vector is given by,
\begin{align} \label{eq:Beam_Solution}
B_1\!=&\!
\cos{\left(\sigma_1\tau\right)}\left[\cos{(\omega_{\text{eff}} \tau)}\sin{(\theta_E\!-\!\chi)}\cos\chi \!+\!\cos{(\theta_E\!-\!\chi)}\sin\chi\right] \nonumber \\
+ &\!\sin{\left(\sigma_1\tau\right)}\sin{(\omega_{\text{eff}} \tau)}\sin{(\theta_E\!-\!\chi)}, \\
B_2\!=&\!
-\!\cos{\left(\sigma_1\tau\right)}\sin{(\omega_{\text{eff}} \tau)}\sin{(\theta_E\!-\!\chi)} \nonumber \\
&\!\!\!\!\!\! + \sin{\left(\sigma_1\tau\right)}\left[\cos{(\omega_{\text{eff}} \tau)}\sin{(\theta_E\!-\!\chi)}\cos\chi\!+\!\cos{(\theta_E\!-\!\chi)}\sin\chi\right], \nonumber \\
B_3\!=&\! -\!\cos{(\omega_{\text{eff}} \tau)}\sin{(\theta_E\!-\!\chi)}\sin\chi\!+\!\cos{(\theta_E\!-\!\chi)}\cos\chi. \nonumber
\end{align}

For pulsars moving along the world-lines of arbitrary static Killing observers, not restricted to the equatorial plane, the precession frequency vector $\hat{\Omega}_p$, in the FS frame, does not lie along the $\hat{e}_3$ leg of their respective spatial FS tetrads. It is still a constant vector that can be expressed as a linear combination of their FS tetrad's $\hat{e}_1$ and $\hat{e}_3$ legs, as can be seen from (\ref{eq:Omega_p}). To obtain the equivalent expression for $\hat{B}$, one starts off with a coordinate transformation (denoted by $U_0$, say) that corresponds to a rotation change of axes which sets $\hat{\Omega}_p$ to $(0,0,1)$. One can then follow through the entire routine from (\ref{eq:U1}), finally ending with the inverse coordinate transformation $U_0^T$ to obtain the beam vector $\hat{B}(\tau)$. Heuristically, it is clear that the same divergence manifests itself for pulsars held fixed spatially off the equatorial plane and close to the ergosurface of a Kerr BH or NS, for Kerr static Killing observers.

\end{appendix}

\bsp	
\label{lastpage}
\end{document}